\documentclass[twocolumn,apj,numberedappendix,twocolappendix]{openjournalb}

\usepackage{amsfonts, amssymb,amsmath,bm}

\usepackage{xspace} 
\usepackage{booktabs}
\usepackage{graphicx}
\usepackage{color}
\usepackage{ulem}
\usepackage{multirow}
\usepackage[dvipsnames]{xcolor}

\usepackage{graphicx}	
\usepackage{amsmath}	

\DeclareRobustCommand{\VAN}[3]{#2}
\let\VANthebibliography\thebibliography
\def\thebibliography{\DeclareRobustCommand{\VAN}[3]{##3}\VANthebibliography}

\definecolor{darkraspberry}{rgb}{0.53, 0.15, 0.34}
\definecolor{radpink}{HTML}{ff3ce4}
\definecolor{darkblue}{rgb}{0.0, 0.0, 0.55}
\definecolor{dragonsbreath}{HTML}{cc0044}

\newcommand{\Quijote}{\textsc{Quijote} }
\newcommand{\Molino}{\textsc{Molino} }
\newcommand{\lin}{\mathrm{L}}
\newcommand{\nlin}{\mathrm{NL}}

\usepackage{hyperref}
\hypersetup{
    colorlinks=true,
    linkcolor=darkblue,
    citecolor= blue,   
    filecolor=blue,      
    urlcolor=magenta,
    pdftitle={Overleaf Example},
    pdfpagemode=FullScreen,
}

\begin{document}

\title{Cosmology on point: modelling spectroscopic tracer one-point statistics}
\author[0009-0003-0217-8566]{Beth McCarthy Gould$^{1,\star}$}
\author[0000-0002-2324-7335]{Lina Castiblanco$^{1,2}$}
\author[0000-0001-7831-1579]{Cora Uhlemann$^{2,1}$}
\author[0000-0001-6120-4988]{Oliver Friedrich$^{3,4}$}

\affiliation{$^{1}$ School of Mathematics, Statistics and Physics, Newcastle University, Herschel Building, NE1 7RU Newcastle-upon-Tyne, U.K.}
\affiliation{$^2$ Fakultät für Physik, Universität Bielefeld, Postfach 100131, 33501 Bielefeld, Germany}
\affiliation{$^{3}$ Universitäts-Sternwarte, Fakultät für Physik, Ludwig-Maximilians Universität München, Scheinerstr. 1, 81679 München, Germany}
\affiliation{$^4$ Excellence Cluster ORIGINS, Boltzmannstr.\ 2, 85748 Garching, Germany}

\thanks{$^\star$\href{mailto:b.gould2@newcastle.ac.uk}{b.gould2@newcastle.ac.uk}}




\begin{abstract}
	
{The 1-point matter density probability distribution function (PDF) captures some of the non-Gaussian information lost in standard 2-point statistics. The matter PDF can be well predicted at mildly non-linear scales using large deviations theory. This work extends those predictions to biased tracers like dark matter halos and the galaxies they host. We model the conditional PDF of tracer counts given matter density using a tracer bias and stochasticity model previously used for photometric data. We find accurate parametrisations for tracer bias with a smoothing scale-independent 2-parameter Gaussian Lagrangian bias model and a quadratic shot noise. We relate those bias and stochasticity parameters to the one for the power spectrum and tracer-matter covariances. We validate the model against the \Quijote suite of N-body simulations and find excellent agreement for both halo and galaxy density PDFs and their cosmology dependence. We demonstrate the constraining power of the tracer PDFs and their complementarity to power spectra through a Fisher forecast. We focus on the cosmological parameters $\sigma_8$ and $\Omega_m$ as well as linear bias parameters, finding that the strength of the tracer PDF lies in disentangling tracer bias from cosmology. Our results show promise for applications to spectroscopic clustering data when augmented with a redshift space distortion model}. 
\end{abstract}

\section{Introduction}


The cosmic large-scale structure contains valuable information for understanding the evolution and content of the Universe. Galaxy clustering is one of the primary probes for testing the standard cosmological model and inferring its parameters. The statistical properties of the galaxy distribution observed in spectroscopic  surveys are sensitive to the cosmological parameters such as the matter density, the amplitude of matter fluctuations and the dark energy equation of state. Current and upcoming observations from Stage-IV spectroscopic galaxy surveys as for example the Dark Energy Spectroscopic Instrument \citep[][DESI]{DESI:2016fyo} and Euclid \citep{EUCLID:2011zbd} will increase the statistical power. To realise the maximum potential from the observable data and achieve precise measurements of the cosmological parameters, accurate modelling of the statistics of the galaxy distribution is required. 

The galaxy 2-point correlation function and its Fourier transform, the galaxy power spectrum, are the standard tools for extracting cosmological information.  Accurate modeling of these statistics requires accounting for non-linear gravitational evolution, baryonic physics, and galaxy bias and stochasticity. Two-point predictions from N-body simulations can explore a wide range of scales but are computationally expensive, limiting their volume and number of realizations and thus sampling the parameter space requires building models \citep[such as \textsc{halofit},][]{Takahashi2012} or emulators \citep[see e.g.][]{EuclidEmulator2,MATRYOSHKA2022}. Alternatively, perturbation theory models like `Standard' Eulerian or Lagrangian Perturbation Theory \citep{Bernardeau2002review}, and their extension in terms of the Effective Field Theory of Large-Scale Structure \citep[EFTofLSS,][]{Carrasco:2012cv,Porto:2013qua} provide accurate predictions on mildly non-linear scales. This has enabled the full-shape analysis of two-point galaxy clustering from the SDSS-III BOSS data in both real and redshift space \citep{GilMarin2015,Alam2017,Beutler2017,Satpathy2017,Sanchez2017}. The one-loop EFTofLSS including a tracer bias expansion, redshift-space, and the damping of the Baryon-Acoustic Oscillation (BAO) peak has been applied to SDSS-III BOSS data, achieving precise estimates of cosmological parameters such as the matter fluctuation amplitude $\sigma_8$, the matter density fraction $\Omega_m$ and the Hubble constant $H_0$ \citep{DAmico:2019fhj,Ivanov:2019pdj,Colas:2019ret}. Including higher-order perturbative calculations could allow to probe slightly further into the nonlinear regime.

While 2-point statistics offer valuable insights into the Universe's density distribution, they do not capture the full non-Gaussian statistical properties of the observed galaxy distribution. To extract the maximum amount of information, additional statistical tools are required to incorporate higher-order correlations. Joint analyses with 2-point and 3-point statistics have demonstrated strong potential for extracting information from available observational data, as that provided from the BOSS survey \citep{DAmico:2022osl}. 


The one-point Probability Density Function (PDF) is a promising beyond two-point statistic to describe the variation of tracer counts in cells across the density distribution. The shape of the PDF depends on a whole series of integrals of higher-order $N$-point correlations, and thus carries significant non-Gaussian information. The underlying dark matter PDF is efficient in extracting information from mildly non-linear scales, improving the estimation of cosmological parameters \citep{uhlemann_fisher_2020}. The one-point PDF of finding $N$ objects in spherical cells of fixed radius $R$ is related to the $k$-nearest neighbour statistics \citep[kNN][]{Banerjee2021kNN}, which instead quote the cumulative probability that at fixed $k$, the $k$-th nearest neighbour is at a distance $r$. Simulation-based Fisher forecasts have shown that kNN statistics of matter and massive halos can yield at least a factor of two in constraining power compared to two-point statistics, in line with findings for the PDFs. The one-point matter PDF can be modelled analytically based on large-deviations theory \citep{BernardeauReimberg2015,Uhlemann2015}, and has demonstrated to provide accurate predictions for variations in cosmological parameters and neutrino masses \citep{uhlemann_fisher_2020}, primordial non-Gaussianity parameters \citep{Friedrich:2019byw,Coulton2023kNN_pNG} as well as dynamical dark energy and modified gravity \citep{Cataneo2022}. 

Accurately modelling the relationship between observable tracer and predictable dark matter densities is crucial for extracting cosmological information from galaxy clustering. To analyse spectroscopic galaxy survey data using the PDF and take advantage of theoretical predictions for the matter PDF, it is necessary to relate tracer to dark matter densities in cells. Going beyond phenomenological fits \citep[see e.g.][]{ColesJones1991,Bel2016,Clerkin2017,Hurtado_Gil_2017} requires simple yet accurate parametrisations for tracer bias and stochasticity. Simple linear bias and stochasticity models have been applied to extract information from one-point statistics in galaxy survey data. \cite{Friedrich2018,Gruen2018} constrained cosmology, linear galaxy bias and stochasticity and the matter density skewness from the signals of lensing-in-cells split by photometric galaxy counts,  showing that the parameters reproduce the photometric tracer PDFs from DES and SDSS.  \cite{repp_galaxy_2020} constrained linear galaxy bias and $\sigma_8$ from SDSS galaxy counts with a simplified bias and redshift-space distortion model along with Poisson shot noise. Leveraging the full power of Stage-IV surveys will require improved modelling of tracer bias and stochasticity which ideally should also be compatible with two-point statistics. Weak lensing is complementary to galaxy clustering and the one-point statistics of the convergence, aperture mass and wavelet $\ell_1$-norm have been modelled in \cite{Barthelemy_2020,Thiele_2020,Barthelemy_2021,boyle_nuw_2020,Barthelemy_2024,Castiblanco_2024,sreekanth_2024}.

In this work, we extend the matter PDF predictions to include biased tracers. The goal of our study is to use the predictions of the matter PDF together with a simple and accurate model for bias and stochasticity suitable for analyzing spectroscopic survey data. 

In Section \ref{ssec:Quijote} we detail the method for extracting  PDFs from the \Quijote simulations.
In Section~\ref{sec:matterPDF} we briefly review how to obtain predictions for the matter PDF from large-deviations theory. In Section~\ref{sec:bias_stochasticity} we describe our bias and stochasticity model for the tracer PDF and its link to the power spectrum. In Section~\ref{sec:tracerPDF_cosmology} we validate predictions for the cosmology-dependence of the tracer PDF with simulations and perform a Fisher forecast to assess the constraining power of the tracer PDF compared to the power spectrum. In Section~\ref{sec:conclusions} we summarise our results and provide an outlook on future work. Appendix~\ref{app:tracer_matter} discusses the tracer-matter connection link between the PDF and the power spectrum. Appendix~\ref{app:validations} presents further validations of our theoretical models and forecasts.

\section{Extracting statistics from the Quijote Simulations}
\label{ssec:Quijote}
In this section, we describe the  \Quijote simulation suite and detail the method for measuring the matter and tracer PDFs, which we will use to refine and validate our theoretical model. The \Quijote N-body simulation suite \citep{villaescusa-navarro_quijote_2019} contains 15,000 independent N-body simulations at a fiducial  $\Lambda$CDM cosmology with the parameters: $\sigma_8=0.834$, $\Omega_{\rm m}=0.3175$, $\Omega_{\rm b}=0.049$, $n_s=0.9624$, $h=0.6711$. Each simulation box has a volume of 1 $(h^{-1}~\textrm{Gpc})^3$ and traces $512^{3}$ CDM particles. 
Snapshots are available at five redshifts and we focus on the lowest three $z=0.0,0.5,1.0$ relevant for galaxy surveys. Part of the \Quijote simulations were designed with Fisher forecasts in mind, and therefore offer results for a large number of single-parameter variations on their fiducial models. A set of 500 simulations are available for each of the cosmological parameters with the following increments/decrements with respect to the fiducial values $\Delta\sigma_8=\pm 0.015$, $\Delta\Omega_{\rm m}=\pm 0.010$, $\Delta\Omega_{\rm b}=\pm 0.002$, $\Delta n_s=\pm 0.020$, $\Delta h= \pm0.020$. 

\subsection{Matter PDFs from the \Quijote Simulations}
Matter PDFs in 99 logarithmic bins from $1+\delta=10^{-2}$ to $10^2$ are pre-computed using the publicly-available \href{https://github.com/franciscovillaescusa/Pylians3}{\textsc{Pylians3}} python library. They rely on a Cloud-in-Cell mass-assignment scheme to deposit the particles on a grid of $512$ and $1024$ cells per side, respectively. The matter density field is then smoothed with a spherical top-hat filter of radius $R=20,25,30$Mpc$/h$ through a multiplication in Fourier space.

\subsection{Halo statistics from the \Quijote Simulations}
\label{ssec:QuijoteTracers}
We make use of the \Quijote simulations to parameterise the tracer bias and stochasticity required to model the tracer PDF given a matter PDF.

We use Friends-of-Friends \citep[FoF][]{davis_evolution_1985} halo catalogues and extract our statistics for the $N_h^\mathrm{tot}$ most massive halos for each realisation. The numbers are chosen to be close to the maximum possible in a given halo catalogue (and thus varying across redshifts) to limit the impact of shot noise. For the fiducial cosmology, the lowest halo mass bin considered is $1.44\times10^{13}M_\odot/h$, while the highest halo masses are $\{7.9,4.3,2.7\}\times 10^{15} M_\odot/h$ for redshifts $z=\{0,0.5,1\}$. Given the size of the simulation box $1$ Gpc$/h$, this corresponds to number densities of $n_\mathrm{tot}(z)\sim (1.5-3.9) \times 10^{-4} (h/$Mpc$)^3$ for $z\in[0,1]$. Shot noise becomes significant if the average tracer count in cells approaches $\overline{N_t}(z,R)\sim 10$, as already happens for $z=1$ and a radius of $R=25$ Mpc$/h$, and more pronounced for $R=20$Mpc$/h$ where $\overline{N_t}(z,R)\sim 5$. We picked a fixed number of halos instead of a fixed mass threshold in order to avoid fluctuations of number density across different realisations. 
For isolating the change of the halo PDF with respect to cosmology, we select a number of halos tuned to leave the halo bias as constant as possible across cosmologies, with values displayed in Table~\ref{tab:Ntot}. This amounts to changing the minimum halo mass threshold with cosmology to compensate the change in bias due to a larger/smaller number of massive halos, most prominent when changing $\sigma_8$.

\begingroup
\renewcommand{\arraystretch}{1.3} 
\begin{table}
  \centering
    \begin{tabular}{c|c||c||c|c||c|c}
    \hline
    tracer & $z$ & fiducial & $\sigma_8^{-}$ &  $\sigma_8^{+}$ & $\Omega_m^{-}$ & $\Omega_m^{+}$\\  \hline \hline
    halos & $0.0$ & 358364 & 390000 & 329930 & 361020 &  355876\\ \hline
    halos & $0.5$ & 275253 & 300000 & 252965 & 269741 & 276134\\ \hline
    halos & $1.0$ & 165107 & 180000 & 151694 & 162795 & 167293\\ \hline \hline
    galaxies & $0.0$ & 156800 & 156316 & 157189 & 151641 & 162033 
    \end{tabular}
  \caption{Number of tracers $N_t$ selected to leave bias constant across cosmology. }
  \label{tab:Ntot}
\end{table}
\endgroup

The processing of the halo catalogues to extract PDFs makes use of the same functions in the \href{https://github.com/franciscovillaescusa/Pylians3}{\textsc{Pylians3}} library as for the matter PDF. We use the Cloud-in-Cell mass-assignment scheme to deposit the halos on a grid of 500 cells per side. We choose to extract number-weighted PDFs (as opposed to weighting the tracers by their masses) because it most closely corresponds to observable galaxy counts. While a mass-weighting is known to increase the correlation between the matter and tracer densities  \citep[see e.g.][]{Seljak2009,Hamaus2010,Jee2012,uhlemann_question_2018}, our simulation contains only relatively massive halos such that we expect this impact to be small. As number counts are integers, number weighting provides us with an unambiguous binning for the tracer PDF. The halo density field is smoothed with a spherical top-hat filter of radius $R=20,25,30$Mpc$/h$ through a multiplication in harmonic space. The binning for the tracer PDF is adapted so that the bins correspond to multiples of the tracer density contributed by a single tracer in a sphere.\footnote{Note that the Cloud-in-Cell (CiC) mass assignment scheme can lead to non-integer tracer counts. We compute the histogram $\mathcal P_t(N_t)$ in bins centered on integer counts $N_i$ with edges at $N_i\pm 0.5$. We find the difference between tracer counts from CiC and a nearest-grid-point (NGP) assignment to be small and opt to keep CiC to treat matter and tracers with the same mass assignment.} 

\subsection{Mock galaxy statistics from the \Molino Suite}

We also use the \Molino suite of galaxy catalogues \citep{Hahn:2020lou} suitable for Fisher forecasts. Molino uses the standard Halo Occupation Distribution (HOD) model \citep{zheng_galaxy_2007} to populate the \Quijote dark matter halo catalogues. The galaxy catalogue contains $\sim 150000$ galaxies and is available for redshift $z=0.0$. The halos are populated according to the probability of a halo of mass $M_h$ to host $N_g$ number of galaxies. Halos are occupied by central and satellite galaxies, in the standard HOD model the mean number of galaxies is given by their sum
\begin{subequations}
\label{eq:HOD}    
\begin{equation}
\label{eq:HOD_total}
    \langle N_g (M_h)\rangle = \langle N_{\rm cen}(M_h)\rangle + \langle N_{\rm sat}(M_h)\rangle \,.
\end{equation}
The mean occupation of central galaxies is parameterised by
\begin{equation}
\label{eq:HOD_cental}
    \langle N_{\rm cen}(M_h)\rangle  = \frac{1}{2} \left[1 - {\rm erf}\left(\frac{\log (M_h/M^{\rm min}_{h})}{\sigma_{\log M }}\right)\right]\,,
\end{equation}
where $M^{\rm min}_{h}$ is the minimum mass for which half of the halos host a central galaxy above the luminosity threshold and $\sigma_{\log M}$ is related to the scatter of the central galaxy luminosity in halos of mass $M_h$. The mean occupation of satellites follows a power law as
\begin{equation}
\label{eq:HOD_satellite}
    \langle N_{\rm sat}(M_h)\rangle  =  \langle N_{\rm cen}(M_h)\rangle \left(\frac{M_h - M_0}{M_1}\right)^{\alpha}\,.
\end{equation}
\end{subequations}
with $M_0$ is the halo mass cut-off for satellite occupation, $M_1$ is given such that $M_h = M_0 + M_1$ is the typical mass scale for halo to host one satellite and $\alpha$ is the slope at high halo mass.  
In the \Molino suite the fiducial HOD parameters are based on the best-fit for the high luminosity galaxies in the Sloan Digital Sky Survey (SDSS) and are given by $\log (h M_{\rm min}/M_\odot) = 13.65 $, $\sigma_{\log M}=0.2$, $\log (h M_0/M_\odot) = 14.0 $, $\alpha = 1.1$ and $\log (h M_1/M_\odot) = 14.0$.  We follow the same procedure explained above for halos to extract the \Molino galaxy PDFs for radii $R=25,30$ Mpc$/h$. The most massive galaxy sample has a mass range from $3.33 \times 10^{13}$ to $4.39 \times 10^{15}M_\odot/h$. We use 1000 realisations at the fiducial cosmology and 500 realisations for changes in parameters.

\section{Matter PDF from Large Deviations Theory}
\label{sec:matterPDF}
Large deviations theory provides a theoretical framework for the calculation of the matter PDF smoothed with a spherical top hat filter in the mildly non-linear regime.

The theory of large deviations reconstructs the probability distribution based on the decay rate of the probabilities with deviations from the mean as some characteristic parameter of the system decreases rapidly to zero \citep{BernardeauReimberg2015}. A probability density function (PDF)  $\mathcal {P}(\delta_L)$ of some random variable $\delta_L$ can be said to satisfy a \textit{large deviation principle} (LDP) if the limit
\begin{equation}
\label{eq:rate_function}
	-\lim_{\epsilon \to 0}\epsilon\ln{\mathcal P}(\delta_L) = \psi(\delta_L)
\end{equation}
exists, where the \textit{rate function} $\psi(\delta_L)$ is the leading order term of the log of the density PDF and $\epsilon$ is the driving parameter. The exact form of the PDF can be written as
\begin{equation}
	{\mathcal P}(\delta_L) = \exp\Big[-\frac{\psi(\delta_L)}{\epsilon}+\mathcal{O}(\epsilon)\Big]
\end{equation}
with $\mathcal{O}(\epsilon)$ encapsulating all terms above linear in $\epsilon$.

Within the context of 1-point statistics of the cosmic large-scale structure, our random variable comes from smoothing the relative matter density contrast $\delta$ on some scale. In the case of Gaussian initial conditions, the PDF of the linear matter density contrast $\delta_\lin$ in spheres of radius $r$ is fully specified by a rate function $\psi_L(\delta_L)=\delta_L^2/2$ and the linear variance at that scale defined by
\begin{equation}
    \label{eq:linvar}
\sigma_\lin^2(r,z)=\int\frac{\mathrm{d}^3k}{(2\pi)^3}P_\lin(k,z)W_\mathrm{3D}(kr)^2\,,
\end{equation}
where $W_\mathrm{3D}(kr)$ is the spherical top-hat kernel in Fourier space and $P_\lin(k,z)$ is the linear power spectrum at redshift $z$.
The Gaussian PDF can be written as 
\begin{equation}
     {\mathcal{P}_{\rm L}}(\delta_L) = \frac{1}{\sqrt{2\pi\sigma^2_L(r,z)}}\exp\left[-\frac{\psi_{\rm L}(\delta_L)}{\sigma_L^2(r,z)}\right] .
\end{equation} 
We can see that the PDF is exponentially decreasing with a decay-rate function $\Psi_L(\delta_L;r,z)=\psi_{\rm L}(\delta_L)/\sigma_L^2(r,z)$ that is the ratio of the rate function and the variance.

For the nonlinear density contrast $\delta_m$ in spheres of radius $R$ we can find an LDP with a rate function 
\begin{equation}
\label{eq:rate_function_delta}
	\psi_m(\delta_m) = -\lim_{\sigma^2\to 0}\sigma^2\log\mathcal{P}_m(\delta_m) \,.
\end{equation} 
This is in the limit of a vanishing non-linear variance $\sigma^2=\sigma^2_{\rm NL}(R,z)$, which is defined by the non-linear power spectrum in analogy to Equation~\ref{eq:linvar}. A consequence of a LDP is the \textit{contraction principle} which allows us to compute the rate function in terms of that of a different random variable via
\begin{equation}
\label{eq:contraction_principle}
	\psi_{m}(\delta_m) = \inf_{\delta_L:\zeta(\delta_L)=\delta_m} \psi_{L}(\delta_L) 
\end{equation}
where $\zeta$ is some continuous mapping. Due to the exponential decay, there will be one dominant, most likely contribution.  

The most probable evolution of densities in spheres can be approximated by spherical collapse such that $\zeta(\delta_m)=\delta_\mathrm{L}^\mathrm{SC}(\delta_m)$, where $\delta_{\rm L}$ and $\delta_m$ are the linear and nonlinear density contrast with initial and final radii related by $r=R(1+\delta_m)^{1/3}$ via mass conservation. For the late-time rate function we have that
\begin{equation}
\label{eq:rate_function_nonlinear}
    \psi_m(\delta_m)=\frac{\sigma_{\rm L}^2(R,z)}{\sigma_{\rm L}^2(R(1+\delta_m)^{1/3},z)} \psi_{\rm L}(\delta_L^{\rm SC}(\delta_m))
\end{equation}
At small variances $\sigma^2< 1$, the decay rate function can be robustly extrapolated from the rate function by restoring the variance like $\Psi_m(\delta_m;R,z)=\psi_m(\delta_m)/\sigma^2_{\rm NL}(R,z)$ \citep{Uhlemann2015,BernardeauReimberg2015}. 
The PDF of the late-time matter density then has exponential behaviour described by
\begin{equation}
\label{eq:matterPDF_exp}
    \mathcal{P}_m(\delta_m) \sim \exp\left[-\frac{\psi_m(\delta_m)}{\sigma_\nlin^2(R,z)}\right] 
           = \exp\big[-\Psi_m(\delta_m;R,z)\big]\,.
\end{equation}
The prefactor of this expression can be calculated via the cumulant generating function. The scaled cumulant generating function $\varphi_m(\lambda)$ (encoding the reduced cumulants) is obtained from the Legendre-Fenchel transform of the rate function
\begin{equation}
\label{eq:scgf}
    \varphi_m(\lambda)=\sup_{\delta_m}\big[\lambda \delta_m-\psi{_m}(\delta_m)\big]\,,
\end{equation}
which is then converted to the cumulant generating function by restoring the nonlinear variance
\begin{equation}
\label{eq:cgf}
     \phi_m(\lambda;R, z) = \frac{1}{\sigma_{\rm NL}^2(R,z)}\  \varphi_m\left(\lambda\ \sigma_{\rm NL}^2(R,z);R, z\right)\,. 
\end{equation}
The matter PDF $\mathcal{P}_m(\delta_m)$ is then obtained from an inverse Laplace transform of the cumulant generating function
\begin{equation}
\label{eq:matterPDF_as_Laplace_transform}
\mathcal P_m(\delta_m; R, z)\!=\!\!\int_{-\infty}^{\infty} \frac{\text{d}\lambda}{2\pi}\exp\left[\phi_m(i\lambda; R, z)-i\lambda\delta_m\right]\ .
\end{equation}

It can be approximated well by a suitable saddle point approximation~\citep{Uhlemann2015}, but we do not rely on this here. Theoretical predictions used in this work are computed with the above theory using the publicly available code CosMomentum\footnote{\href{https://github.com/OliverFHD/CosMomentum}{https://github.com/OliverFHD/CosMomentum}} \citep{Friedrich:2019}. While spherical collapse dynamics effectively predict the nonlinear rate function~\eqref{eq:rate_function_nonlinear} and the generating function of reduced cumulants~\eqref{eq:scgf}, the non-linear variance entering equations~\eqref{eq:matterPDF_exp}~and~\eqref{eq:cgf} cannot be accurately inferred. CosMomentum instead predicts this using the non-linear power spectrum $P_\nlin$ from the revised \textsc{halofit} fitting function from \cite{Takahashi2012} building upon the original model \citep{Smith2003halofit}. The \textsc{halofit} model uses a functional form for the matter power spectrum based on the halo model with parameters fit to N-body simulations. For an alternative approach to computing the PDF with a perturbation theory around spherical collapse following principles of Effective Field Theory see \cite{Ivanov_2019,chudaykin2023}.

\begin{figure}
	\centering
	\includegraphics[width=\columnwidth]{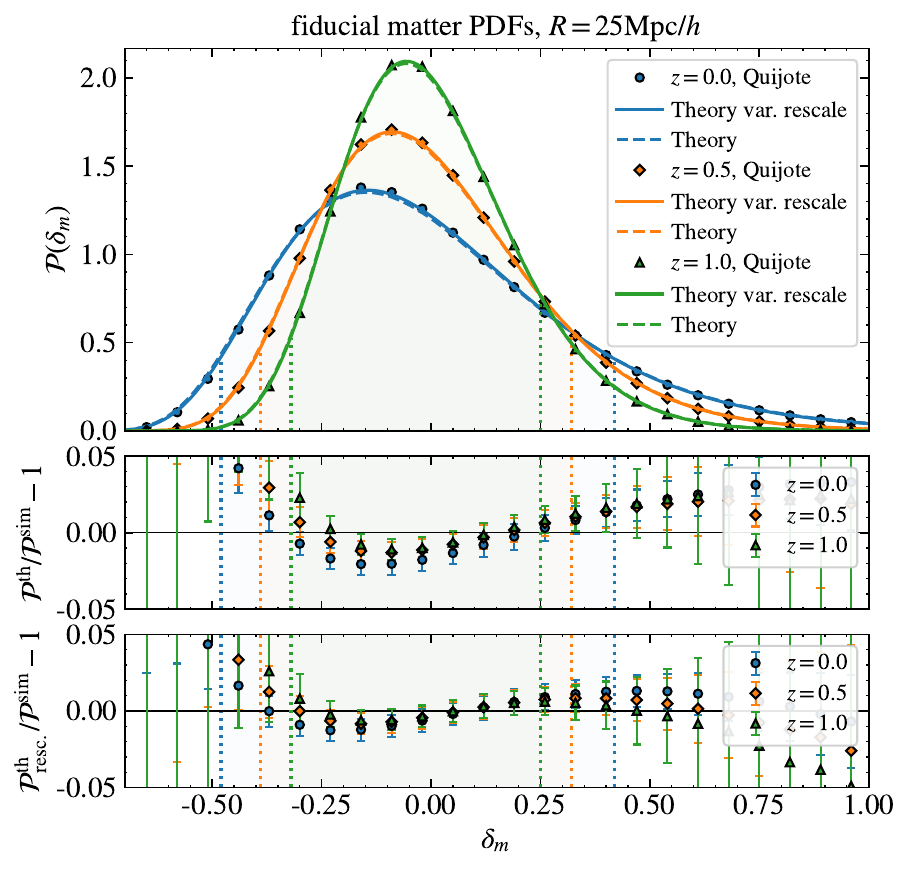}
	\caption{Comparison between matter PDFs extracted from \Quijote (data points) with those from theory at smoothing scale $R=25.0$Mpc/$h$ for redshifts $z=0.0,0.5,1.0$. Theory predictions are shown using the original \textsc{halofit} variance (dashed lines) and with the non-linear variance rescaling (solid lines), with corresponding residuals in the middle and lower panels. Errors on the residuals in the lower panels are from the standard deviation of 100 realisations of the \Quijote fiducial cosmology.} 
	\label{fig:Quijote_CosM_compare_m}
\end{figure}

Figure~\ref{fig:Quijote_CosM_compare_m} compares matter density PDFs at smoothing scale $R=25$Mpc$/h$ and redshifts $z=0.0,0.5$ and 1.0 extracted from \Quijote to those computed using CosMomentum. Due to the inaccuracies in predicting the non-linear matter variance in the CosMomentum code, here we rescale $\sigma_\mathrm{NL}^2$ with the measured value from \Quijote. The figure shows theory lines and residuals both with and without this rescaling. We find that this rescaling ratio is constant across cosmologies, so we use the fiducial ratio for all theoretical predictions in this work. The errors shown in the lower two panels are calculated from the standard deviation across 100 realisations of the \Quijote fiducial cosmology. In later sections we will focus our analysis on the bulk of the PDF and thus exclude the highest 10\% and lowest \{10,5,3\}\% of density bins for scales $R=20,25,30$Mpc/$h$ respectively, similar to \citep{uhlemann_fisher_2020}.  The predicted PDFs agree with the measured ones to within 2\% around the bulk of the PDF. For a validation of the matter PDF derivatives with respect to cosmological parameters see Appendix~\ref{app:matterPDF}.

\section{Parameterising tracer bias and stochasticity}
\label{sec:bias_stochasticity}

Various studies have aimed to develop accurate and simple models to capture the one-point tracer-matter density relationship. For example, \cite{uhlemann_question_2018,Uhlemann2018cylinders} considered mass-weighted subhalo densities and demonstrated that abundance matching using a quadratic mean bias model for log-densities is sufficient to obtain accurate PDFs for tracers in spheres and cylinders. However, neglecting the scatter between galaxy and matter densities fails to capture the stochastic nature of this relationship, which could lead to inaccuracies in more realistic scenarios. The first approach to modelling the non-Poissonian stochasticity between the galaxy field and the matter density field within the PDF framework was presented by \cite{Friedrich2018,Gruen2018} for projected densities. They explored two models to describe shot noise in conjunction with a linear bias: one with a free parameter that encodes the correlation between the matter and tracer fields, and another using a generalized Poisson distribution with two parameters. Building on this, \cite{Friedrich2022} proposed a quadratic Lagrangian bias expansion for photometric galaxy clustering. They showed that at fixed order the Lagrangian model provides a better fit for the conditional mean than the Eulerian bias expansion. The authors validated their Lagrangian bias expansion against standard consistency relations between Eulerian and Lagrangian perspectives, confirming that their approach is robust and consistent with established two-point statistics. Their analysis also confirms that shot noise deviates from the expected Poisson distribution. The current approach for modelling tracer kNN statistics uses Hybrid Effective Field Theory that combines a perturbative Lagrangian bias model with $N$-body dynamics for the displacements of dark matter and tracers \citep{Banerjee2022kNN-HEFT}. Here we will take advantage of theoretical predictions for the matter PDF and augment them with a parameterisation of the conditional tracer given matter density PDF relying on suitable bias and stochasticity models.

\begin{figure}
    \centering
    \includegraphics[width=\columnwidth]{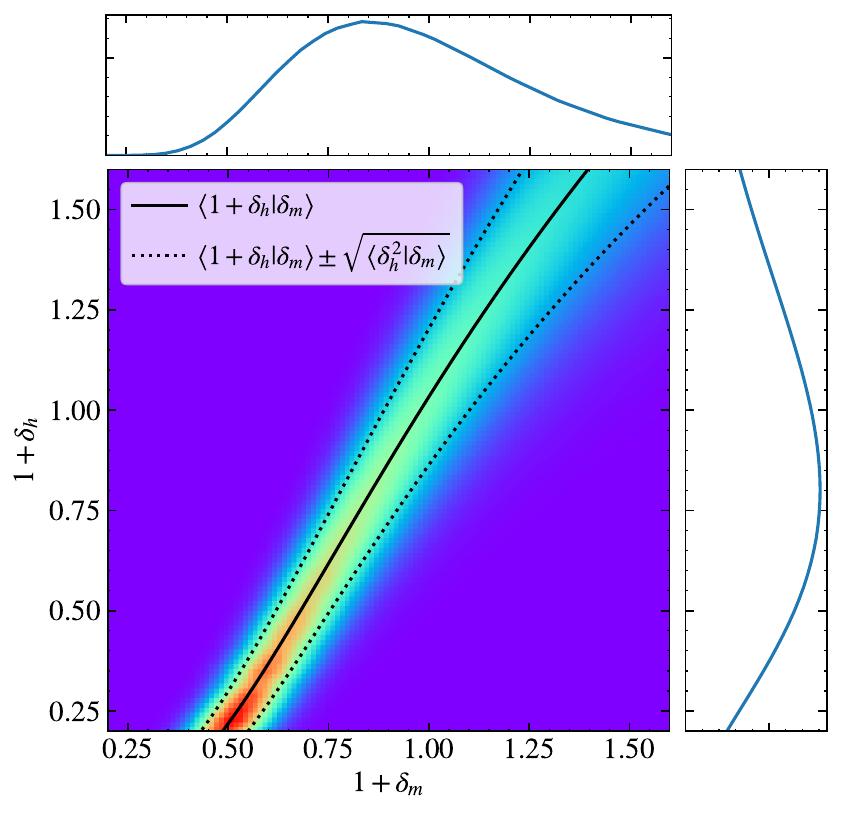}
    \caption{Conditional PDF $\mathcal P(\delta_h|\delta_m)$ using 8000 realisations of the \Quijote fiducial cosmology at redshift $z=0.0$ and smoothing scale $R=25$Mpc$/h$. The black line shows the conditional mean roughly following the `ridge' of the conditional PDF, while there is also some scatter around the conditional mean, i.e. shot noise (black dashed lines).}
    \label{fig:jointPDF}
\end{figure}

\subsection{Conditional tracer given matter density PDF}
Having predicted the matter PDF, we have the first ingredient for the joint PDF of tracer and matter densities, which we can write as a product of the conditional PDF of tracer counts given matter density and the matter PDF
\begin{equation}
    \label{eq:tracer_matter_jointPDF}
    \mathcal P(N_t,\delta_m)=\mathcal P(N_t|\delta_m) \mathcal P_m(\delta_m)\,.
\end{equation}
A tracer density contrast can be defined through $\delta_t=N_t/\bar N_t -1$  with the mean number of tracers per cell $\bar N_t$. Then we can write $\mathcal P(N_t|\delta_m)=\mathcal P(\delta_t|\delta_m)/\bar N_t$.
In Figure~\ref{fig:jointPDF} we show the conditional PDF $\mathcal{P}(\delta_h|\delta_m)$ of halo densities in spheres given a certain matter density in spheres. This is extracted from 8000 realisations of the \Quijote fiducial cosmology. We clearly see that there is a strong trend and correlation between matter and tracer densities in cells (with correlation coefficients of around $0.9$, see Appendix~\ref{app:crosscov} for details), but also some scatter. We describe this conditional PDF with two ingredients, the conditional mean $\overline{N_t}(\delta_m)=\langle N_t|\delta_m\rangle$ and the conditional variance $\langle N_t^2|\delta_m\rangle_c$ of tracer counts at fixed matter density contrast. The conditional mean -- shown by the solid black line -- follows the `ridge' of the joint PDF, while the conditional variance captures the scatter around the expectation value -- indicated by the dotted lines. For a Poisson distribution, the conditional variance agrees with the conditional mean, and as such we express the model in terms of the ratio $\alpha(\delta_m)=\langle N_{\rm t}^2|\delta_m\rangle_c/\langle N_{\rm t}|\delta_m\rangle$.
This model was introduced for density-split statistics in \cite{Friedrich2018,Gruen2018}, generalised in \cite{Friedrich2022} and used for studying HODs in \cite{Britt2024}. We apply it to 3D densities for the first time.  
The conditional distribution of galaxy counts at fixed matter density with a bias model $\overline{N_{\rm t}}(\delta_m)=\bar N_{\rm t} [1+\delta_t(\delta_m)]$ and a shot noise model $\alpha(\delta_m)$ is given as equation~(23) in \cite{Friedrich:2019}
\begin{align}
    \label{eq:condPDF}
\mathcal{P}(N_{\rm t}|\delta_{\rm m})&=\frac{1}{\alpha(\delta_{\rm m})}\exp\left(-\frac{\overline{N}_{\rm t}(\delta_{\rm m})}{\alpha(\delta_{\rm m})}\right)\\
&\times\left[\Gamma\left(\frac{N_{\rm t}}{\alpha(\delta_\mathrm{m})}+1\right)\right]^{-1}\left(\frac{\overline{N}_{\rm t}(\delta_{\rm m})}{\alpha(\delta_{\rm m})}\right)^{\frac{N_{\rm t}}{\alpha(\delta_{\rm m})}}\notag\,.
\end{align}
This can be viewed as a remapping of the continuation of the discrete Poisson distribution where $\alpha=1$
\begin{equation}
\mathcal P(N_t|\delta_m) = \mathcal P_{\rm P}\left(\tilde N_t=\!\frac{N_t}{\alpha(\delta_m)};\overline{\tilde N_t}=\frac{\overline{N_t}(\delta_m)}{\alpha(\delta_m)}\right) \frac{1}{\alpha(\delta_m)}\,,    
\end{equation}
where the normalisation $\alpha(\delta_m)^{-1}$ comes from the Jacobian $d\tilde N_t/dN_t$ and the Gamma function is the generalisation of the factorial to non-integers. This distribution produces the input conditional mean $\langle N_t|\delta_m\rangle=\overline{N_{\rm t}}(\delta_m)$ and the conditional variance $\langle N_{\rm t}^2|\delta_m\rangle_c = \alpha(\delta_m) \overline{N_{\rm t}}(\delta_m)$. In the limit of fine sampling $\bar N_t\rightarrow \infty$, this is well approximated by a Gaussian of mean $\overline{N_t}(\delta_m)$ and variance $\alpha(\delta_m)\overline{N_t}(\delta_m)$ as we show in more detail in Appendix~\ref{sec:fine_sampling}.

In the spectroscopic case, the relevant observable is the tracer PDF $\mathcal P(N_t)$, which is obtained as a marginal of this joint PDF by integrating over the matter densities
\begin{equation}
\label{eq:tracerPDF}
    \mathcal{P}_t(N_t)=\int \mathcal P(N_t|\delta_m) \mathcal P_m(\delta_m) d\delta_m\,.
\end{equation} In the photometric case, the joint one-point PDF of finding $N$ tracers and a matter overdensity $\delta_m$ in cylindrical cells \citep{Friedrich2022} can be translated to an observable joint PDF between the tracer count and the weak lensing convergence. The joint PDF is also related to the corresponding cross-correlation $(k_1,k_2)$-NN statistics \citep{Banerjee2021kNN-cross}, which have been extended to the correlations of tracers with a continuous field in \cite{Banerjee2023kNN-tracerfield}.

\subsection{Conditional mean bias model}
\label{subsec:condmean}

\begin{figure}
	\centering
	\includegraphics[width=\columnwidth]{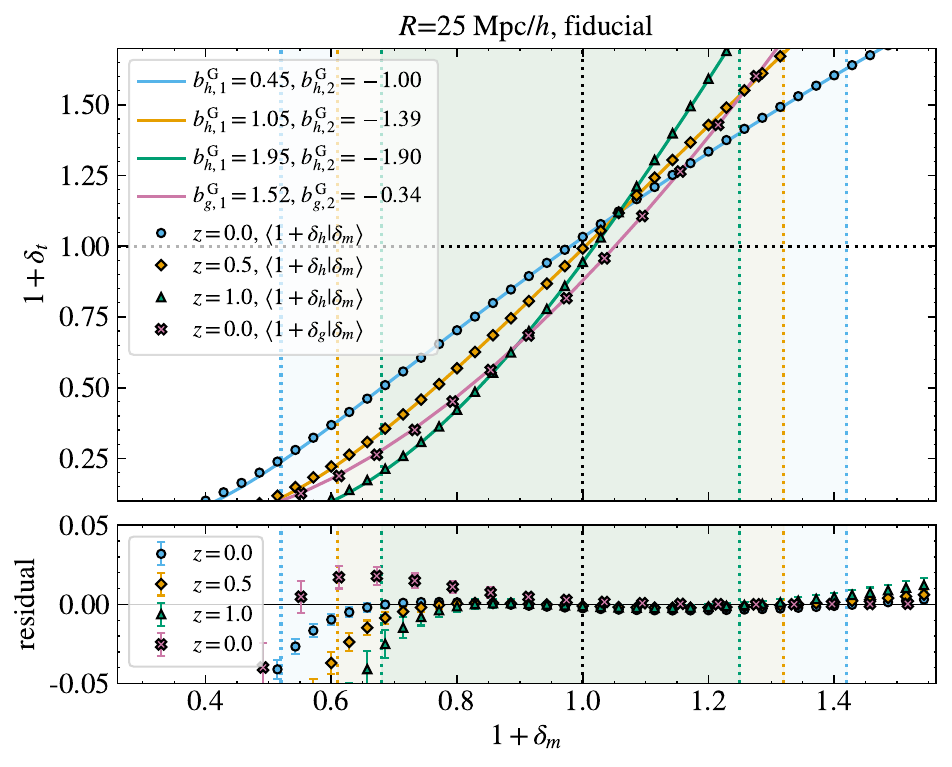}
	\caption{(Upper panel) Conditional mean $\langle\delta_t|\delta_m\rangle$ from \Quijote matter and halos and \Molino galaxies in spheres of radius $R$ (data points) along with a renormalised Gaussian Lagrangian bias fit~\eqref{eq:bias_L_Gauss} of $b_1^\mathrm{G}$ and $b_2^\mathrm{G}$ (solid lines).  (Lower panel) Residual defined as $(1+\langle\delta_t|\delta_m\rangle_{\rm fit})/(1+\langle\delta_t|\delta_m\rangle_{\rm sim})-1$ with error bars indicating the standard deviation across 500 realisations.
    }
	\label{fig:bias_fit_condmean}
\end{figure}

The conditional mean encodes a local tracer bias model $\langle N_t|\delta_m\rangle=\bar N_t (1+\langle \delta_t|\delta_m\rangle)$. Figure~\ref{fig:bias_fit_condmean} shows the conditional mean calculated from 500 realisations of the \Quijote and \Molino fiducial cosmology at redshifts $z=0.0,0.5,1.0$ and $z=0.0$ respectively, and smoothing scale $R=25$Mpc/$h$ (data points). To perform cosmological inference, a simple yet accurate tracer bias parameterisation is desirable. We use local bias models for densities in cells (with a long history including works by \cite{FryGaztanage93,Manera2011MNRAS,Salvador_2018,repp_galaxy_2020} and reviews in \cite{Bernardeau2002review,DesjacquesJeongSchmidt18} ) and their combination with Lagrangian bias principles described in \cite{Friedrich2022}. 

Of course, tracer formation is in principle non-local, e.g. halos of mass $M$ are expected to form at peaks of the initial density field smoothed over the Lagrangian size $R_*(M)=(3M/4\pi)^{1/3}$ \citep{DesjacquesJeongSchmidt18}. Our tracer selection contains a wide range of halo masses with Lagrangian radii between $1.5$ and $15-25$ Mpc$/h$ (decreasing with increasing redshift) with a mean of around 3 Mpc$/h$. Hence, we can expect the bias to be close to local on our smoothing radii of $R=20-30$ Mpc$/h$.

\subsubsection{Eulerian bias models}
In a quadratic Eulerian bias model, one would parameterise the tracer bias in terms of the two Eulerian bias parameters $b_1^E$ and $b_2^E$, 
\begin{equation}
\label{eq:bias_E_quad}
\langle\delta_h|\delta_m\rangle = b_1^E\delta_m +  \frac{b^E_2}{2}(\delta_m^2-\sigma_m^2) ,
\end{equation}
where $\delta_{h/m}$ are measured in the spherical cells and $\sigma^2_m$ is the matter variance at the same scale that ensures $\langle\delta_h\rangle=0$. The bias generally becomes more linear with increasing scale and the linear Eulerian bias $b_1^E$ approaches the linear bias obtained from the large-scale power spectrum on scales of about $R\sim 60$ Mpc/h \citep{Manera2011MNRAS,Friedrich2022}. We find this model to be instructive for a qualitative understanding of the impact of nonlinear bias, but insufficient for a percent-level description of the conditional mean bias.

The Sheth-Mo-Tormen (hereafter SMT) model \citep{SMT_2001} uses a mass function calculated from an extension to the Press-Schechter excursion set approach \citep{PressSchechter1974} generalised using ellipsoidal collapse equations and fitted to numerical simulations. This allows for statistical predictions of the bias for halos and HOD galaxies. For halos, one can use the SMT model to predict the Eulerian bias parameters $b_n^{\rm E}$ as a function of halo mass $M_h$. We can then obtain bias parameters for our halo selection from an average over all selected masses weighted by their probability $\mathcal P(M_h)$ via
\begin{equation}
\label{eq:bias_SMT}
    \bar b_{{\rm h},n}= \int \mathrm{d}M_h b_n^{\rm E}(M_h) \mathcal P(M_h) \,.
\end{equation}

In our case, we determine $ \mathcal P(M_h)$ from the halo mass function measured in \Quijote considering a fixed number of the most massive halos, such that the first halo mass bin $M^{\rm min}_{h}$ is re-weighted according to the leftover number of halos.

\begin{figure}
	\centering
	\includegraphics[width=\columnwidth]{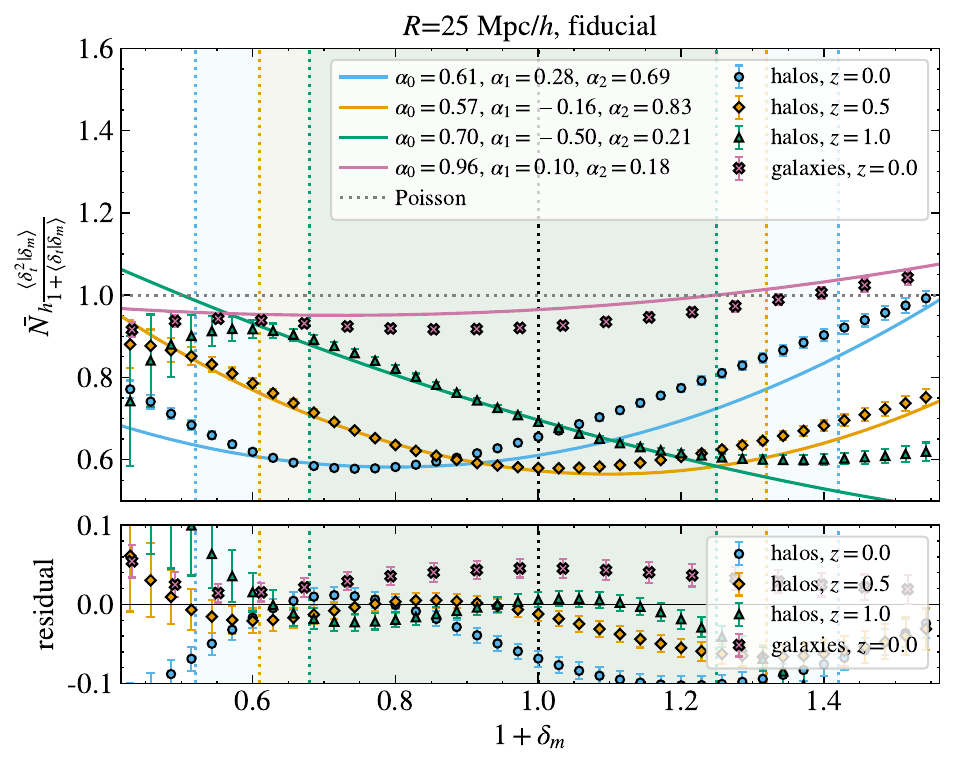}
   \caption{Quadratic shot noise~\eqref{eq:SN_quad} fitted to the ratio of the conditional variance to the conditional mean of the halo counts in the shaded regions corresponding to each redshift, and the error bars are calculated from the standard deviation across 500 realisations of the fiducial cosmology.}
	\label{fig:SN_fit_condmean}
\end{figure}

\begingroup
\renewcommand{\arraystretch}{1.3} 
\begin{table}
	\centering
	\begin{tabular}{c||c||c|c||c|c}
		\hline
		tracer & $z$ & $b_1^\mathrm{E,fit}$ & $\bar{b}_1^\mathrm{SMT}$ & $b_2^\mathrm{E,fit}$ & $\bar{b}_2^\mathrm{SMT}$ \\
        \hline \hline
		& 0.0 & 1.56 & 1.54 & -0.56 & -0.28 \\
		halos & 0.5 & 2.11 & 2.08 & 0.38 & 0.37 \\
		& 1.0 & 2.88 & 2.84 & 3.11 & 2.21 \\
		\hline
		galaxies & 0.0 & 2.24 & 2.44 & 2.29 & 2.48 
	\end{tabular}
    \caption{Eulerian bias predictions using SMT model and the \Quijote halo mass function~\eqref{eq:bias_SMT} using the most massive \{358364,275253,165107\} halos for redshift $z=0,0.5,1$ for halos and $\{156800\}$ galaxies at redshift $z=0.0$ in comparison to Eulerian bias fits of the conditional mean~\eqref{eq:bias_E_quad} for spheres of radius $R=25$Mpc$/h$.}
    \label{tab:SMT}
\end{table}
\endgroup

The bias for HOD galaxies is obtained from a re-weighting given by \cite{zheng_galaxy_2007}
\begin{equation}
\label{eq:bias_SMT_HOD}
    \bar{b}_{{\rm g}, n} = \frac{\int  \, dM_{h} \langle N_g (M_h)\rangle  b_n^{\rm E}(M_h)\mathcal P(M_h)}{\int \, dM_{h}\langle N_g (M_h)\rangle \mathcal P(M_h)}\,,
\end{equation}
where the expected number of galaxies per halo mass $\langle N_g (M_h)\rangle$ is determined by the HOD~\eqref{eq:HOD}. Note that the SMT model is insufficiently precise for real data and is only used here for consistency checks of our analysis.
Table~\ref{tab:SMT} shows these Eulerian bias predictions for halos and galaxies using the \Quijote and \Molino fiducial halo and galaxy mass functions.  

\subsubsection{Lagrangian bias models}
Alternatively, one can implement a local Lagrangian bias model at the field level, which is then evaluated along the saddle-point relevant for computing the tracer PDF. This corresponds to a relationship between Eulerian densities in cells with the following functional dependence
\begin{equation}
\label{eq:bias_L}
1\!+\!\langle\delta_h|\delta_m\rangle = (1+\delta_m)f_{L}(\delta_{\rm L}(\delta_m))-\langle(1+\delta_m)f_{L}(\delta_{\rm L}(\delta_m))\rangle,
\end{equation}
where $\delta_L(\delta_m)$ is the inverse spherical collapse mapping relating nonlinear and linear densities and
the constant second term is ensuring a zero mean for $\langle\delta_h|\delta_m\rangle$ that is generated in the computation of the PDF \citep[for details see][]{Friedrich2022}. For a local quadratic Lagrangian bias, we have that
\begin{equation}
\label{eq:bias_L_quad}
f_{L}(\delta_{\rm L})=1+b_1^{\rm L}\delta_{\rm L} + \frac{b^{\rm L}_2}{2}\delta_{\rm L}^2\,.
\end{equation}
The Eulerian bias parameters are related to the Lagrangian ones as $b_1^{\rm E}\approx 1+b_1^{\rm L}$ and $b_2^{\rm E}=(1-\nu^{-1}) b_1^{\rm L} + b_2^{\rm L}$, where commonly one assumes $\nu=21/13$ in line with the spherical collapse approximation \citep{Wagner2015,Lazeyras_2016,DesjacquesJeongSchmidt18}. For projected densities studied in \cite{Friedrich2022} the quadratic Lagrangian model outperformed the quadratic Eulerian model. However they argued that these findings may not generalise, and indeed we find no improvement here.

Recently, a Gaussian Lagrangian bias model was proposed in \citep{Stuecker2024,Stuecker2024_cumulantbias} which takes the unrenormalised form $\tilde f_{\rm L}(\delta_L)=\exp(\tilde b^{\rm G}_1\delta_{\rm L} + \tilde b^{\rm G}_2 \delta_{\rm L}^2/2)$ with scale-dependent parameters $\tilde b_n^{\rm G}$. This bias model was designed for the ratio between the Lagrangian galaxy density environment distribution and the background density distribution $f(\delta_{m,\rm L})=\mathcal P(\delta_{m,\rm L}|g)/\mathcal P(\delta_{m,\rm L})$. For Gaussian initial conditions, $P(\delta_{m,\rm L})$ is Gaussian and empirically, the Lagrangian galaxy density environment distribution is close to a Gaussian as well. This model corresponds to a cumulant- rather than moment-based bias expansion \citep{Stuecker2024_cumulantbias}, which matches the spirit of our PDF predictions. A perturbative expansion of the Gaussian bias relation suggests that the leading order terms in this model correspond to a quadratic model with $b_1^{\rm L}\approx \tilde b_1^{\rm G}$ and $b_2^{\rm L}\approx (\tilde b_1^{\rm G})^2+\tilde b_2^{\rm G}$. After renormalisation through a peak-background split, this Gaussian Lagrangian model becomes
\begin{equation}
    \label{eq:bias_L_Gauss}
f_{L}(\delta_{\rm L})=\frac{\exp\left[-\frac{(b_1^{\rm G})^2}{2b_2^{\rm G}}\right]}{\sqrt{1+b_2^{\rm G}\sigma_m^2}} \exp\left[\frac{b_2^{\rm G}\left(\frac{b_1^{\rm G}}{b_2^{\rm G}}+\delta_L\right)^2}{2(1+b_2^{\rm G}\sigma_m^2)}\right]\,,
\end{equation}
where the renormalised parameters $b_n^{\rm G}$ are now expected to be scale-independent. The scale-dependent parameters are then $\tilde b_n^{\rm G} = b_n^{\rm G}/(1+b_2^{\rm G}\sigma_m^2(R))$. This bias model~\eqref{eq:bias_L_Gauss} is fitted to the conditional mean data points and shown as solid lines in Figure \ref{fig:bias_fit_condmean}. The fit is performed in the matter density range corresponding to the bulk of the PDF shown by the shaded regions, with errors calculated from the standard deviation across realisations. We find that the parts of the conditional mean outside of this region do not significantly impact the predicted halo PDF, and so are justifiably excluded from the fit. 

The fitted values for the renormalised Gaussian bias for the fiducial cosmology are displayed in the middle two columns of Table~\ref{tab:bias_fits}. As a consequence of the renormalised bias model, there is very little variation between the fitted bias parameters across different smoothing scales. For this reason we perform a fit on all smoothing scales simultaneously ($R=20,25,30$ Mpc/$h$ for halos and $R=25,30$ Mpc/$h$ for galaxies) and these are the values used in Figure~\ref{fig:bias_fit_condmean}. The variation across redshift is driven by the formation of halos from high to low redshift such that with an almost fixed minimum halos mass of order $10^{13} M_\odot/h$, halos at higher redshift are rarer and thus more biased. 

Given a model for the conditional mean of tracer densities given matter densities, one can obtain the cross-covariance between tracer and matter densities as described in Appendix~\ref{app:crosscov}.

\subsection{Conditional variance stochasticity model}
\label{subsec:condvar}

The conditional variance models the tracer stochasticity or shot noise. We focus on the ratio  
\begin{equation}
\label{eq:SN_def}
\alpha(\delta_m)=\frac{\langle N_t^2|\delta_m\rangle_c}{\langle N_t|\delta_m\rangle}= \frac{\bar N_t\langle \delta_t^2|\delta_m\rangle_c}{1+\langle \delta_t|\delta_m\rangle} \,,
\end{equation} 
which would be unity in the case of Poisson sampling and was already used in the photometric clustering case in \cite{Friedrich2018,Friedrich2022}. As before, density contrasts $\delta_{t/m}$ are within spheres of radius $R$. Note that this shot noise is the stochasticity with respect to a deterministic local bias model described by the conditional mean $\langle\delta_h|\delta_m\rangle$. Figure~\ref{fig:SN_fit_condmean} shows this ratio at redshifts $z=0.0,0.5,1.0$ for halos and $z=0.0$ for galaxies, which possesses a clear deviation from the Poissonian expectation including a density-dependence. 

We parameterise the shot noise ratio with a quadratic density-dependence
\begin{equation}
\label{eq:SN_quad}
\alpha(\delta_m)= \alpha_0 + \alpha_1\delta_m + \alpha_2\delta_m^2 \,.
\end{equation} 
As before, the fits are performed using the data points and errors extracted from the  \Quijote and \Molino realisations in the shaded regions and the results are shown as solid lines.
\begingroup
\renewcommand{\arraystretch}{1.3} 
\begin{table}
  \centering
  \begin{tabular}{c|c|c||c|c||c|c|c}
    \hline
    tracer & $z$ & $R$ & $b_1^\mathrm{G}$ & $b_2^\mathrm{G}$ & $\alpha_0$ & $\alpha_1$ & $\alpha_2$ \\ \hline \hline
    & ~ & 20 & 0.449 & -0.991 & 0.57 & 0.22 & 0.8\\
    halos & 0.0 & 25 & 0.443 & -1.019 & 0.65 & 0.42 & 0.82\\
    & ~ & 30 & 0.440 & -1.018 & 0.72 & 0.56 & 0.76\\
    & ~ & All & 0.446 & -1.001 & 0.61 & 0.28 & 0.69\\
    \hline
    & ~ & 20 & 1.058 & -1.381 & 0.56 & -0.19 & 0.87\\
    halos & 0.5 & 25 & 1.051 & -1.399 & 0.58 & -0.08 & 1.06 \\
    & ~ & 30 & 1.045 & -1.396 & 0.61 & -0.0 & 1.17 \\
    & ~ & All & 1.053 & -1.390 & 0.57 & -0.16 & 0.83 \\
    \hline
    & ~ & 20 & 1.962 & -1.87 & 0.69 & -0.49 & 0.23\\
    halos & 1.0 & 25 & 1.946 & -1.918 & 0.69 & -0.51 & 0.52\\
    & ~ & 30 & 1.934 & -1.935 & 0.71 & -0.51 & 0.74\\
    & ~ & All & 1.951 & -1.895 & 0.70 & -0.50 & 0.21 \\
    \hline
     &  & 25 & 1.54 & -0.34 & 0.92 & 0.09 & 0.31 \\
    galaxies & 0.0 & 30 & 1.51 &  -0.30& 1.02 & 0.13 & 0.32\\
    &~ & All & 1.52 & -0.34 & 0.96 & 0.10 & 0.18
  \end{tabular}
	\caption{Values of renormalised Gaussian Lagrangian tracer bias $b_n^\mathrm{G}$ and quadratic shot noise $\alpha_n$ parameters at different redshifts and smoothing scales $R\;[\mathrm{Mpc}/h$] from fits to the conditional mean and variance for the fiducial cosmology.}
	\label{tab:bias_fits}
\end{table}
\endgroup
The fitted values for the shot noise parameters for the fiducial cosmology are displayed in the last three columns of Table~\ref{tab:bias_fits}. The conditional variances for different sphere radii possess a moderate scale dependence. Therefore, we perform the fits for the tracer stochasticity for all smoothing scales simultaneously. 

While the modelling of stochasticity for the PDF and the power spectrum follow somewhat different principles, we discuss their connection in Appendix~\ref{app:crosscov}. When using a fitted bias function, it might be beneficial to modify the defined ratio in equation~\eqref{eq:SN_def} as follows
\begin{equation}
\label{eq:SN_fit_bias_fit}
    \alpha_{\rm fit}(\delta_m)=\frac{\bar N_t\langle\delta_t^2|\delta_m\rangle}{1+\langle\delta_t|\delta_m\rangle_{\rm fit}}=\alpha(\delta_m) \frac{1+\langle\delta_t|\delta_m\rangle}{1+\langle\delta_t|\delta_m\rangle_{\rm fit}}\,.
\end{equation}
to avoid a propagation of inaccuracies in fitting the conditional mean to the conditional variance. In practice we find that thanks to the good accuracy of the Gaussian Lagrangian bias parametrisation this makes little difference to the accuracy of the halo PDFs at the fiducial cosmology. 

\subsection{Validating the fiducial tracer PDFs}

We compute theoretical predictions for the PDF using the publicly available code CosMomentum\footnote{\href{https://github.com/OliverFHD/CosMomentum}{https://github.com/OliverFHD/CosMomentum}} \citep{Friedrich:2019} that implements equation~\eqref{eq:tracerPDF}. We calculate the theory PDFs with a tracer density $n_h=N_h^\mathrm{tot}\times(\mathrm{Gpc}/h)^{-3}$ given values from the third column of Table~\ref{tab:Ntot} and with $\Lambda$CDM cosmological parameters corresponding to the \Quijote fiducial cosmology. 

\begin{figure}
	\centering
	\includegraphics[width=\columnwidth]{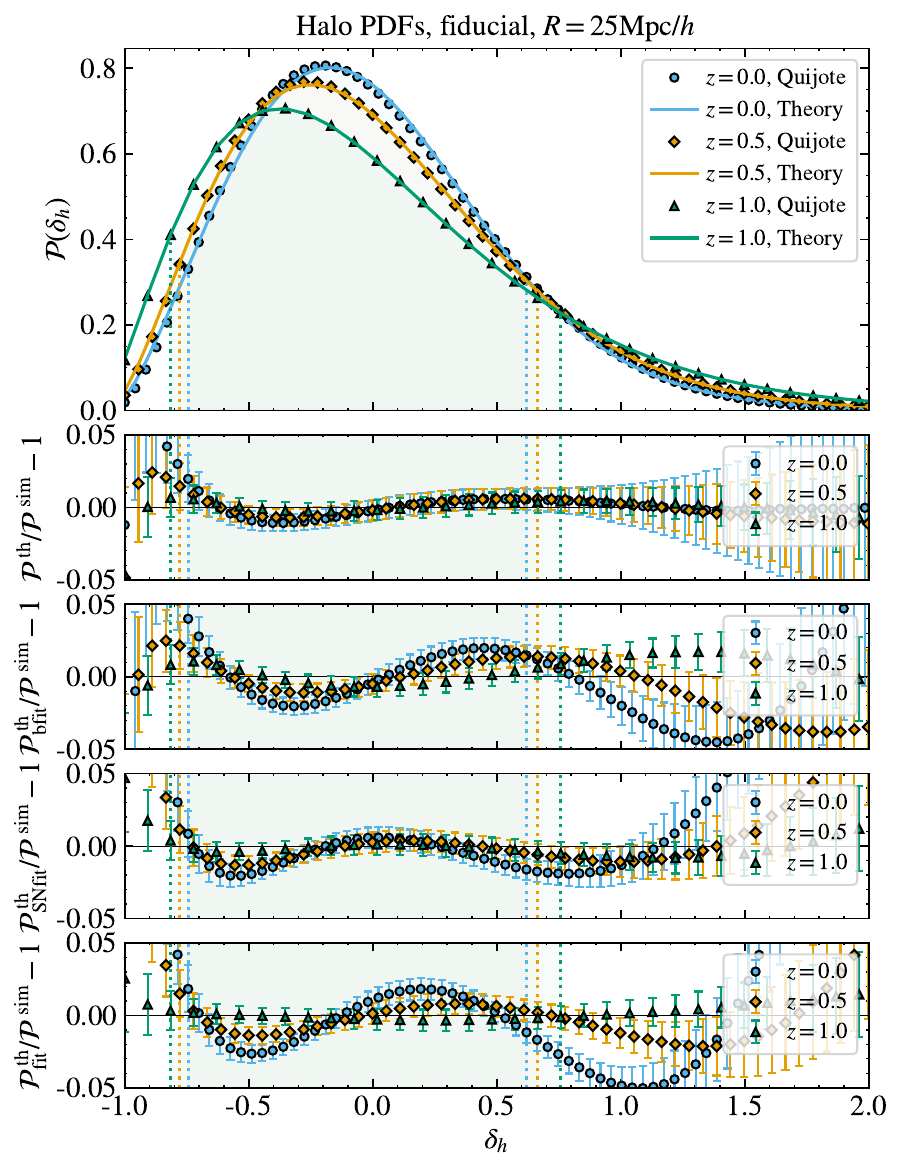}
	\caption{[Upper panel] Comparison of halo PDFs from \Quijote (data points) and predictions using the Gaussian Lagrangian bias and quadratic shot noise fits (solid lines) at smoothing scale $R=25$ Mpc/$h$ for redshifts $z=0.0,0.5,1.0$. [Lower panels] Residuals using full functional forms of the bias and shot noise,
	fitted bias, fitted shot noise, and fitted bias and shot noise. Errors are the standard deviation of 15000 fiducial \Quijote realisations.} 
	\label{fig:Quijote_CosM_compare}
	\includegraphics[width=\columnwidth]{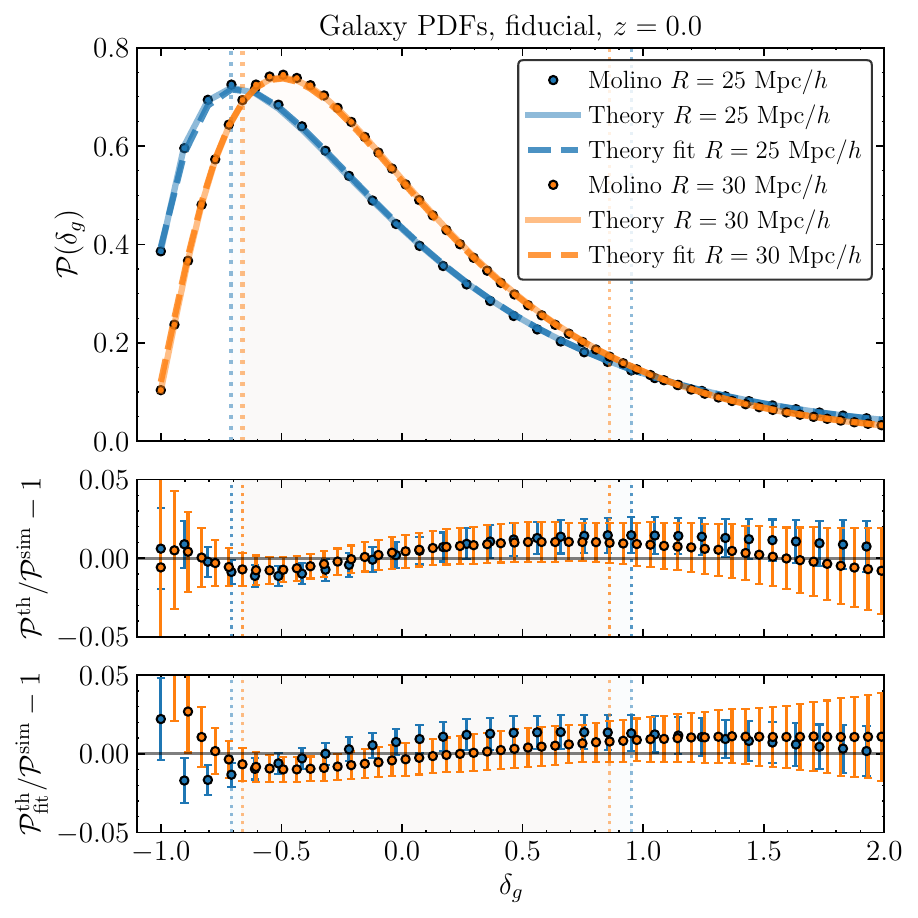}
	\caption{[Upper panel] Comparison of Galaxy PDFs from \Molino (data points) and  theory using the full functional forms of bias and shot noise (solid line) or their fits (dashed line) at smoothing scales  $R=25,30$ Mpc/$h$ for redshift $z=0.0$. [Lower panels] Residuals of the theory with full functional forms or using the bias and shot noise fits.}
	\label{fig:Molino_CosM_comparison}
\end{figure}

Figure \ref{fig:Quijote_CosM_compare} compares the halo PDFs from theory with those extracted from simulations. In contrast with the matter PDFs in Figure~\ref{fig:Quijote_CosM_compare_m}, the halo PDFs become more non-Gaussian with a higher variance as the redshift increases. As can be seen from Table~\ref{tab:bias_fits}, halos are increasingly biased at higher redshift. The variance of the halo density field scales with the linear bias like $\sigma_h^2 \sim b^2 \sigma_m^2$ so although the matter variance still decreases, this increasing bias leads to a mildly increasing halo variance with redshift. As for the matter PDF in Figure~\ref{fig:Quijote_CosM_compare_m}, we focus our analysis on the bulk of the PDF and thus exclude the highest 10\%  and lowest \{10,5,3\}\% of density bins for scales $R=20,25,30$Mpc$/h$ respectively. When we use the full functional forms of the conditional mean and variance first residual panel, the residual of the halo PDF has a very similar profile to that of the matter PDF, indicating that the inaccuracy of our theoretical model stems from the imperfections in the matter PDF theory\footnote{We find that using the measured matter PDF leads to sub-percent inaccuracies in the predicted halo PDF, confirming this assertion.}. We also show a theoretical model with the renormalised Gaussian Lagrangian bias parameterisation~\eqref{eq:bias_L_Gauss} but the full functional form of the stochasticity (second residual panel), which shows increased residuals at lower redshifts indicating the limitations of the two-parameter bias model. When using the full functional form of the bias and the quadratic shot noise model~\eqref{eq:SN_quad}, again residuals increase with redshift but with slightly smaller amplitude indicating limitations of the three-parameter stochasticity model (third residual panel). Finally, when using both parameterisations of bias and stochasticity (lower panel) residuals from the two middle residual panels combine. Since the renormalised bias fits result in very little variation of $b_1^\mathrm{G}$ and $b_2^\mathrm{G}$ across smoothing scales, we do a joint fit across three smoothing scales and use the same values for all scales. The third and fourth panels show the residuals when using the parameterised bias and the full shot noise, and vice versa. While the parameterisation of both bias and shot noise increases the errors, but we still see good agreement, with the residual plots showing difference of less than 3\% around the bulk of the PDF. 

The method illustrated for the halos is also applicable to a galaxy sample. In Figure~\ref{fig:Molino_CosM_comparison} we compare the measured PDF from the \Molino mock galaxy catalogues with the theory predictions from CosMomentum. We present PDF prediction computed by using the full functional form of the conditional mean and conditional variance (Theory) and using the renormalised Gaussian Lagrangian bias and the quadratic shot noise fits (Theory fit). In both cases the theory predicts the measured galaxy PDF to within 3\% accuracy around the bulk of the PDF as shown in the middle and lower panel respectively. We note that the galaxy PDF presents a higher degree of non-Gaussianity compared to the halo PDF at the same redshift due to its larger linear bias.

\subsection{Power spectrum bias and stochasticity}
\label{sec:pktheory}

\begin{table}[]
    \centering
    \begin{tabular}{c||c||c|c||c}
        tracer & $z$ & $b_1^{\rm E}$ & $b_{2,k^2}^{\rm E}$ & $\bar \alpha_0$  \\ \hline \hline
         & 0.0 &  1.44 & -0.01 & 0.83\\ 
        halos & 0.5 & 2.03 & 0.17 & 0.67\\
         & 1.0 & 2.89 & 0.41 & 0.75 \\ \hline
        galaxies & 0.0 & 2.46 & 0.67 & 1.29\\
    \end{tabular}
    \caption{Values of the fitted Eulerian bias~\eqref{eq:bias_Pk} and stochasticity~\eqref{eq:shotnoise_Pk} parameters for the tracer power spectra.}
    \label{tab:bias_SN_powerspectrum}
\end{table}

\begin{figure}
    \centering
    \includegraphics[width=1\columnwidth]{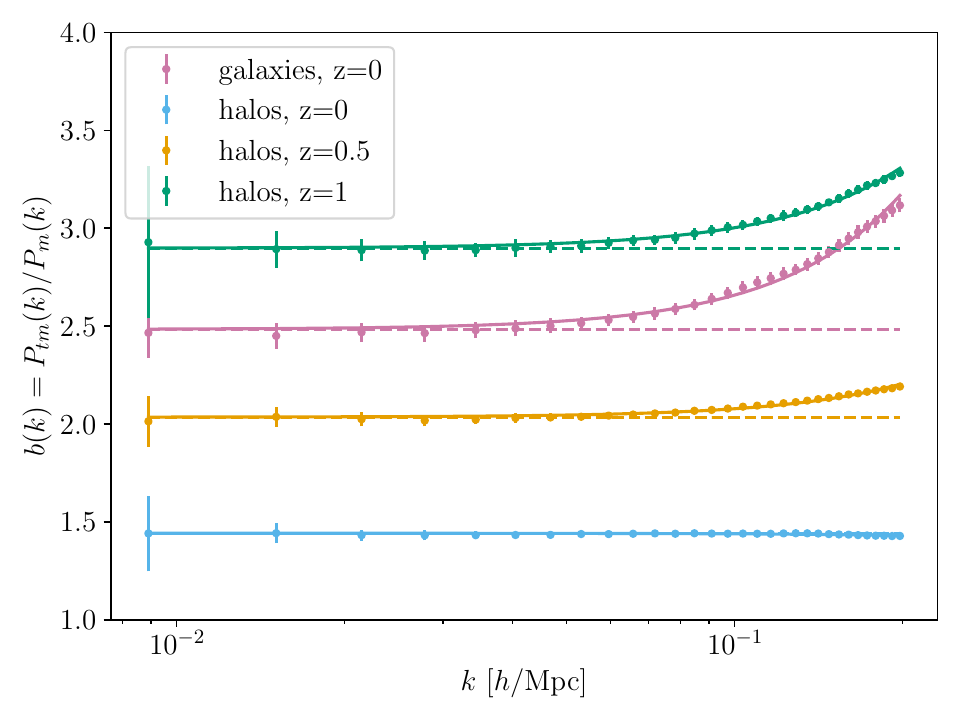}
    \caption{Halo and galaxy bias $b(k)$ obtained from the ratio of the cross-power spectrum and the matter power spectrum (data points with error bars) along with the corresponding quadratic fit~\eqref{eq:bias_Pk} (solid lines) and its linear part (dashed lines).}
    \label{fig:haloPk_bias}
\end{figure}

Following our approach for the tracer PDF, we want to adopt a description of bias and shot noise that is independent of the theoretical modelling for dark matter. Hence, we choose to determine the functional forms of the bias and the shot noise for the tracer power spectrum from the ratios of the tracer-matter cross-power spectra and the matter auto spectrum. The linear, but potentially scale-dependent bias is obtained as the ratio of the cross-power spectrum to the matter power spectrum
\begin{equation}
    b(k)=\frac{P_{tm}(k)}{P_m(k)} \approx b_1 +b_{2,k^2}\frac{k^2}{k_{\rm max}^2}\,,
\label{eq:bias_Pk}
\end{equation}
shown in Figure~\ref{fig:haloPk_bias} for the fiducial cosmology with fit values given in Table~\ref{tab:bias_SN_powerspectrum}. We find that the linear Eulerian bias parameter $b_1^{\rm E}$ is close to the expectation from the Gaussian Lagrangian bias from the conditional mean $b_1\simeq 1+b_1^{\rm G}$. We picked a simple parameterisation for linear bias beyond $b_1$ that is proportional to $k^2$ \citep[in analogy to the case of projected densities][]{Friedrich2022} and a reference scale at our $k_{\rm max}=0.2 h/$Mpc.  Scale-dependent bias of this form can originate from the nonlinear and non-local nature of tracer formation. We briefly touch on quadratic bias terms in Appendix~\ref{app:crosscov}. Halos form preferentially at peaks of the initial density field leading to a term of the form $(kR_{\rm pk})^2$ with some characteristic peak scale $R_{\rm pk}$ \citep{DesjacquesJeongSchmidt18}. As we consider a selection of halos with a wide range of masses, which form not only on different scales, but also at different times, it is hard to predict the combined value for $b_{2,k^2}$. We suspect that larger values of $b_{2,k^2}$ for the halo samples at higher redshift and the mock galaxies are responsible for the slight scale dependence of the renormalised Gaussian Lagrangian bias parameters $b_1^{\rm G}$ we obtained from a fitting the conditional mean that were summarised in Table~\ref{tab:bias_fits}.

\begin{figure}
    \centering
    \includegraphics[width=1\columnwidth]{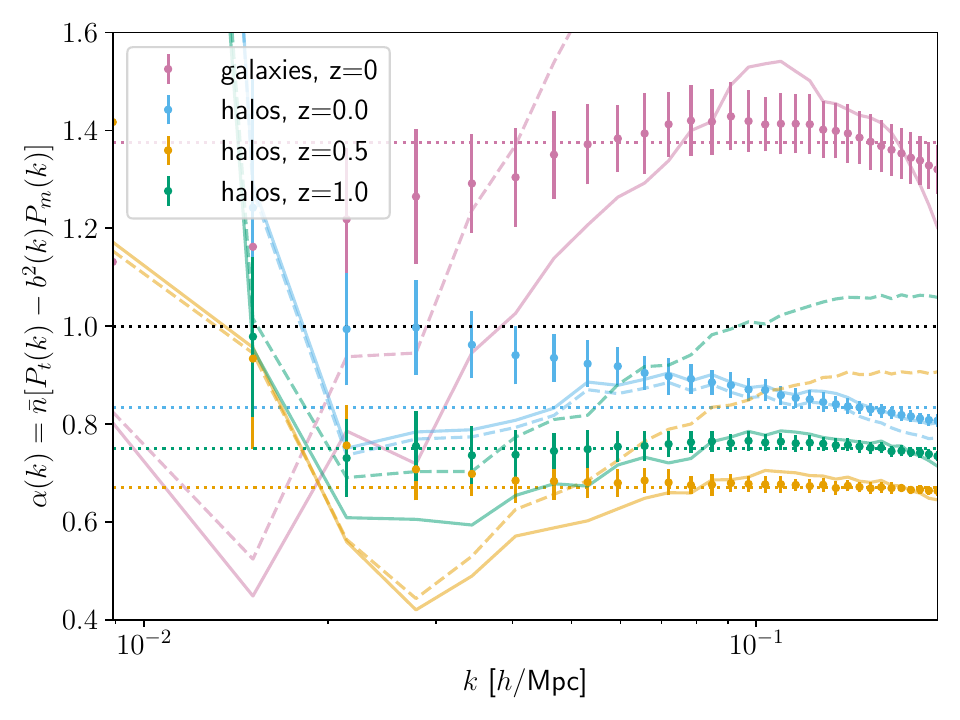}
    \caption{Halo and galaxy shot noise $\alpha(k)$ from equation~\eqref{eq:shotnoise_Pk} obtained from the cross-power spectra (data points) or the modelled bias $b(k)$ with  quadratic fits~\eqref{eq:bias_Pk} (faint solid lines) and their linear parts (faint dashed lines). We also show white noise fits at constant $\bar\alpha_0$ (coloured dotted lines) which deviate from the Poisson expectation (black dotted).}
    \label{fig:haloPk_shotnoise}
\end{figure}

We can define a shot noise with respect to the linear bias as 
\begin{equation}
    \frac{\alpha(k)}{\bar n }
    = P_{t}(k)-b^2(k)P_m(k)\approx \frac{\bar\alpha_0}{\bar n}\,,
    \label{eq:shotnoise_Pk}
\end{equation}
with $b(k)$ from equation~\eqref{eq:bias_Pk}. The result is shown in Figure~\ref{fig:haloPk_shotnoise} for halos and galaxies. Fitting a constant $\bar\alpha_0$ corresponding to white noise (coloured dotted lines) leads to decent agreement on the more nonlinear scales where the shot noise term is most important, while it fails to fit the larger more linear scales. We quote values for the fitted stochasticity parameters at the fiducial cosmology in Table~\ref{tab:bias_SN_powerspectrum}. Note that (overly) simplistic choices for the bias function can change the obtained shot noise. This happens for smaller $k$ ranges for our two-parameter scale-dependent bias fit~\eqref{eq:bias_Pk} (translucent solid lines) and more significantly for a  scale-independent bias with the same $b_1$ (translucent dashed lines) for which $\alpha(k)$ increases significantly for larger $k$. If our linear scale-dependent bias model~\eqref{eq:bias_Pk} would be exact, the shot noise obtained from a scale-independent modelling would change as $\Delta\alpha(k_{\rm max})=b_{2,k^2}(2b_1+b_{2,k^2})\bar nP_m(k_{\rm max})$. 
The amplitude of the power spectrum shot noise agrees well with a shot noise amplitude defined using cross-covariances of the smoothed density~\eqref{eq:SN_var} around the relevant scales $k\sim \pi R^{-1}$, which can be related to the $\alpha_n$ parameters describing the conditional variance as described in Appendix~\ref{app:crosscov}. 

To test the convergence of the simulated derivatives for the mildly nonlinear halo power spectrum, we adopt a simplistic model for the tracer power spectrum in terms of the nonlinear matter power spectrum 
\begin{equation}
    P_t(k,z)=b(k)^2 P_m(k,z) + \frac{\alpha(k)}{\bar n(z)}\,,
    \label{eq:haloPk_model}
\end{equation}
where we use a quadratic model for $b(k)$ following equation~\eqref{eq:bias_Pk} with two Eulerian bias parameters $b_1$ and $b_{2,k^2}$ and a constant shot noise amplitude $\alpha(k)=\bar\alpha_0$ in relation to the inverse of the number density $\bar n(z)$ as expected for Poisson sampling.\footnote{A comparison with perturbative halo power spectrum models from Eulerian Standard Perturbation Theory, Lagrangian Perturbation Theory are Effective Field Theory is beyond the scope of this work, as it would require a careful selection of scales and matching of free parameters to fairly compare the different nonlinear regimes.} We will use \textsc{halofit} predictions for the non-linear matter power spectrum, which also determine the non-linear matter variance in the PDF predictions. On large scales, the matter power spectrum will reduce to the linear prediction. In Appendix~\ref{app:Pk_validation} we show that despite the simplicity of our model, the fiducial signal is well reproduced as shown in Figure~\ref{fig:haloPk_model}. Additionally, the derivatives with respect to cosmological parameters shown in Figure~\ref{fig:halo_Pkderivs} are well captured leading to decent agreement between the Fisher contours. Having validated our predictions, we will later use them to forecast the constraining power at fixed tracer number density and bias.


\section{Probing cosmology with tracer PDFs}
\label{sec:tracerPDF_cosmology}

In this section we determine the sensitivity of the tracer PDF to changing cosmological and bias parameters. We will focus on $\sigma_8$ and $\Omega_m$ as cosmological parameters and one effective bias parameter $\beta$ per redshift. As described in the previous section, the tracer PDF~\eqref{eq:tracerPDF} depends on the underlying cosmology that determines the matter PDF~\eqref{eq:matterPDF_as_Laplace_transform}, and additionally on the conditional PDF of tracer density given matter density. We build the conditional PDF~\eqref{eq:condPDF} from two functions: the mean tracer count given matter density $\overline{N_t}(\delta_m)=\bar N_t(1+\langle\delta_t|\delta_m\rangle)$ and stochasticity captured through the ratio $\alpha(\delta_m)=\langle N_t^2|\delta_m\rangle_c/\overline{N_{\rm t}}(\delta_m)$. The mean number of tracers per cell is determined by the tracer number density $\bar n$ and the cell radius $R$ as $\bar N_t=4\pi R^3 \bar n /3$. We have described effective parameterisations for the two crucial functions in terms of two Gaussian Lagrangian bias parameters $b_n^{\rm G}$ and three shot noise parameters $\alpha_n$. From theory arguments we know that the halo bias~\eqref{eq:bias_SMT} and HOD-based galaxy bias~\eqref{eq:bias_SMT_HOD} carry a cosmology dependence through the halo mass function $\mathcal P(M_h)$. Similarly, the total number density of tracers will be cosmology-dependent. We do not seek to extract information from the halo mass function through the tracer PDF here, and hence decide to select tracers across different cosmologies in a way that keeps the tracer bias fixed. This comes at the price of changing number densities, but we mitigate the main impact of this by looking at PDFs of tracer density contrasts $\mathcal P_{\delta_t}(\delta_t) =\bar N_t \mathcal P_t(N_t)$. This approach leads to tighter constraints on $\sigma_8$, however this effect exists for both the PDF and the power spectrum so the relative strength of the two probes is comparable.

\subsection{Cosmology \& mass dependence of bias}
\label{sec:bias_mass_relation}

\begingroup
\renewcommand{\arraystretch}{1.4} 
\begin{table*}
    \centering
    \begin{tabular}{l|c||c|c|c|c||c|c|c|c|c|c||c|c|c|c||c|c} \hline 
        & &  \multicolumn{2}{|c|}{$b_1^\mathrm{G}(\beta)$}&  \multicolumn{2}{|c||}{$b_2^\mathrm{G}(\beta)$}&  \multicolumn{2}{|c|}{$\alpha_0(\beta)$}&  \multicolumn{2}{|c|}{$\alpha_1(\beta)$} &  \multicolumn{2}{|c||}{$\alpha_2(\beta)$}& \multicolumn{2}{c|}{$b_1^\mathrm{E}(\beta)$}&  \multicolumn{2}{|c||}{$b_{2,k^2}^\mathrm{E}(\beta)$}&  \multicolumn{2}{|c}{$\bar\alpha_0(\beta)$} \\ \cline{3-18}
        tracers & $z$  & $\beta^{\rm fid}$ & $\beta^-$ & $\beta^{\rm fid}$ & $\beta^-$ & $\beta^{\rm fid}$ & $\beta^-$ & $\beta^{\rm fid}$ & $\beta^-$ & $\beta^{\rm fid}$ & $\beta^-$ & $\beta^{\rm fid}$ & $\beta^-$ & $\beta^{\rm fid}$ & $\beta^-$ & $\beta^{\rm fid}$ & $\beta^-$ \\ \hline \hline
         
     & 0.0 & 0.446 & 0.277 & -1.001 & -1.068 & 0.61 & 0.88 & 0.28 & 0.63 & 0.69 & 0.27 & 1.44 & 1.28 & -0.01 & -0.09 & 0.83 & 1.14 \\
    halos & 0.5 & 1.053 & 0.842 & -1.390 & -1.519 & 0.57 & 0.75 & -0.16 & 0.27 & 0.83 & 0.70 & 2.03 & 1.84 & 0.17 & 0.07 & 0.67 & 0.91 \\
     & 1.0 & 1.951 & 1.613 & -1.895 & -2.082 & 0.70 & 0.80 & -0.50 & -0.09 & 0.21 & 0.51 & 2.89 & 2.59 & 0.41 & 0.26 & 0.75 & 0.89 \\ \hline
    galaxies & 0.0 &1.524 &1.409 & -0.344 & -0.348& 0.96 & 0.87 & 0.09 &0.02 & 0.19 & 0.42 & 2.46 & 2.36 & 0.67 & 0.60 & 1.29 & 1.15 \\
    \end{tabular}
    \caption{Values of PDF tracer bias and stochasticity parameters from a joint fit of all smoothing scales along with power spectrum bias and stochasticity, using the fiducial and modified tracer selection prioritising lower halo masses and satellite galaxies, respectively.}
    \label{tab:bias_mostleast}
\end{table*}
\endgroup

 We extract tracer PDFs in such a way as to leave the bias as constant as possible by tailoring the number of halos selected for each cosmology, i.e. selecting the only the most massive $N_h^\mathrm{tot}$ halos. The numbers used to achieve this for each cosmology and redshift are shown in Table~\ref{tab:Ntot}.

To obtain the values for the $N_h$ cut across different cosmologies we follow the SMT bias predictions~\eqref{eq:bias_SMT} relying on the cosmology-dependent halo mass function $\mathcal P(M_h)$ measured from the simulations. Since fine-tuning $\bar b_n$ by changing the minimum mass $M_h^{\rm min}$ is challenging due to the wide bins of the halo mass function (set by the mass resolution), we instead select a total number $N_h^\mathrm{tot}$ which we use to cut the mass function by emptying the lowest required mass bin by the necessary number of halos such that the integral evaluates to the desired bias value. One can then find the value of $N_h^\mathrm{tot}$ that minimises the difference between $\bar b_1$ and the desired linear bias of the fiducial cosmology. Note that if we kept the selected number of halos fixed across cosmologies this would significantly alter the shape of the derivatives. For example, a change in $\sigma_8$ would cause the matter variance and the linear bias to change in opposite directions such that that the halo variance $\sigma_h^2\approx b_1^2\sigma_m^2$ remains almost unchanged.

Similarly, to keep the linear galaxy bias fixed across different cosmologies, we follow the procedure outlined above. We extract the quantity $\langle N_g (M_h) \rangle P(M_h)$ from the \Molino suite and search for the number of galaxies that minimises the difference between the measured and the fiducial bias $\bar{b}_{{\rm g}, 1}$. Once the number of galaxies is determined for each cosmology the galaxy PDF is measured from the \Molino suite considering the corresponding number of galaxies in the most massive halos. As we consider the most massive halos this selection includes most of the satellite galaxies. We measure the conditional mean and conditional variance of galaxy counts given matter density from the \Molino suite, considering the corresponding number of galaxies for each cosmology.

\subsection{Response to cosmology and tracer selection} 
We want to quantify the response of the tracer density PDF to changes in cosmology and the tracer selection. While we extract the PDFs of the tracer number count $\mathcal P(N_t)$, we convert them to PDFs of the tracer density contrast $\mathcal P(\delta_t)$ to avoid extracting cosmological information from the variation of the tracer number density across cosmologies. To this end, we compute the parameter derivatives from finite differences between the PDFs of incremented and decremented parameters $\theta^\pm$ , i.e.
\begin{equation}
    \frac{\partial \mathcal P}{\partial\theta} = \frac{\mathcal P(\theta^+)-\mathcal P(\theta^-)}{\theta^+-\theta^-}\,.
\end{equation}
The simulations are with symmetric increments such that $\theta^\pm=\theta_{\rm fid}\pm \Delta \theta$ with $\Delta \sigma_8=0.015$ and $\Delta\Omega_m=0.01$.

We vary the tracer selection at fixed number density by changing the fiducial selection of the most massive halos to the least massive halos. We parameterise this change by introducing the parameter $\beta$ as the fractional difference from the fiducial value of the linear Eulerian bias parameter
\begin{equation}
    \beta = \frac{b_1}{b_1^\mathrm{fid}}-1 .
\end{equation}
with values of $b_1$ taken from the power spectrum fits described in Section~\ref{sec:pktheory}, which closely resemble the Eulerian linear bias for the PDF. The $\beta$ derivative is a one-sided derivative representing the effect of changing the whole conditional PDF $\mathcal P(\delta_t|\delta_m)$ parameterised through the bias and the shot noise at fixed cosmology. For the simulation derivatives it is calculated by extracting PDFs not with the $N_t^\mathrm{tot}$ most massive halos/galaxies as described in the previous subsection, but instead prioritising lower mass objects. In practice this means taking the $N_h^\mathrm{tot}$ lowest mass halos or $N_g^\mathrm{tot}$ galaxies where satellites are preferentially selected. For halos the least massive sample corresponds to a mass range from $1.0 \times 10^{13}$ to approximately $\{8.4,6.8,4.4\}\times 10^{13} M_\odot/h$ for redshifts $z=\{0,0.5,1\}$ respectively. This is a very different sample from the most massive case that includes halos from $1.44\times10^{13}$ to $\{7.9,4.3,2.7\}\times 10^{15} M_\odot/h$. For galaxies the less massive sample that prioritises satellite galaxies has a mass range from $1.31 \times 10^{13}$ to $4.39\times10^{15}M_\odot/h$, and the more massive sample is in the range $3.33 \times 10^{13}$ to $4.39 \times 10^{15}M_\odot/h$. We then take the derivative to be
\begin{equation}
    \frac{\partial \mathcal P}{\partial\beta} = \frac{\mathcal P(\beta^\mathrm{fid})-\mathcal P(\mathrm{\beta^-})}{\beta^\mathrm{fid}-\beta^-} \,.
\end{equation}

\begin{figure}
    \centering
    \includegraphics[width=\columnwidth]{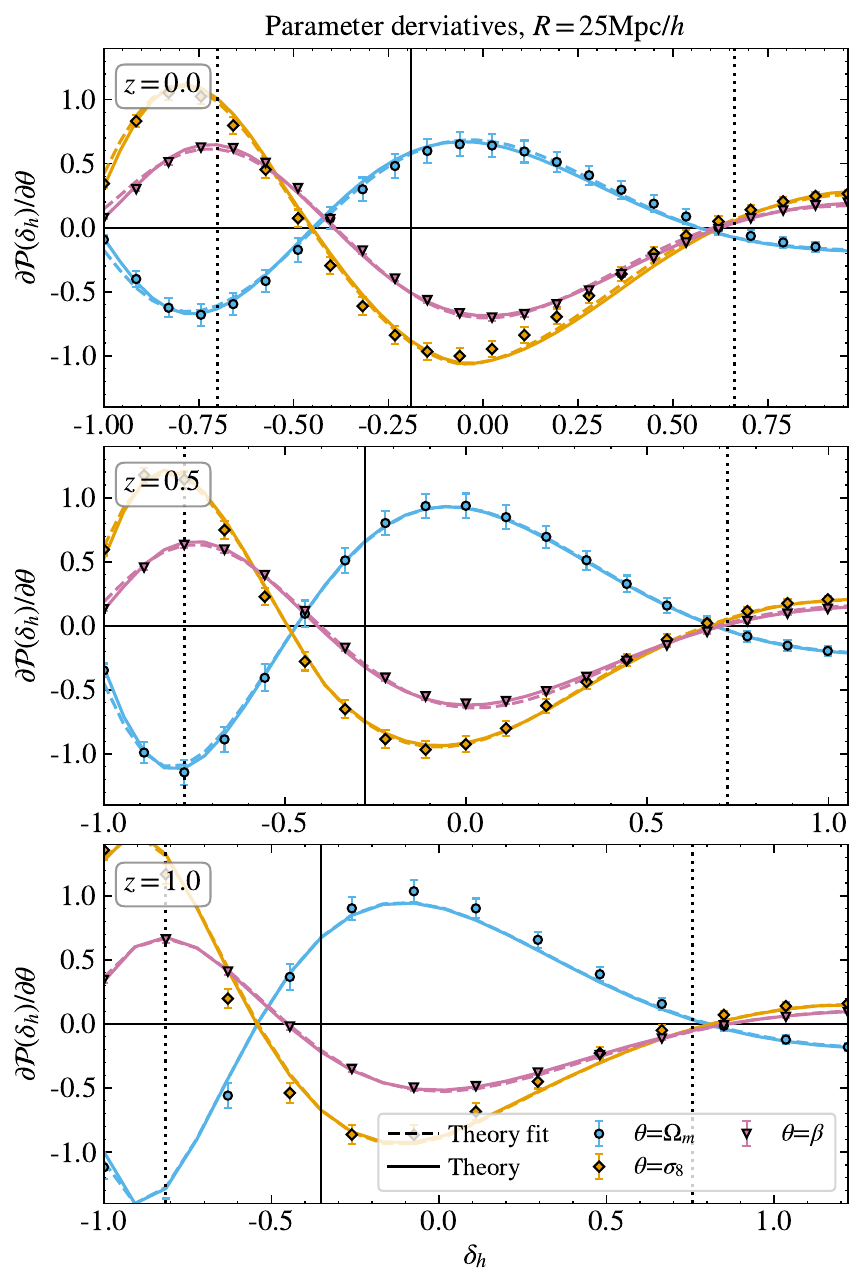}
    \caption{Halo PDF derivatives with respect to the cosmological parameters from \Quijote (data points) and theory (solid lines) and parameterised theory (dashed lines) using the renormalised Gaussian Lagrangian bias and quadratic snot noise combined smoothing scale fits. Derivatives are shown at smoothing scale $R=25.0$ Mpc/$h$ for redshifts $z=0.0,0.5,1.0$. Fiducial bias and shot noise is used for all cosmologies. Error bars on data points are from 500 realisations of each \Quijote cosmology. The solid vertical line indicates the location of the peak of the PDF. The dotted vertical lines bookend the range of $\delta_h$ bins used for the Fisher analysis in the next section. 
    }
    \label{fig:derivatives}
\end{figure}
\begin{figure}
	\includegraphics[width=\columnwidth]{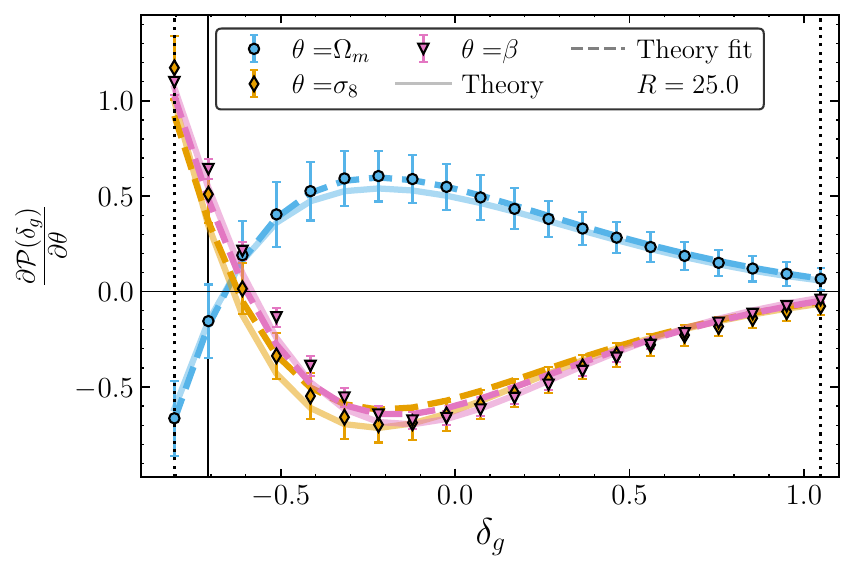}
	\caption{Galaxy PDF derivatives. Derivatives of the galaxy PDFs with respect to the cosmological parameters from \Molino (data points), theory using the full functional form of the conditional mean and conditional variance (solid lines) and parameterised theory  using the joint fit for the Gaussian Lagrangian bias and the quadratic shot noise (dashed lines). Fiducial bias and shot noise fits are used for all cosmologies.} 
	\label{fig:Molino_derivatives}
\end{figure}

The changes in the bias and shot noise parameters as a response to this tracer selection are shown in Table~\ref{tab:bias_mostleast}. While this results in larger step sizes than is ideal for the use of finite difference derivatives, we only use this to cross-validate theory and simulations and use the same step size for both.

The relevant PDFs $\mathcal P(\theta^\pm)$ can be computed in CosMomentum from the same model described previously and tracer power spectra can be constructed following our model~\eqref{eq:haloPk_model}. Since we constructed halo samples from \Quijote in such a way that the bias does not significantly change across cosmology, the fiducial bias can be used for all cosmologies. Similarly, the shot noise can be regarded as effectively constant across cosmologies, although it does still change slightly. Then the only changes going into the theory go into the underlying matter field and a change in tracer density $n_h$ coming from the $N_h^\mathrm{tot}$ cuts in Table~\ref{tab:Ntot}. Since we consider PDFs of density contrasts rather than number counts, the slight changes do not significantly change the derivatives.

Figure \ref{fig:derivatives} shows the derivatives of the \Quijote (data points) and theory halo PDFs (solid lines) with respect to the cosmological parameters $\Omega_m$ and $\sigma_8$, as well as the tracer selection parameterised by $\beta$. For the simulated derivatives, the errors on the data points are the standard deviation of the results of this computation over 500 realisations. The dashed lines (labelled `theory fit') come from the theoretical model described in the previous section using the renormalised Gaussian Lagrangian bias \eqref{eq:bias_L_Gauss} and quadratic shot noise \eqref{eq:SN_quad}, and fit the parameters $\{b_1^\mathrm{G},b_2^\mathrm{G},\alpha_0,\alpha_1,\alpha_2\}$ using the fiducial cosmology for each redshift where all smoothing scales have been fitted simultaneously. We find good agreement between the predicted derivatives compared to those measured from \Quijote. Without a contribution due to the change of bias, cosmological derivatives of the halo PDF closely resemble those of the matter PDF (shown in Appendix Figure~\ref{fig:matterderivs}). The $\beta$ derivative captures the response of the PDF to a change in tracer selection, and is similar in form to the linear combination of the derivatives with respect to the full set of bias and shot noise parameters $\{b_1^\mathrm{G},b_2^\mathrm{G},\alpha_0,\alpha_1,\alpha_2\}$ (shown in the Appendix Figure~\ref{fig:haloPDF_derivatives_bias}).

Figure \ref{fig:Molino_derivatives} shows the galaxy PDF derivatives for the same set of parameters we consider for the halo PDF derivatives in Figure \ref{fig:derivatives}. The galaxy PDF derivatives show the same behaviour as the matter and halo PDF derivatives. As the number of galaxies in the \Molino suite is less than the number of halos in the \Quijote simulations, the effect of shot noise is stronger. In comparison to the halo case we notice a larger variation of the conditional variance among different cosmologies, especially for variations of $\sigma_8$. We employ the measured conditional mean from the fiducial cosmology and shot-noise from the varied cosmologies (solid lines) and the corresponding joint renormalised Gaussian Lagrangian bias and joint quadratic shot noise fits for the fiducial (dashed lines) as input to compute the theoretical predictions with CosMomentum. Compared to the halo PDF derivatives, there is a visible difference between the theory lines computed from the full conditional mean and variance (solid) and the fits (dashed), which stems from inaccuracies in the joint quadratic shot noise fit displayed as pink line in Figure~\ref{fig:SN_fit_condmean}, which stems from a stronger scale dependence.
While we focused on the tracer PDFs here, we show the corresponding derivatives of the mildly nonlinear tracer power spectra in Appendix~\ref{app:Pk_validation}.

\subsection{Tracer PDF Covariance}
\label{sec:PDF_covariance}
\begin{figure}
    \centering
    \includegraphics[width=0.97\columnwidth]{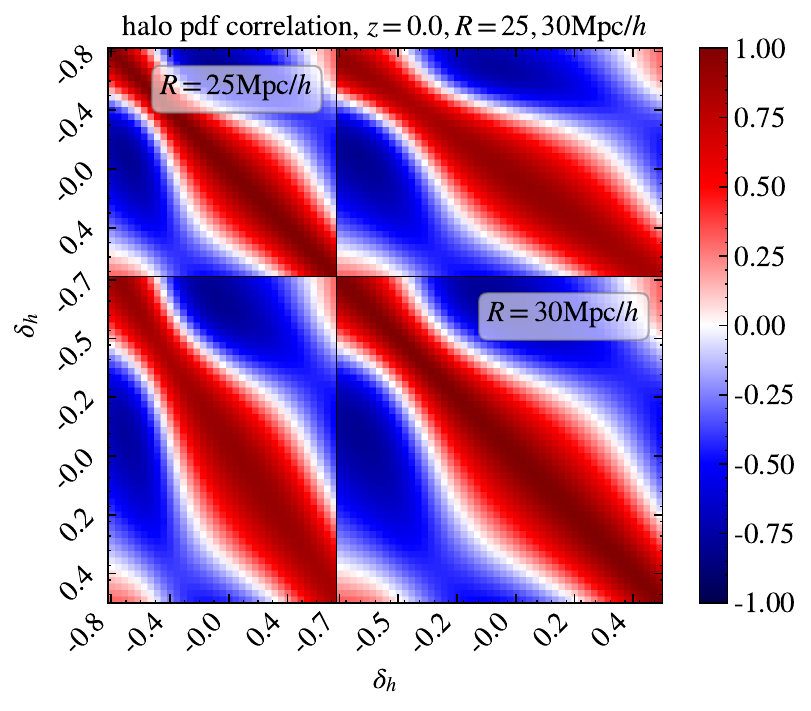}
    \caption{Correlation matrix for the halo PDF in density bins at redshift $z=0$ and radii $R=25,30{\rm Mpc}/h$, extracted from all 15000 realisations of the \Quijote fiducial cosmology. }
    \label{fig:covcorr_mats}
\end{figure}

The error bars shown in Figure~\ref{fig:Quijote_CosM_compare} indicate how accurately the tracer PDF bins can be measured in the simulation volume given the grid of overlapping cells. While overlaps are desirable to reduce the overall error of the PDF measurement, they induce strong correlations between the PDF measurements in neighbouring bins \citep{Uhlemann2022cov}. Figure \ref{fig:covcorr_mats} shows the correlation matrix for the halo density PDF in delta bins, using 15000 realisations of the \Quijote fiducial cosmology at redshift $z=0$ and $R=25,30{\rm Mpc}/h$, where the combined data vector is the concatenation of the different scales in ascending order. The strong correlations adjacent to the diagonal are induced by the correlation between densities in overlapping cells. Intermediate under- and overdensities are as expected anti-correlated, while in the corners one can observe a positive correlation of more extreme low and high densities. Due to the normalisation of the PDF, an anticorrelation of the peak and the surrounding central region (which together contain the bulk of the probability) is expected. As the density contrast has a zero mean, the tails of very low and very high densities can be expected to be positively correlated. The PDFs of halo densities at two subsequent sphere radii are strongly correlated, as the density in the larger cell will be similar to the density in the enclosed smaller cell. While it is possible to reorganise the information to describe densities in a central sphere and surrounding spherical shells \citep{bernardeauvalageas2001,Bernardeau_2014,Uhlemann2015,Codis_2016}, we opt to have a simpler data vector while taking account of the cross-correlations.

\subsection{Fisher Forecast}
\label{sec:Fisher_forecast}

We quantify the information content of the tracer PDF on key $\Lambda$CDM parameters and the halo bias using the Fisher matrix formalism. Within this formalism we also further validate our theoretical model using the \Quijote suite of simulations.

In this section we will introduce the elements that go into the formalism, explain the contents of our data vector, and discuss how combinations of redshifts $z$ and scales $R$ can break degeneracies thus extract the maximum information from the halo density field.

The Fisher matrix is defined by 
\begin{equation}\label{key}
	F_{ij} = \sum_{\alpha,\beta} \frac{\partial S_\alpha}{\partial\theta_i}C_{\alpha\beta}^{-1}\frac{\partial S_\beta}{\partial\theta_j}
\end{equation} where $ S_i $ and $ \theta_i $ are elements of some statistic $ \vec{S} $ and some set of parameters $ \vec{\theta} $ respectively. 
$ C $ is the data covariance matrix defined by 
\begin{equation}
\label{eq:covariance}
	C_{\alpha\beta} = \big\langle (S_\alpha-\langle S_\alpha\rangle)(S_\beta-\langle S_\beta\rangle)\big\rangle .
\end{equation}
Here we construct the data vector from the values of the halo one-point PDFs in bins of halo density contrast $\delta_h$, combining three different smoothing scales $R$=$20,25,30$Mpc/$h$. We discount the effects of the tails of the PDFs by performing a CDF cut between {0.1,0.05,0.03} for $R=20,25,30$ Mpc/$h$ and 0.9 following the spirit of \citep{uhlemann_fisher_2020}. The derivatives $\partial S_\alpha/\partial\theta_i$ are as discussed in the previous subsection and shown in for $R=25$ Mpc/$h$ in Figure~\ref{fig:derivatives}. When including the parameter $\beta$, it is treated as separate parameters $\beta_z$ for each redshift in the derivatives. The covariance matrix of the halo PDF is the same as previously discussed in Section~\ref{sec:PDF_covariance}. For the Fisher analysis we use a covariance computed from all $N_\mathrm{sim}=15,000$ realisations of the \Quijote fiducial cosmology. When the covariance matrix is inverted, noise present in the estimation of $C$ will lead to bias in the elements of $C^{-1}$. To correct for this we multiply $C^{-1}$ with the Kaufman-Hartlap factor \citep{Kaufman67,Hartlap:2006kj} defined by
\begin{equation}
    h = \frac{N_\mathrm{sim}-2-N_S}{N_\mathrm{sim}-1} 
\end{equation}
where $N_S$ is the length of the data vector. Since $N_\mathrm{sim}=15,000$ is much larger than $N_S$ (between 47 and 85 for the PDFs at different redshifts), this factor is always close to unity. We assume no correlation between the density fields at different redshifts\footnote{This stems from the fact that in a survey different redshifts probe different parts of the lightcone, which is in contrast to simulations where different snapshots correspond to the same volume. This also holds for redshift bins that are closer to each other, as long as they are much thicker than the redshift uncertainty.}, so the Fisher matrices from the three redshifts considered ($z=0.0,0.5,1.0$) can simply be linearly combined.  
Once the Fisher matrix is known, the marginalised error on the parameter $ \theta_i $ is given by 
\begin{equation}\label{keyb}
	\delta(\theta_i) \geq \sqrt{(F)^{-1}_{ii}} .
\end{equation}

\begin{figure}
    \centering
    \includegraphics[width=\columnwidth]{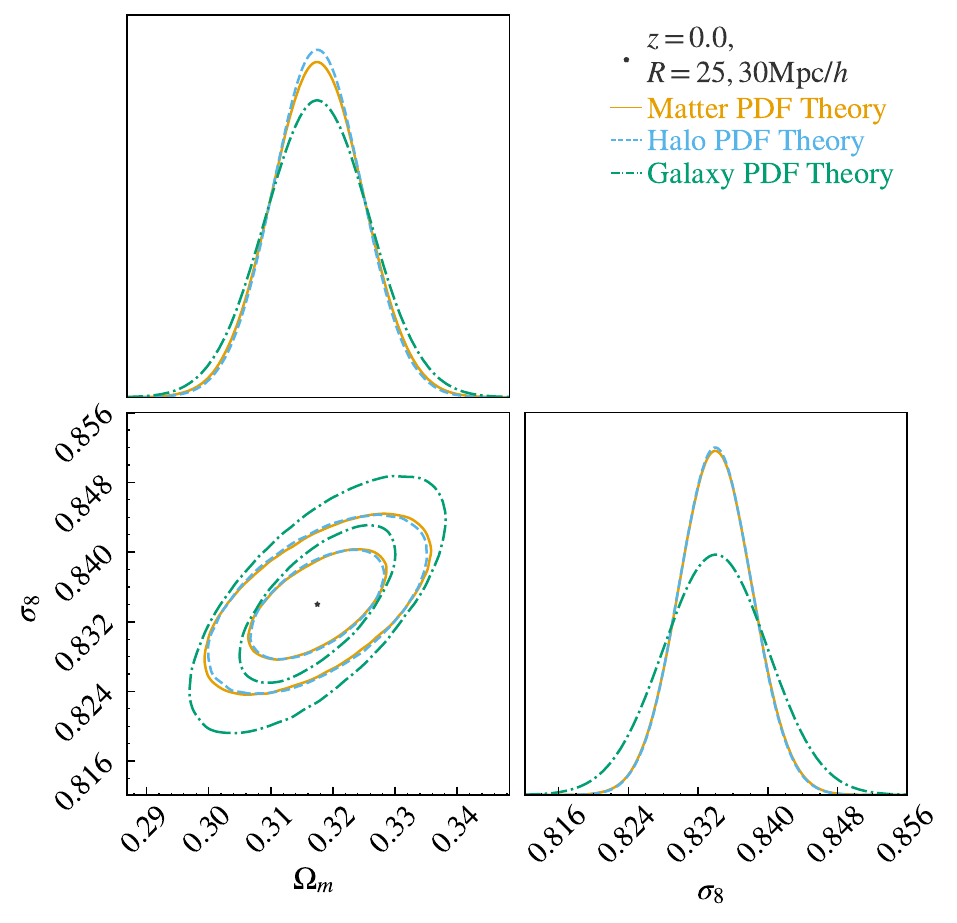}
    \caption{Constraints from theory matter PDFs (orange, solid lines), halo PDFs (blue, dashed lines) and  galaxy PDFs (green, dot-dashed) at redshift $z=0.0$ and radii $R=25,30$Mpc/$h$. Halo and galaxy bias theory using Gaussian Lagrangian bias and quadratic shot noise fits from the fiducial cosmology. All other parameters, including bias, are fixed.
    }
    \label{fig:fisher_matterhalo}
    \centering
    \includegraphics[width=\columnwidth]{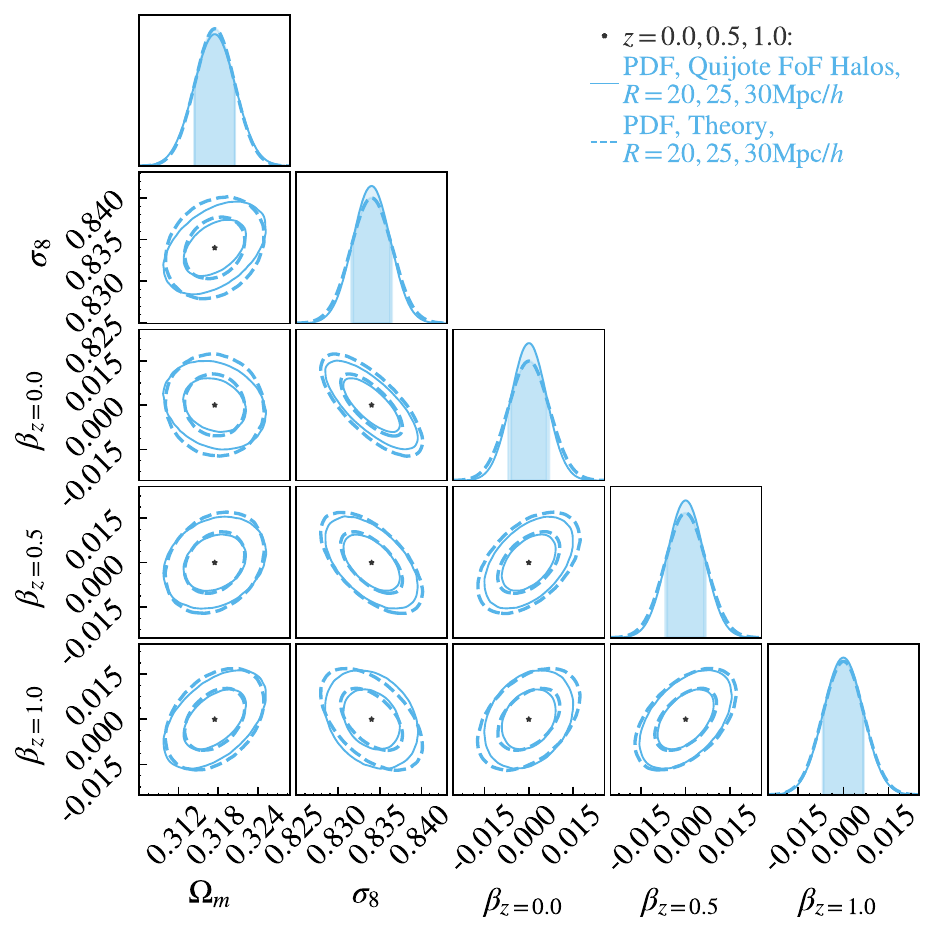}
    \caption{Comparison of halo PDF constraints from \Quijote (solid, thinner lines) and from theory (dashed, thicker) using Gaussian bias and quadratic shot noise parameters from fits for the fiducial cosmology. }
    \label{fig:fisher_val}
\end{figure}

\subsubsection{Validation of tracer PDF constraining power}

Before we proceed to making forecasts for the constraining power of the tracer PDF in comparison to the power spectrum, we perform several validations of our theoretical model.

When keeping the tracer bias and number density fixed across cosmologies, the matter and tracer PDFs carry a comparable or lower amount of cosmological information as we show in Figure~\ref{fig:fisher_matterhalo}. This is expected as for the simple case of linear Eulerian bias and no shot noise there is a simple one-to-one relation between the matter and halo PDFs $\mathcal P_h(\delta_h)\simeq  \mathcal P_m(\delta_m) d\delta_m/d\delta_h =\mathcal P_m(\delta_m=\delta_h/b_1)/b_1$. This is focused on a single redshift $z=0$ and two scales $R=25, 30$Mpc$/h$, where we have validated matter (orange solid), halo (blue dashed) and galaxy PDF (green dot-dashed) predictions. The presence of shot noise can lead to a loss of information through the convolution in equation~\eqref{eq:tracerPDF}, so it is expected that the tracer PDF constraints are weakened with decreasing number density from the halos to the galaxies.

When considering a single redshift but extending the parameter space by one tracer selection parameter $\beta$, the agreement of the Fisher forecasts for theoretical and simulated tracer PDFs is not very good. This can be attributed to the similarity between derivatives w.r.t. the cosmological parameters and the combined bias and stochasticity parameter $\beta$ paired with small residuals in the derivatives. For the galaxies there is a striking similarity between $\sigma_8$ and $\beta$ derivatives. For the halos, there is more similarity between the shape of the $\Omega_m$ and $\beta$ PDF derivatives. In the more realistic case of combining several redshifts the degeneracy between the cosmological and the combined bias and stochasticity parameter $\beta$ is lifted, such that parameter constraints will become more robust against residual modelling uncertainty.

In Figure~\ref{fig:fisher_val} we show a comparison of the Fisher forecast for the theoretically predicted and the simulated derivatives finding excellent agreement of the two approaches even after marginalising over one bias parameter $\beta$ per redshift. This reassures us that it is safe to use the theoretical predictions to further explore degeneracy breaking brought about by combining the tracer PDF at different redshifts and adding in the tracer power spectrum, for which we discuss results separately in Appendix~\ref{app:Pk_validation}.

\subsubsection{Degeneracy breaking with different redshifts}
\label{subsec:PDF_diffz}

Having validated our theoretical models for the tracer PDF and power spectrum, we proceed to forecast parameter constraints at fixed number density to avoid degeneracy breaking arising from different levels of shot noise across the cosmologies. For an initial assessment we consider only one combined tracer bias and stochasticity parameter for both the halo PDFs and the power spectrum.

\begin{figure}
	\centering
 \includegraphics[width=0.97\columnwidth]{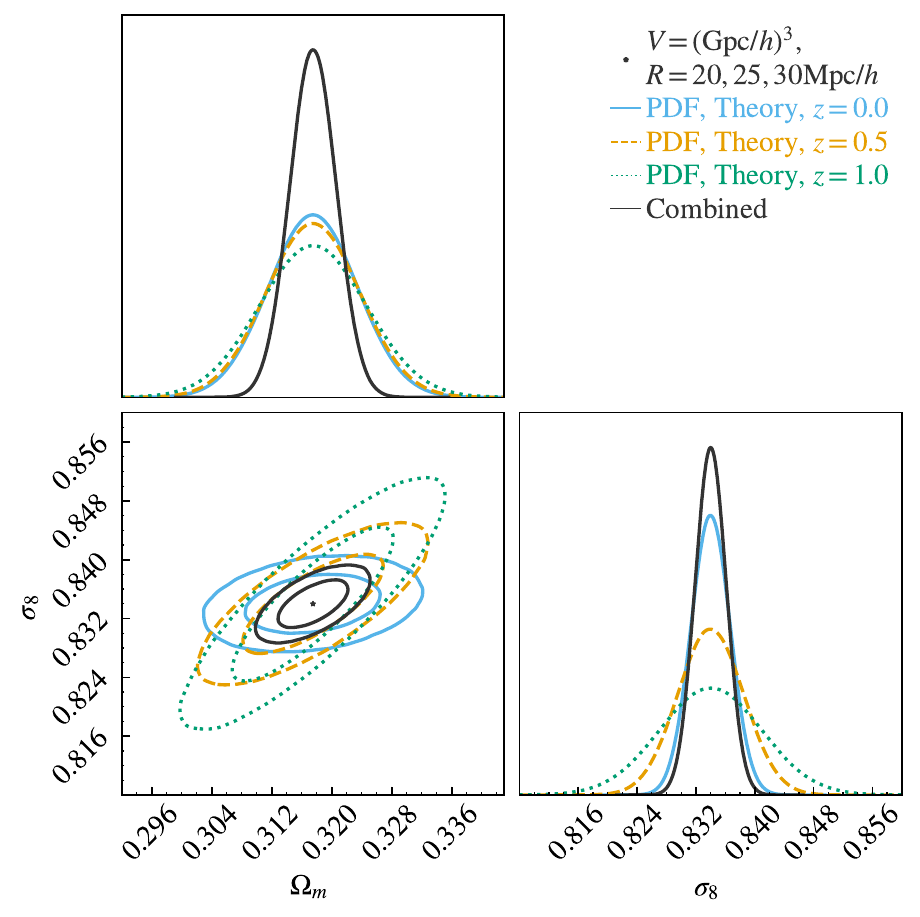}
 \includegraphics[width=0.96\columnwidth]{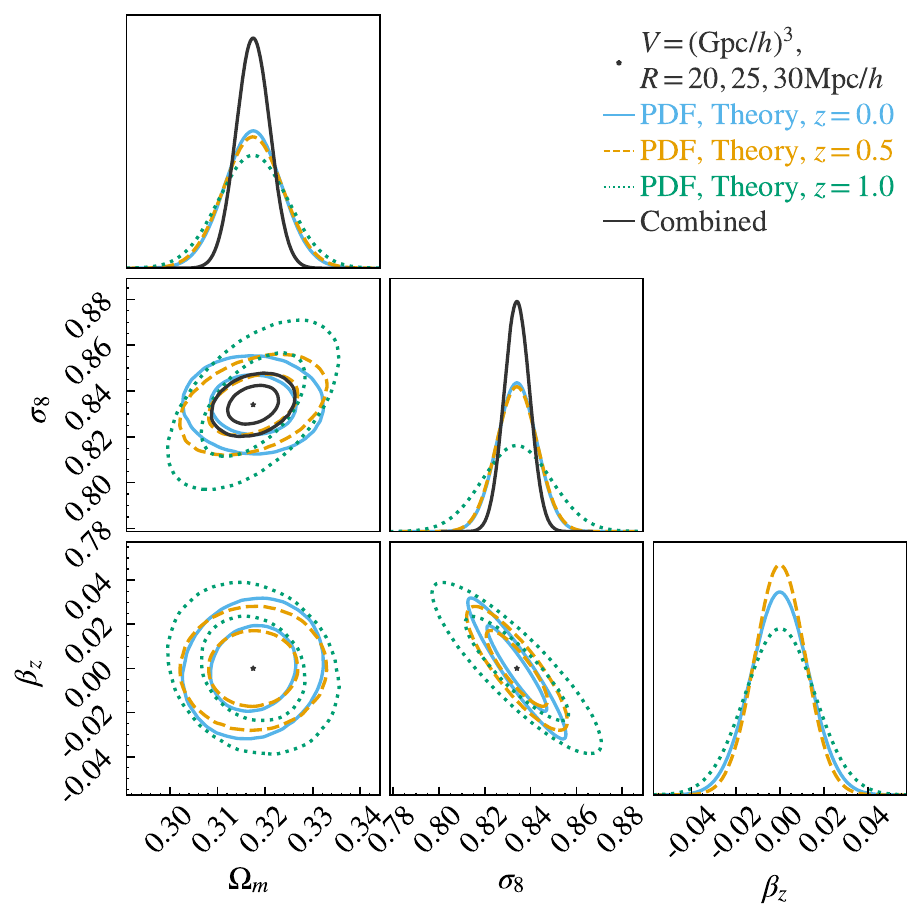}
	\caption{(Upper panel) Fisher forecast for $\{\Omega_m, \sigma_8\}$ for the theory halo PDF at smoothing scales $R=20,25,30{\rm Mpc}/h$ and redshifts $z=0.0,0.5,1.0$ separately and then combined with fixed bias. (Lower panel) also includes the individual bias parameters $\beta_z$ for each redshift. Theory PDFs use the Gaussian Lagrangian bias fits and quadratic shot noise fits for the fiducial cosmology.    
 }
	\label{fig:fisherz}
\end{figure}

The upper panel of Figure~\ref{fig:fisherz} shows the Fisher forecast for the $\Lambda$CDM parameters $\{\Omega_m,\sigma_8\}$ using data vectors constructed from the theoretical PDFs in $\delta_h$ bins with radii $R=20,25,30$ Mpc/$h$ with the bias fixed. Constraints are shown for each redshift $z=0.0,0.5,1.0$ individually in different line style and colour, and then combined in black via a linear combination of the Fisher matrices as appropriate for independent probes. The lower panel also includes a bias parameter $\beta_z$ for each redshift. The constraints from these are naturally not combined since these are different parameters.

We can see that the PDFs at different redshifts can break degeneracies present at individual redshifts. As one can see in Figure~\ref{fig:derivatives}, the PDF derivatives with respect to $\Omega_m$ and $\sigma_8$ have similar but opposite profiles. This tells us that raising $\Omega_m$ and lowering $\sigma_8$ have similar effects, which leads to the diagonal alignment of the contours in the upper panel of Figure~\ref{fig:fisherz} with the different slopes set by the different amplitude ratios between the two parameter derivatives. This correlation is strongest at $z=1.0$, where as explained for the derivatives above, the response of the PDF to a change in both $\Omega_m$ and $\sigma_8$ is close to that of rescaling the non-linear variance. On the other hand, at lower redshift the skewness change induced by $\Omega_m$ makes the effect of the two parameters more distinguishable. This accounts for the rotation of the ellipses as $z$ decreases.

\begin{figure}
	\centering
 \includegraphics[width=0.97\columnwidth]{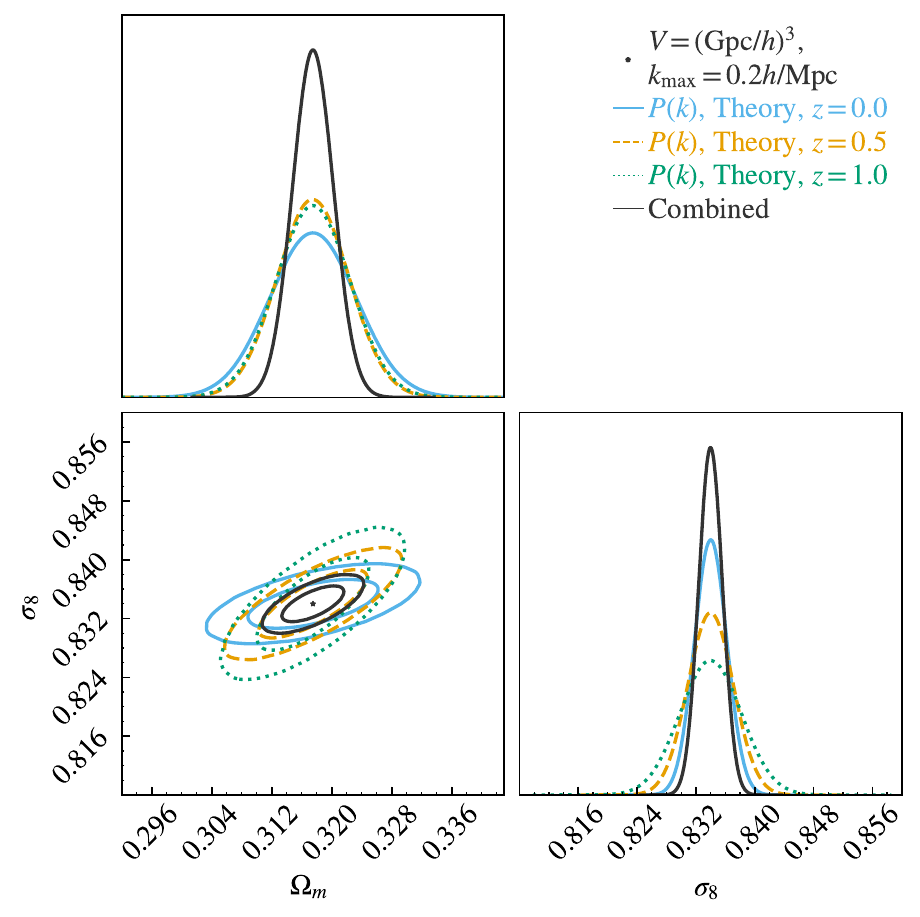}
 \includegraphics[width=0.975\columnwidth]{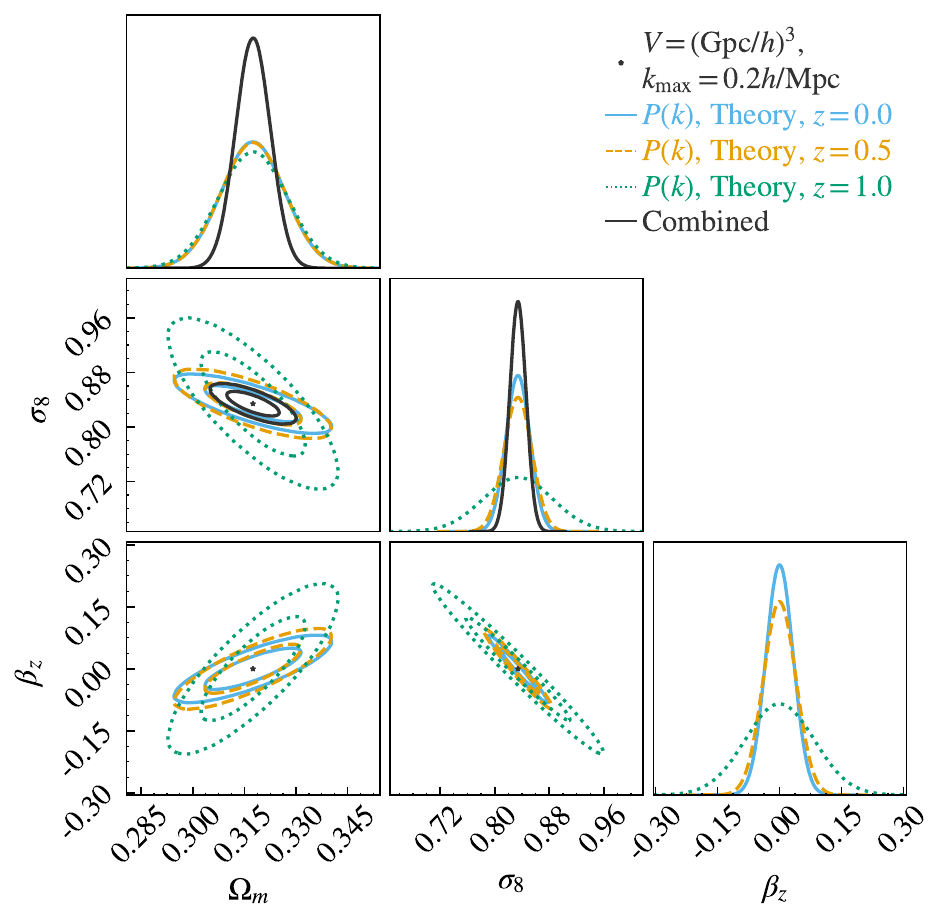}
	\caption{Fisher forecast for $\{\Omega_m, \sigma_8\}$ for the theory halo power spectra at redshifts $z=0.0,0.5,1.0$ separately and then combined. Upper panel with fixed bias, and the lower panel shows also the contours from the individual bias parameters $\beta_z$.
 }
	\label{fig:fisherz_Pk}
\end{figure}

Figure~\ref{fig:fisherz_Pk} shows the equivalent plots for the halo power spectrum theory. We see in the upper plot that the $\Omega_m$-$\sigma_8$ contours have different orientations at different redshifts. As explained in Appendix Section~\ref{app:Pk_validation}, $\Omega_m$ and $\sigma_8$ affect the amplitude of the halo power spectrum in opposite ways, but can be distinguished by the Baryon Acoustic Oscillation signature seen in the $\Omega_m$ derivative. Furthermore, as in the halo PDF case, the sensitivity of $\Omega_m$ to the growth rate accounts for the twisting of the contour orientations across redshift. The power spectrum constraints are less affected by  shot noise which lowers the constraining power of the PDF at increasing redshift. The lower panel shows the significant impact of marginalising over the bias which flips the $\sigma_8$-$\Omega_m$ contour and significantly widens constraints. This is because the profiles of the $\sigma_8$ and $\beta$ derivatives seen in Figure~\ref{fig:halo_Pkderivs} are very similar, since these both affect the amplitude in similar ways. On the other hand, $\beta$ and $\Omega_m$ affect the amplitude in opposite ways. This accounts for the diagonal contours of different orientation seen in the $\beta_z$-$\Omega_m$ and $\beta_z$-$\sigma_8$ panels. When marginalised over the bias parameter $\beta$ the $\sigma_8$-$\Omega_m$ contours are very similarly diagonally orientated, as the affects can no longer be well distinguished. This can be understood by the similar $\sigma_8$ and $\beta$ derivative profiles leading to the remaining part of the $\sigma_8$ derivative becoming less distinguishable from $\Omega_m$. \vfill

\subsubsection{Complementarity of tracer PDF and power spectrum}

The halo PDF is expected to be complementary to the halo power spectrum as it extracts additional non-Gaussian information encoded in the shape around its peak. To assess the constraining power more quantitatively, we look at Fisher forecasts for the two probes individually and their combination. 

As seen in the upper panels of Figures~\ref{fig:fisherz}~and~\ref{fig:fisherz_Pk}, at fixed bias the halo power spectrum outperforms the PDF. When the cosmology is fixed the constraints on the bias parameters from both probes are very similar and hence not shown separately. As seen in the lower panels of Figures~\ref{fig:fisherz}~and~\ref{fig:fisherz_Pk}, when jointly constraining cosmology and bias the PDF outperforms the power spectrum. While the $\sigma_8$-$\beta$ contours are similarly aligned for PDF and power spectrum, the PDF constraints are much tighter. This demonstrates that the additional non-Gaussian information captured in the PDF can break the degeneracy between $\sigma_8$ and bias that is present in the variances and correlation function. When using the PDF, combining different scales effectively captures some compressed information from the power spectrum, namely the variances at the different scales, which is augmented by the non-Gaussian shape around the peak. 

\begin{figure}
    \centering
    \includegraphics[width=0.9\columnwidth]{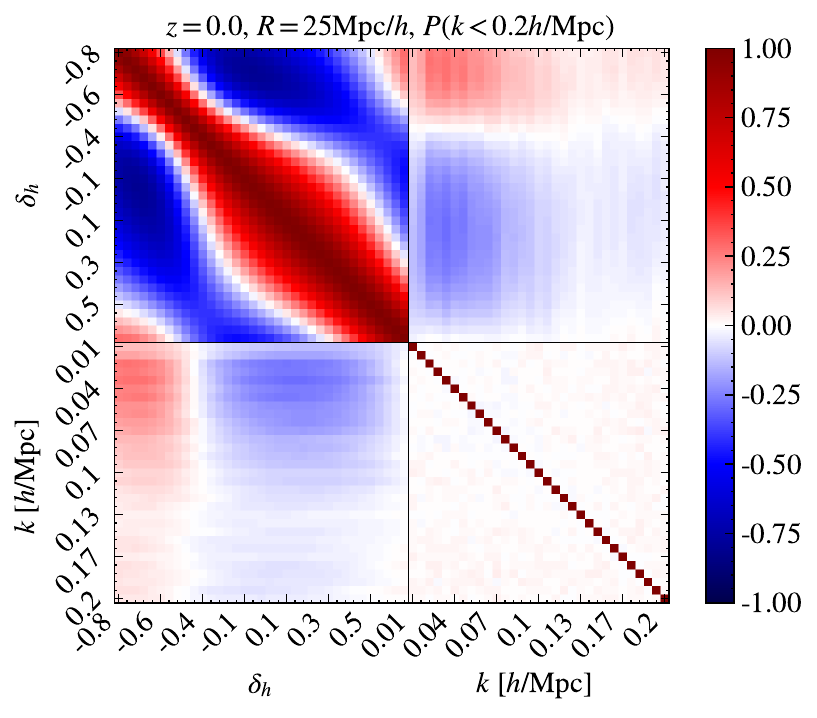}
    \caption{Cross correlation matrix of halo PDF at smoothing scale $R=25{\rm Mpc}/h$ and halo power spectrum at redshift $z=0$.}
    \label{fig:covcorr_mats_Pk}
\end{figure}

\paragraph{Covariance between probes \& super-sample covariance} For the combined constraints of the tracer PDFs and the power spectrum  we take their full cross-correlation matrix into account, an example of which is shown in Figure~\ref{fig:covcorr_mats_Pk} for halos at a single redshift, with the PDF at a single scale and the power spectrum. If we focus on mildly nonlinear scales, the tracer power spectrum correlation-matrix is diagonal. The cross-correlation between the PDF and the power spectrum stems from the correlation of the PDF variance with the amplitude of the power spectrum. The band-like structure is caused by the broadness of the smoothing kernel covering all mildly nonlinear $k$-scales shown. 

\begin{figure*}
    \centering
    \includegraphics[width=0.68\textwidth]{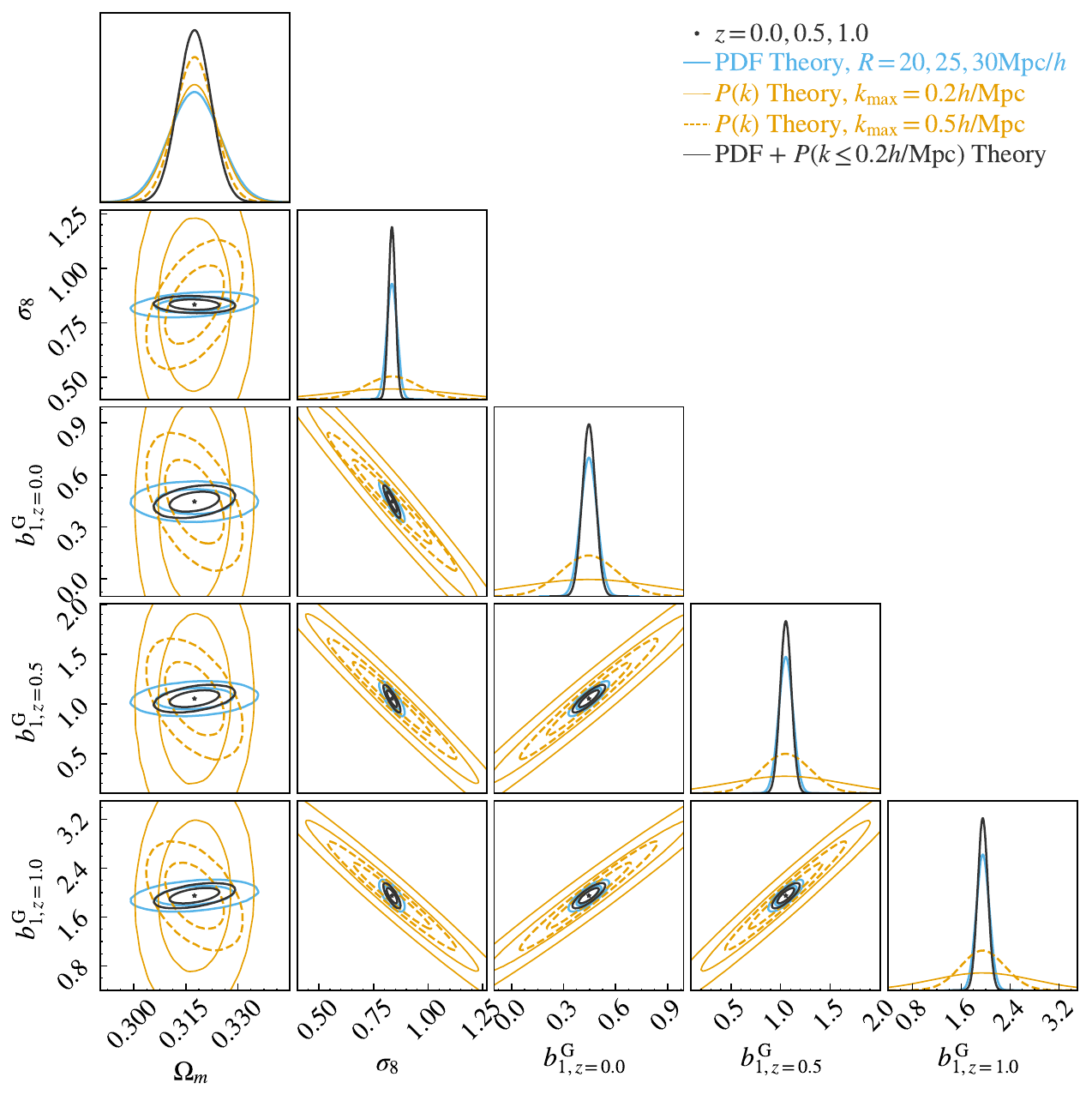}
    \caption{Comparison of constraints from halo PDF theory (blue lines) using fiducial renormalised Gaussian Lagrangian bias fits for all cosmologies and halo power spectra from theory (orange) and their combination (black), shown for the cosmological parameters $\{\Omega_m,\sigma_8\}$ and 1 bias parameter $b^\mathrm{G}_1=b^\mathrm{G}_1-1$ for each redshift. Constraints are marginalised over all other bias and shot noise parameters. Results for the halo power spectrum at different $k_{\rm max}$ (orange solid, dashed) illustrate that only information beyond the mildly nonlinear regime can break the degeneracies present in the perturbative regime.}
    \label{fig:fisher_PDF_Pk}
\end{figure*}

To estimate the impact of super-sample covariance effect that is driven by the background density, we use the separate-universe style `DC' runs of the Quijote simulation suite emulating a background density contrast $\delta_b=\pm 0.035$ through changed cosmological parameters and simulation snapshot times from the separate universe approach \citep{Sirko2005}.
The super-sample covariance between two data vector entries $D_i$ and $D_j$ can be estimated by 
\begin{equation}
\label{eq:cov_SSC_SU}
  \text{cov}^{\rm SSC}_{\rm SU}(D_i,D_j)
  = \sigma_b^2 \frac{\partial D_i}{\partial \delta_b}
               \frac{\partial D_j}{\partial \delta_b},
\end{equation}
where $\sigma_b^2$ is the variance of $\delta_b$ (here just $\delta_b^2$), and the other two
terms encode the linear response of the data vector, which can be determined from the simulations using finite differences. We add this super-sample covariance term to the measured covariance for both the PDFs and the power spectra. 

\paragraph{Marginalising over bias and shot noise parameters} To perform more realistic forecasts between the halo PDF and power spectrum, we have to consider a larger set of independent tracer bias and stochasticity parameters. We proceed with a forecast that varies two cosmological parameters along with one linear bias parameter per redshift, shared between the PDF $b_1^{\rm G}$ and the power spectrum $b_1=1+b_1^{\rm G}$.

Figure~\ref{fig:fisher_PDF_Pk} shows the Fisher forecast for the halo PDF and power spectrum from theory for the cosmological parameters $\{\Omega_m,\sigma_8\}$ and one shared linear bias parameter $b_{1,z}^{\rm G}$ for each redshift. The analysis combines redshifts $z=0.0,0.5,1.0$ and for the PDFs the smoothing scales $R=20,25,30$ Mpc/$h$. Those constraints are marginalised over one set of additional bias and shot noise parameters per redshift, $\{b_2^{\rm G},\alpha_0,\alpha_1,\alpha_2\}$ for the PDFs across all three scales, and $\{b_{2,k^2},\bar\alpha_0\}$ for the power spectrum. The contours are shown for the halo PDFs and two sets of contours for the power spectrum with $k_\mathrm{max}=0.2,0.5 h/$Mpc respectively. We observe that at mildly nonlinear scales, the halo power spectrum suffers from the strong degeneracy between $\sigma_8$ and linear bias $b_1$ and demonstrate that including more nonlinear information through an increase $k_{\rm max}$ increasingly breaks degeneracies. The combined constraints between the mildly nonlinear halo PDF and power spectrum (with the lower $k_\mathrm{max}$) are shown in black.
In the three right-hand columns of Figure~\ref{fig:fisher_PDF_Pk}, we see that the halo PDF is much better at constraining the bias parameters and indeed dominates the combined constraints. In order to constrain two bias and three shot-noise parameters for the PDF per redshift, combining different scales is crucial. Additionally, combining different redshifts tightens the cosmological constraints which in turn helps to fix the bias and stochasticity parameters, for both the PDF and the power spectrum.

Including super-sample covariance leads to a mild widening of the contours on both the PDFs and power spectrum. For $\sigma_8$ errors increase by 4\% and 11\% for the halo PDFs and power spectra, respectively. For $\Omega_m$ errors increase by 3\% and less than 1\%. For the three $b^\mathrm{G}_{1,z}$ parameters at z=0,0.5,1 the errors widen by \{16,25,5\}\% and \{11,5,4\}\% for the halo PDF and power spectrum, respectively. This is also accompanied by a minor rotation in the degeneracy directions.

\section{Conclusions}
\label{sec:conclusions}

\subsection{Summary}
In this paper we have adapted a model of the conditional PDF of tracer counts given matter density~\eqref{eq:condPDF} (introduced for photometric data in \cite{Friedrich2022}) to a spectroscopic setting. We have shown that a 2-parameter Gaussian Lagrangian bias model recently proposed in \cite{Stuecker2024,Stuecker2024_cumulantbias} and a quadratic shot noise model provide accurate parameterisations of its two ingredients for both halos and HOD-populated galaxies. We validated the theoretical predictions against the tracer PDFs for dark matter halos extracted from the \Quijote suite of N-body simulations and galaxies from the associated \Molino suite. In both cases we find excellent agreement of no more than 2\% around the central region of the PDF. 
We related our conditional PDF parameters to the power spectrum bias and stochasticity parameters. The Eulerian linear bias in the power spectrum and first Gaussian Lagrangian parameter are roughly related by $b_1^\mathrm{E} \approx 1+b_1^\mathrm{G}$. The renormalised Gaussian Lagrangian bias model we adopt gives bias parameters that are very close to scale-dependent across different scales. The quadratic shot noise parameters show some scale-dependence, although we found its impact on the tracer PDFs to be mild. The power spectrum shot noise amplitude $\bar\alpha_0$ is close to the shot noise amplitude obtained from variances that can be linked to the quadratic shot noise model as discussed in Appendix~\ref{app:bias_stochasticity_link}. This is a promising first step towards allowing a joint analysis of the tracer PDF and power spectrum with shared bias and stochasticity parameters.

We validated the response of the halo and galaxy PDF to changes in cosmology (at fixed bias) and tracer selection by comparing parameter derivatives with respect to $\{\Omega_m,\sigma_8\}$ and one effective bias parameter $\beta_z$ per redshift in Figure~\ref{fig:derivatives}. We also validated the constraining power of the theory on the same parameters when combining different redshifts with a Fisher forecast in Figure~\ref{fig:fisher_val}.

After validating the constraining power of the theory, we performed Fisher forecasts at fixed number density with the theoretical model comparing the tracer PDF with the power spectrum. Figure~\ref{fig:fisher_PDF_Pk} shows the constraints for the parameters \{$\Omega_m,\sigma_8,b_{1,z}^{\rm G}$\} for both halo PDF and power spectrum theory and their combination across three redshifts. In our model, the Eulerian linear bias of the power spectrum is tied to the renormalised Lagrangian Gaussian bias of the PDF as $b_1=1+b_1^{\rm G}$. The strength of the PDF is disentangling the effect of changing bias to that of changing $\sigma_8$. 
The combined constraints are dominated by the PDF, but tightened by additional degeneracy breaking. While our adopted tracer power spectrum model (combining the \textsc{halofit} matter power spectrum with a linear but scale-dependent bias and a white noise stochasticity) is simplistic, state-of-the-art perturbation-theory based models will likely contain more free parameters.

The spectroscopic one-point tracer PDF is a promising probe of cosmology on mildly non-linear scales and complementary to standard two-point statistics.

\subsection{Outlook}

In this work we focused on biased tracers in real space. For an application to spectroscopic clustering survey data, we will need to understand their one-point statistics in redshift space. Therefore, future work will need to model the effect of redshift space distortions. Previous work on incorporating the impact of redshift-space distortions through a modification of the matter variance \citep{repp_galaxy_2020} or absorbing it in an effective bias model \citep{uhlemann_question_2018} will be useful starting points for this. To extract additional information from redshift space distortions, the concept of a 2D-kNN statistics  distinguishing between radial and angular distances \citep{Yuan2023kNN-2D} could be translated to the PDF by using spheroidal cells.

Here we use simulated covariance matrices for the tracer PDFs and power spectra calculated from a large number of realisations at a fiducial cosmology. \cite{Uhlemann2022cov} shows that it is possible to predict covariances for the 3D matter PDF from theory, including the predictions for the super-sample covariance. Future work could extend this to tracer PDFs and their cross-correlation with tracer power spectra.

Our tracer PDF is parameterised by two bias parameters $\{b_{1}^\mathrm{G},b_{2}^\mathrm{G}\}$ and three shot noise parameters $\{\alpha_{0},\alpha_{1},\alpha_{2}\}$ per redshift, where the latter can vary across smoothing scale. Our Fisher analysis only includes one shared linear bias parameter between halo PDFs of all scales and the power spectrum at each redshift, and marginalises over one set of nonlinear bias and stochasticity parameters for the PDFs across three scales and the power spectrum, respectively. Better understanding of effective tracer bias parametrisations \citep[see e.g.][]{Banerjee2022kNN-HEFT} and stochasticity parametrisations \citep[including HOD-informed parameter bounds][]{Britt2024} will be required to leverage the power of theoretical one-point statistic models for jointly constraining cosmology and astrophysical parameters from data. 

\section*{Acknowledgements}
We thank Francisco Villaescusa-Navarra and the \Quijote team for making the data accessible via Globus and the Binder. We thank Alexandre Barthelemy, Sandrine Codis, Neal Dalal, Daniel Gr\"un and Alex Eggemeier for useful discussions. The figures in this work were created with \textsc{matplotlib} \citep{matplotlib} making use of the \textsc{numpy} \citep{numpy}, \textsc{scipy} \citep{2020SciPy-NMeth}, and \textsc{ChainConsumer} \citep{ChainConsumer} Python libraries.  
BMG is supported by a PhD studentship of the Robinson Endowment at Newcastle University. LC and CU were supported by the STFC Astronomy Theory Consolidated Grant ST/W001020/1 from UK Research \& Innovation. CU was also supported by the European Union (ERC StG, LSS\_BeyondAverage, 101075919). BMG, LC and CU are grateful for the hospitality of Perimeter Institute where the part of this work was carried out. Research
at Perimeter Institute is supported in part by the Government of Canada through the Department of Innovation, Science and Economic Development and by the Province of Ontario through the Ministry of Colleges and Universities. CU's research was also supported in part by the Simons Foundation through the Simons Foundation Emmy Noether Fellows Program at Perimeter Institute. OF was supported by a Fraunhofer-Schwarzschild-Fellowship at Universitätssternwarte München (LMU observatory) and by DFG's Excellence Cluster ORIGINS (EXC-2094 – 390783311). This work was supported by the Deutsche Forschungsgemeinschaft (DFG, German Research Foundation) via the PaNaMO project (project number 528803978).

\section*{Data Availability}

The data products were extracted with the public code \href{https://github.com/franciscovillaescusa/Pylians3}{\textsc{Pylians3}} from the publicly available Quijote simulations \citep{villaescusa-navarro_quijote_2019} and the Molino catalogs \citep{Hahn:2020lou}. The measured tracer PDFs, conditional moments and power spectra are made available on the \href{https://quijote-simulations.readthedocs.io/en/latest/access.html}{\Quijote Globus}.

\newpage
\bibliography{references.bib} 

\appendix

\section{Tracer-matter connection}
\label{app:tracer_matter}
\subsection{Fine sampling limits for tracer PDFs}
\label{sec:fine_sampling}
In the limit of fine sampling $\bar N_t\rightarrow \infty$, the conditional PDF~\eqref{eq:condPDF} and the integral to obtain the tracer count PDF~\eqref{eq:tracerPDF} with the matter PDF can be simplified. In the spirit of large-deviations theory, we start by writing the conditional PDF for $N_t=\bar N_t (1+\delta_t)$ given matter density $\delta_m$ in terms of a function that is exponentially decaying with increasing $\bar N_t$
\begin{subequations}
    \begin{align}
\label{eq:pre_rate_function_cond}
-\frac{\ln \mathcal P(N_t|\delta_m)}{\bar N_t}=&\frac{1+b(\delta_m)}{\alpha(\delta_m)}-\frac{1+\delta_t}{\alpha (\delta_m)}\ln\left(\frac{\bar N_t(1+b(\delta_m))}{\alpha(\delta_m)}\right)\notag\\
&+\frac{1}{\bar N_t}\ln\Gamma\left(1+\frac{\bar N_t(1+\delta_t)}{\alpha(\delta_m)}\right)\\
&+\frac{\ln \alpha(\delta_m)}{\bar N_t}\,.\notag
\end{align}
We obtain a rate function~\eqref{eq:rate_function}
 from the limit $\bar N_t\rightarrow \infty$ (corresponding to a driving parameter $\epsilon=1/\bar N_t\rightarrow 0$)\footnote{Note that even given a large $\bar N_t$, the large $N_t$ limit for underdensities $\delta_t<0$ is only achieved for small $|\delta_t|\ll 1$.}
\begin{align}
\label{eq:rate_function_cond}
-\lim_{\bar N_t\rightarrow \infty} \frac{\ln \mathcal P(N_t|\delta_m)}{\bar N_t}
:= \psi_{t|m}(\delta_t;\delta_m)\,.
\end{align}
We use Stirling's formula for large arguments of the Gamma function (reducing to the factorial for integers) \begin{equation}
    \label{eq:Stirling_approx}
    \ln \Gamma(1+x)\approx x(\ln x - 1)+\ln[(2\pi x)^{1/2}]\,,
\end{equation}
where the second term does not affect the exponential decay, but is relevant for the  normalisation. We find
\begin{align}
\psi_{t|m}(\delta_m;\delta_t) = \frac{b(\delta_m)-\delta_t + (1+\delta_t)\ln\left(\frac{1+\delta_t}{1+b(\delta_m)}\right)}{\alpha(\delta_m)}\,.
\end{align}

\subsubsection{Fine sampling limit for the conditional tracer PDF}
To obtain the fine-sampling limit for the conditional PDF $\mathcal P(N_t|\delta_m)$, we can Taylor-expand the function $\psi_{t|m}(\delta_t;\delta_m)$ around its minimum at the most likely value $\delta_t^*$ given $\delta_m$. We have that $\psi_{t|m}(\delta_t^*)=0$ for $\delta_t^*=b(\delta_m)$. The first derivative w.r.t. to $\delta_t$ is
\begin{align}
\psi_{t|m}'(\delta_t)=\frac{\ln\left(\frac{1+\delta_t}{1+b(\delta_m)}\right)}{\alpha(\delta_m)}\,,
\end{align}
and thus zero at $\delta_t^*$ showing that is is indeed an extremum. The second derivative yields
\begin{align}
\psi_{t|m}''(\delta_t)=\frac{1}{\alpha(\delta_m)(1+\delta_t)} > 0\,,
\end{align}
which shows the extremum is in fact a minimum. With this we can write
\begin{equation}
\label{eq:rate_function_cond_Gaussian_t}
\psi_{t|m}(\delta_t)\approx \frac{(\delta_t-b(\delta_m))^2}{2\alpha(\delta_m)(1+b(\delta_m))}\,.
\end{equation}
After having identified the peak of the PDF at $\delta_t^*=b(\delta_m)$ in the limit of $\bar N_t\rightarrow \infty$, we can attempt to restore the normalisation at finite $\bar N_t$. Considering the last term  of the original exponent~\eqref{eq:pre_rate_function_cond} along with the sub-leading term in the Stirling approximation~\eqref{eq:Stirling_approx}, we obtain
\begin{align}
\mathcal P(N_t|\delta_m)
\approx &\left[2\pi N_t \alpha(\delta_m)\right]^{-\frac{1}{2}}\exp\left[-\bar N_t\psi_{t|m}(\delta_t)\right] \notag\,,
\end{align}
where the exponential is quadratic as given by equation~\eqref{eq:rate_function_cond_Gaussian_t} and $N_t=\bar N_t (1+\delta_t)$. This shows that the conditional PDF $\mathcal P(N_t|\delta_m)$ tends to a Gaussian with mean $\bar N_t(1+b(\delta_m))$ and variance $\alpha(\delta_m)\bar N_t(1+b(\delta_m))$.
We can convert this to the conditional PDF of tracer density contrast 
\begin{align}
\label{eq:decay_rate_function_cond_delta}
\mathcal P(\delta_t|\delta_m) 
\approx & \left[2\pi\frac{\alpha(\delta_m)(1+b(\delta_m))}{\bar N_t}\right]^{-\frac{1}{2}}\\
&\times \exp\left[-\bar N_t\frac{(\delta_t-b(\delta_m))^2}{2\alpha(\delta_m)(1+b(\delta_m))}\right]\,, \notag
\end{align}
which is a Gaussian with mean $b(\delta_m)$ and variance $\alpha(\delta_m)(1+b(\delta_m))/\bar N_t$.
\end{subequations}

\subsubsection{Fine sampling limit for the tracer PDF}

Let us now use similar arguments to simplify the tracer PDF~\eqref{eq:tracerPDF} given as $\delta_m$-integral over the product of the conditional PDF~\eqref{eq:condPDF} and the matter PDF. For the $\delta_m$ integral in equation~\eqref{eq:tracerPDF}, $\delta_t$ can be considered a constant.
Evidently, $\psi_{t|m}(\delta_m^*)=0$ for $\delta_m^*$ such that $\delta_t=b(\delta_m^*).$ Let us consider the first derivative of $\psi_{t|m}(\delta_m;\delta_t)$ w.r.t $\delta_m$
\begin{align}
   \psi_{t|m}'(\delta_m)=\frac{b'(\delta_m)}{\alpha(\delta_m)} \left[1-\frac{1+\delta_t}{1+b(\delta_m)}\right] 
   -\frac{\alpha'(\delta_m)}{\alpha(\delta_m)}\psi_{t|m}(\delta_m)\,,
\end{align}
which also vanishes at $\delta_m^*$ making this an extremum. The second derivative is 
\begin{align}
\psi_{t|m}''(\delta_m)&=
\frac{b'(\delta_m)}{\alpha(\delta_m)} \frac{b'(\delta_m)(1+\delta_t)}{(1+b(\delta_m))^2}\\
&+
\left(\frac{b'(\delta_m)}{\alpha(\delta_m)}\right)' \left[\frac{b(\delta_m)-\delta_t}{1+b(\delta_m)}\right]\notag\\
&-\left(\frac{\alpha'(\delta_m)}{\alpha(\delta_m)}\right)'\psi_{t|m}(\delta_m)
-\frac{\alpha'(\delta_m)}{\alpha(\delta_m)}\psi_{t|m}'(\delta_m) \notag \,.
\end{align}
At the extremum $\delta_m^*$ where $\delta_t=b(\delta_m^*)$ only the first term remains
\begin{equation}
\psi_{t|m}''(\delta_m^*)=\frac{b'(\delta_m^*)^2(1+\delta_t)}{\alpha(\delta_m^*)[1+b(\delta_m^*)]^2}=\frac{b'(\delta_m^*)^2}{\alpha(\delta_m^*)(1+\delta_t)} > 0\,,
\end{equation}
which shows that $\delta_m^*$ is a minimum. This leads to the conditional PDF~\eqref{eq:condPDF} resembling a Gaussian function around the saddle point $\delta_m^*(\delta_t)$ 
\begin{align}
\label{eq:rate_function_cond_Gaussian}
   \exp\left[-\bar N_t \psi_{t|m}(\delta_t;\delta_m)\right] &\propto \exp\left[-\frac{[\delta_m-\delta_m^*(\delta_t)]^2}{2\sigma_*^2(\delta_m^*(\delta_t))}\right]\\
   \sigma_*^2(\delta_m^*(\delta_t))&=\frac{\alpha(\delta_m^*(\delta_t))(1+\delta_t)}{\bar N_t b'(\delta_m^*(\delta_t))^2}
   \,.
\end{align}
As before, the normalisation is $[2\pi N_t\alpha(\delta_m^*(\delta_t))]^{-1/2}$ with $N_t=\bar N_t(1+\delta_t)$. When converting the PDF ot the tracer count $\mathcal P_t(N_t)$ to the PDF of the tracer density contrast $\mathcal P_t(\delta_t)$, we obtain an extra factor of $\bar N_t$ such that the prefactor changes to $[2\pi \alpha(\delta_m^*(\delta_t))(1+\delta_t)/\bar N_t]^{-1/2}$. This can be rewritten as $(2\pi \sigma_*^2 )^{-1/2}/b'(\delta_m^*(\delta_t))$.
The integral becomes a convolution of the matter PDF with a zero-mean Gaussian of variance $\sigma_*^2(\delta_t)$
such that
\begin{equation}
\mathcal P_t(\delta_t) \approx \frac{1}{b'(\delta_m^*(\delta_t))}\Big(\mathcal P_G\!\left(\sigma_*^2(\delta_t)\right)\!\,*\mathcal P_m\Big) \left(\delta_m^*(\delta_t)\right)\,.
\end{equation}
We can see that in the asymptotic limit of $\bar N_t\rightarrow \infty$, the Gaussian becomes a delta function and we obtain 
\begin{equation}
    \mathcal P_t(\delta_t) \stackrel{\bar N_t\rightarrow \infty}{\longrightarrow} \frac{1}{b'(\delta_m^*(\delta_t))} \mathcal P_m \left(\delta_m^*(\delta_t)\right)\,,
\end{equation}
where we can replace $\delta_m^*=b^{-1}(\delta_t)$. In this model, the asymptotic limit thus corresponds to a deterministic bias where $\mathcal P_t(\delta_t) = \mathcal P_m(\delta_m(\delta_t)) d\delta_m/d\delta_t$. Note that even if the PDF of $\delta_m^*$ was Gaussian, a nonlinear bias $\delta_t=b(\delta_m^*)$ would imply a non-Gaussian PDF for $\delta_t$.

If the matter PDF was a zero-mean Gaussian with variance $\sigma_m^2$, then the convolution would be another Gaussian function with an increased density-dependent variance $\sigma^2(\delta_t)=\sigma_m^2+\sigma_*^2(\delta_t)$. Thus, the exponent of the PDF for $\delta^*_m(\delta_t)$ would agree with the one of matter PDF evaluated at $\delta_m(\delta_t)=\delta_m^*(\delta_t)/\sqrt{1+\sigma_*^2(\delta_t)/\sigma_m^2}$. Note that even if the PDF of $\delta_m^*$ was Gaussian and the bias was linear, the $\delta_t$-dependence of the variance would imply a non-Gaussian PDF for $\delta_t$.

\subsection{Tracer-matter spectra and covariances}
\label{app:bias_stochasticity_link}
\subsubsection{Tracer-matter cross and auto spectra}
Here we further connect the conditional mean bias and stochasticity models presented in Sections~\ref{subsec:condmean}~and~\ref{subsec:condvar} to the bias and stochasticity in the power spectrum used in Section~\ref{sec:pktheory}. Let us briefly summarise how a Eulerian quadratic bias model manifests in the tracer cross and auto power spectra. We start with the noise-free tracer density
\begin{subequations}
\begin{align}
\label{eq:quad_bias_field}
\tilde\delta_t(\vec{x})=&b_1\delta_m(\vec{x})+\frac{b_2}{2}\left(\delta_m(\vec{x})^2-\langle \delta_m(\vec{x})^2\rangle\right)\\
&+\frac{b_{s^2}}{2} \left(s_{ij}(\vec{x})^2-\langle s_{ij}(\vec{x})^2\rangle\right)
\,,\notag
\end{align}
where $b_{s^2}$ describes the non-local bias associated with the tidal shear $s_{ij}=(\partial_i\partial_j/\partial^2-\delta_{ij}/3)\delta_m(\vec{x})$. In Fourier space this translates to
\begin{align}
\label{eq:quad_bias_field_k}
\tilde\delta_t(\vec{k})=&b_1\delta_m(\vec{k})+\frac{b_2}{2} \overbrace{\int\! d^3q\, \delta_m(\vec{k}-\vec{q})\delta_m(\vec{q})}^{(\delta_m* \delta_m)(\vec{k})}\\
&+\frac{b_{s^2}}{2}\underbrace{\int\! d^3q\, \delta_m(\vec{k}-\vec{q})\delta_m(\vec{q})S_2(\vec{q},\vec{k}-\vec{q})}_{(s_{ij}*s_{ij})(\vec{k})}\,,\notag
\end{align} 
where $S_2(\vec{q_1},\vec{q_2})=\vec{q_1}\cdot\vec{q_2}/(q_1^2q_2^2)-1/3$ and we omitted the $\delta$-function terms that arise from the subtraction of the constant expectation values $\langle \cdot \rangle$. With this we obtain the cross power spectrum
\begin{align}
    P_{tm}(k)&=\langle\delta_t(\vec{k})\delta_m(-\vec{k})\rangle\notag\\
\label{eq:Pk_tracer_cross_Eulerian}
    &=b_1 P_m(k)+\frac{b_2}{2} \int\! d^3q B_m(\vec{q},\vec{k}-\vec{q},-\vec{k})\notag\\
    &+\frac{b_{s^2}}{2} \int\! d^3q S_2(\vec{q},\vec{k}-\vec{q}) B_m(\vec{q},\vec{k}-\vec{q},-\vec{k})\,,
\end{align}
which depends on the matter power spectrum $P_m$ and two integrals over the bispectrum $B_m$, one of which is the skew-spectrum. The bias function~\eqref{eq:bias_Pk} is then obtained as
\begin{align}
\label{eq:bias_Pk_Eulerian}
b(k) = \frac{P_{tm}(k)}{P_m(k)}&= b_1+ \frac{b_2}{2} \frac{\int\! d^3q B_m(\vec{q},\vec{k}-\vec{q},-\vec{k})}{P_m(k)}\\
&+\frac{b_{s^2}}{2} \frac{\int\! d^3q S_2(\vec{q},\vec{k}-\vec{q}) B_m(\vec{q},\vec{k}-\vec{q},-\vec{k})}{P_m(k)}\,,\notag
\end{align}
where the quadratic terms term would typically be evaluated within perturbation theory and lead to either constant terms that combine with $b_1$ or a leading-order scale dependence of $k^2$.
\end{subequations}
For the case of just local bias (thus neglecting the tidal bias term $b_{s^2}$), the tracer auto power spectrum is obtained as
\begin{subequations}
\begin{align}
\label{eq:Pk_tracer}
    P_{t}(k)&=\langle\delta_t(\vec{k})\delta_t(-\vec{k})\rangle\notag\\
    &=b_1^2 P_m(k)+b_1b_2\langle(\delta_m*\delta_m)(\vec{k})\delta_m(-\vec{k})\rangle\\
    &+\frac{b_2^2}{4} \langle(\delta_m*\delta_m)(\vec{k})(\delta_m*\delta_m)(\vec{-k})\rangle + P_\epsilon(k) \notag\\
    &=b_1^2 P_m(k)+b_1 b_2 \int\! d^3q B_m(\vec{q},\vec{k}-\vec{q},-\vec{k})\\
    &+\frac{b_2^2}{4}\!\!\int\!\!\!\!\int\!\!d^3q_1 d^3q_2 T_m(\vec{q_1},\vec{q_2},\vec{k}-\vec{q_1},-\vec{k}-\vec{q_2}) + P_\epsilon(k)\notag\,,
\end{align}
which depends on the matter power spectrum, the skew-spectrum and an integrated version of the trispectrum $T_m$ and the shot noise power spectrum $P_\epsilon(k)=\langle\epsilon(\vec{k})\epsilon(-\vec{k})\rangle$. The tidal bias terms will involve similar integrals with additional weightings with one $S_2$ kernel for the bispectrum and two for the trispectrum term.
The shot noise defined in~\eqref{eq:shotnoise_Pk} would then be given by
\begin{align}
\label{eq:shotnoise_Pk_Eulerian}
\frac{\alpha(k)}{\bar n}
&=\frac{b_2^2}{4} \Bigg[\!\!\int\!\!\!\!\int\!\!d^3q_1 d^3q_2 T_m(\vec{q_1},\vec{q_2},\vec{k}-\vec{q_1},-\vec{k}-\vec{q_2}) \notag\\
&\qquad 
- \frac{\left(\int\! d^3q B_m(\vec{q},\vec{k}-\vec{q},-\vec{k})\right)^2}{P_m(k)} \Bigg] + P_\epsilon(k)\,,
\end{align}
where the terms including the trispectrum and square of the bispectrum are of higher perturbative order. As expected, nonlinear bias does contribute to the shot noise defined relative to the linear bias. This is in contrast to the shot noise defined based on the conditional variance~\eqref{eq:SN_def} which would vanish for any nonlinear but still deterministic local bias.
\end{subequations}

\subsubsection{Tracer-matter covariances from bias and stochasticity}
\label{app:crosscov}

Let us connect the conditional mean bias presented in Section~\ref{sec:bias_stochasticity} to the cross-covariance between tracer and halo densities. We can obtain the covariance between tracer and matter densities from the joint PDF~\eqref{eq:tracer_matter_jointPDF}
\begin{subequations}
\begin{align}
    \sigma_{tm}^2&= \langle\delta_t\delta_m\rangle=\int d\delta_m\!\! \int\! d\delta_t\, \delta_t\delta_m \mathcal P(\delta_t|\delta_m) \mathcal P_m(\delta_m) \notag\\
    &=\int  \langle \delta_t|\delta_m\rangle\delta_m \mathcal P(\delta_m) d\delta_m =\langle b(\delta_m)\delta_m\rangle\,.
    \label{eq:covariance_tracer}
\end{align}
For a quadratic Eulerian 
bias~\eqref{eq:bias_E_quad}, we have that
\begin{equation}
    \sigma_{tm}^2=b_1\sigma_m^2+\frac{b_2}{2}\langle\delta_m^3\rangle\,,
    \label{eq:covariance_tracer_Eulerian}
\end{equation}
where for brevity we wrote $b_n$ instead of $b_n^{\rm E}$ for the Eulerian bias parameters. This closely resembles the result for the  tracer-matter cross power spectrum~\eqref{eq:Pk_tracer_cross_Eulerian}. For a Gaussian field, the quadratic bias term would not affect the tracer-matter covariance.
In analogy to the power spectrum bias~\eqref{eq:bias_Pk}, we can define a bias from the ratio of the tracer-matter covariance and matter variance 
\begin{align}
    \label{eq:bias_cov}
    b_{\times,\sigma^2}&=\frac{\sigma_{tm}^2}{\sigma_m^2}=\frac{\langle b(\delta_m)\delta_m\rangle}{\langle \delta_m^2\rangle} \\
    \label{eq:bias_cov_Eulerian}
    &=b_1 + \frac{b_2}{2} \frac{\langle\delta_m^3\rangle}{\sigma_m^2} =b_1 + \frac{b_2}{2} S_3\sigma_m^2\,.
\end{align}
\end{subequations}
The presence of the second term nicely illustrates how the non-Gaussianity of the late-time matter field increases the relevance of the nonlinearity in the bias.
To obtain the variance of the tracer density $\sigma_t^2$, we will additionally require a description of shot noise. 

After determining how the conditional mean bias sets the covariance between tracer and matter densities~\eqref{eq:covariance_tracer}, let us determine how it propagates into the variance of tracer densities together with the shot noise. 
\label{eq:variance_tracer}
    \begin{align}
\sigma_{t}^2&=\langle\delta_t^2\rangle 
=\int d\delta_m\!\! \int\! d\delta_t\, \delta_t^2\mathcal P(\delta_t|\delta_m) \mathcal P_m(\delta_m) \notag\\
&=\int \overbrace{\left(\langle \delta_t|\delta_m\rangle^2+\langle\delta_t^2|\delta_m\rangle_c\right)}^{\langle \delta_t^2|\delta_m\rangle} \mathcal P(\delta_m) d\delta_m \\
&= \int  \left(\langle \delta_t|\delta_m\rangle^2 +\frac{\alpha(\delta_m)(1+\langle\delta_t|\delta_m\rangle)}{\bar N_t} \right)  \mathcal P(\delta_m) d\delta_m\notag
\end{align}
\begin{subequations}
Let us split the result into to terms. The first term is the variance for a shot-noise free tracer density $\tilde\delta_t$
\begin{align}
    \label{eq:variance_tracer_noSN}
    \tilde\sigma_t^2 &= \langle\tilde\delta_t^2\rangle =\langle b^2(\delta_m)\rangle)\,.
\end{align}
This can be expanded for a quadratic Eulerian bias~\eqref{eq:bias_E_quad}
\begin{align}
    \label{eq:variance_tracer_noSN_Eulerian}
    \tilde\sigma_t^2  &= b_1^2\sigma_m^2+b_1b_2\langle\delta_m^3\rangle + \frac{b_2^2}{4}\left(\langle\delta_m^4\rangle-\sigma_m^4\right)\,.
\end{align}
Even for a Gaussian field, the quadratic bias affects the tracer variance while the tracer-matter covariance was unaffected, similarly as for the tracer power spectra.
\end{subequations}
The second term includes the shot noise 
\begin{subequations}
\begin{align}
\sigma_{\rm \alpha}^2&=\sigma_{t}^2-\tilde\sigma_t^2 =
\frac{\langle\alpha(\delta_m)\rangle+\langle\alpha(\delta_m)b(\delta_m)\rangle}{\bar N_t}\,,
\end{align}
and can be expanded for a quadratic model~\eqref{eq:SN_quad} 
\begin{align}
\label{eq:SN_averaged}
\sigma_{\rm \alpha}^2&=   \frac{\alpha_0+(\alpha_0b_1+\alpha_1)\langle\delta_m\rangle+(\alpha_2+\alpha_1b_1)\langle\delta_m^2\rangle}{\bar N_t} \notag\\
    &+ \frac{(\alpha_1\frac{b_2}{2}+\alpha_2b_1)\langle\delta_m^3\rangle+\alpha_2\frac{b_2}{2}\langle\delta_m^2(\delta_m^2-\sigma_m^2)\rangle}{\bar N_t}\notag\\
    &= \frac{\alpha_0+(\alpha_1b_1+\alpha_2)\sigma_m^2}{\bar N_t}\\
    &+ \frac{(\alpha_1\frac{b_2}{2}+\alpha_2b_1)S_3\sigma_m^4+\alpha_2\frac{b_2}{2}(2\sigma_m^4+S_4\sigma_m^6)}{\bar N_t}\notag\,.
\end{align}
For linear bias and linear shot noise  only the first line remains, which still present an interesting coupling of the linear bias and shot noise parameters at leading order in the dark matter variance $\sigma_m^2$.
\end{subequations}

In analogy to the power spectrum shot noise with respect to linear bias~\eqref{eq:shotnoise_Pk}, one can define a shot noise parameter obtained from the covariance and variances of tracer and matter densities 
\begin{subequations}
\label{eq:SN_var}
\begin{equation}
    \alpha_{\sigma^2}=\bar N_t \left(\sigma_t^2-\frac{\sigma_{tm}^4}{\sigma_m^2}\right)=\bar N_t \left(\sigma_t^2-b_{\times,\sigma^2}^2 \sigma_m^2\right)\,,
\end{equation}
for which we list measured values for the fiducial cosmology in Table~\ref{tab:SN_variances}. For quadratic models for Eulerian bias and shot noise, the stochasticity parameter $\alpha_{\sigma^2}$
is obtained from inserting equations~\eqref{eq:bias_cov_Eulerian}~and~\eqref{eq:variance_tracer}
\begin{align}
\frac{\alpha_{\sigma^2}}{\bar N_t} 
&= \sigma_\alpha^2+\frac{b_2^2}{4}\underbrace{\left(\langle\delta_m^4\rangle-\sigma_m^4-\frac{\langle\delta_m^3\rangle^2}{\sigma_m^2}\right)}_{\sigma_m^4[2+(S_4-S_3^2)\sigma_m^2]}\,.
\end{align}
\end{subequations}
We can see that the size of the $b_2^2$ term is controlled by the square of the variance $\sigma_m^4$ and thus higher order, similarly as for the power spectrum~\eqref{eq:shotnoise_Pk_Eulerian}. In addition, we have several coupling terms between stochasticity and bias parameters $\alpha_k b_{n-k}$ scaling as $\sigma_m^4$. 

We can also obtain an associated linear correlation coefficient
\begin{equation}
    \label{eq:corrcoeff_var}
r=\frac{\sigma_{tm}^2}{\sqrt{\sigma_t^2\sigma_m^2}}
=\left(1+\frac{\alpha_{\sigma^2}}{\bar N_t b_{\times,\sigma^2}^2\sigma_m^2}\right)^{-1/2}\,,
\end{equation}
which is between $0.9$ and $0.95$ for our halo samples across different scales and redshifts.
\begin{table}
  \centering
  \begin{tabular}{c||c||c|c|c}
    \hline
     z & $\bar\alpha_0$ & \multicolumn{3}{|c}{$\alpha_{\sigma^2}(R$ [Mpc$/h$])}  
    \\ \cline{3-5}    
     &  & $R=30$ & $R=25$ & $R=20$ 
    \\ \hline \hline
     0.0 & 0.83 & 0.85 & 0.82 & 0.80 \\
    \hline
     0.5 & 0.67 & 0.65 & 0.64 & 0.63 \\
    \hline
     1.0 & 0.75 & 0.73 & 0.72 & 0.71 \\
  \end{tabular}
	\caption{Values of halo stochasticity parameters fitted from the power spectrum~\eqref{eq:shotnoise_Pk} and measured (co)variances~\eqref{eq:SN_var} 
 for the fiducial cosmology.}
	\label{tab:SN_variances}
\end{table}

\subsubsection{Higher-order tracer moments}
We can also compute higher-order moments for the tracer distribution, for example the third moment
\begin{subequations}
    \begin{align}
    \label{eq:3rdmoment_tracer}
&\langle\delta_t^3\rangle 
=\int d\delta_m\!\! \int\! d\delta_t\, \delta_t^3\mathcal P(\delta_t|\delta_m) \mathcal P_m(\delta_m) \\
&=\int \overbrace{\left(\langle \delta_t|\delta_m\rangle^3+3\langle\delta_t^2|\delta_m\rangle_c\langle \delta_t|\delta_m\rangle+\langle\delta_t^3|\delta_m\rangle_c\right)}^{\langle \delta_t^3|\delta_m\rangle} \mathcal P(\delta_m) d\delta_m \notag\,,
\end{align}
Again, we will split it into a shot-noise free term 
\begin{align}
    \label{eq:3rdmoment_tracer_det}
\langle\tilde\delta_t^3\rangle &= \langle b(\delta_m)^3\rangle\,,
\end{align}
and a shot noise contribution
    \begin{align}
\label{eq:3rdmoment_tracer_SN}
\langle\delta_t^3\rangle-\langle\tilde\delta_t^3\rangle
&=
3 \frac{\langle \alpha(\delta_m)b(\delta_m)\rangle+\langle\alpha(\delta_m)b(\delta_m)^2\rangle}{\bar N_t}\\
&\quad+ \frac{\langle \alpha(\delta_m)^2\rangle+ \langle \alpha(\delta_m)^2 b(\delta_m) \rangle}{\bar N_t^2}\notag\,,
\end{align}
where we used that $\langle N_t^3|\delta_m\rangle_c=\alpha(\delta_m)^2\bar N_t(\delta_m)$. We now consider the contributions from a quadratic Eulerian bias
\begin{align}
    \label{eq:3rdmoment_tracer_det_Eulerian}
\langle\tilde\delta_t^3\rangle 
&=b_1^3\langle\delta_m^3\rangle +\frac{3b_1b_2}{2}\left(b_1\langle\delta_m^4\rangle+\frac{b_2}{2} \langle\delta_m^5\rangle\right) + \frac{b_2^3}{8} \langle\delta_m^6\rangle\,,
\end{align}
and the coupling of linear stochasticity and bias
\begin{align}
\label{eq:3rdmoment_tracer_Eulerian_lin}
\langle\delta_t^3\rangle-\langle\tilde\delta_t^3\rangle &\stackrel{\text{lin}}{\approx}\, 3 \frac{(\alpha_1b_1+\alpha_0b_1^2)\sigma_m^2+\alpha_1 b_1^2\langle\delta_m^3\rangle}{\bar N_t}\\
&\quad +\frac{\alpha_0^2+\alpha_1(\alpha_1+2\alpha_0b_1)\sigma_m^2+\alpha_1^2b_1\langle\delta_m^3\rangle}{\bar N_t^2}\notag \,.
\end{align}
Those results could be used to approximate the tracer PDF by a shifted lognormal distribution by imposing the right variance and skewness through the lognormal shift parameter. 
\end{subequations}

\section{Additional Details \& Validations}
\label{app:validations}

\subsection{Matter PDF derivatives}
\label{app:matterPDF}

\begin{figure}[t]
    \centering
    \includegraphics[width=\columnwidth]{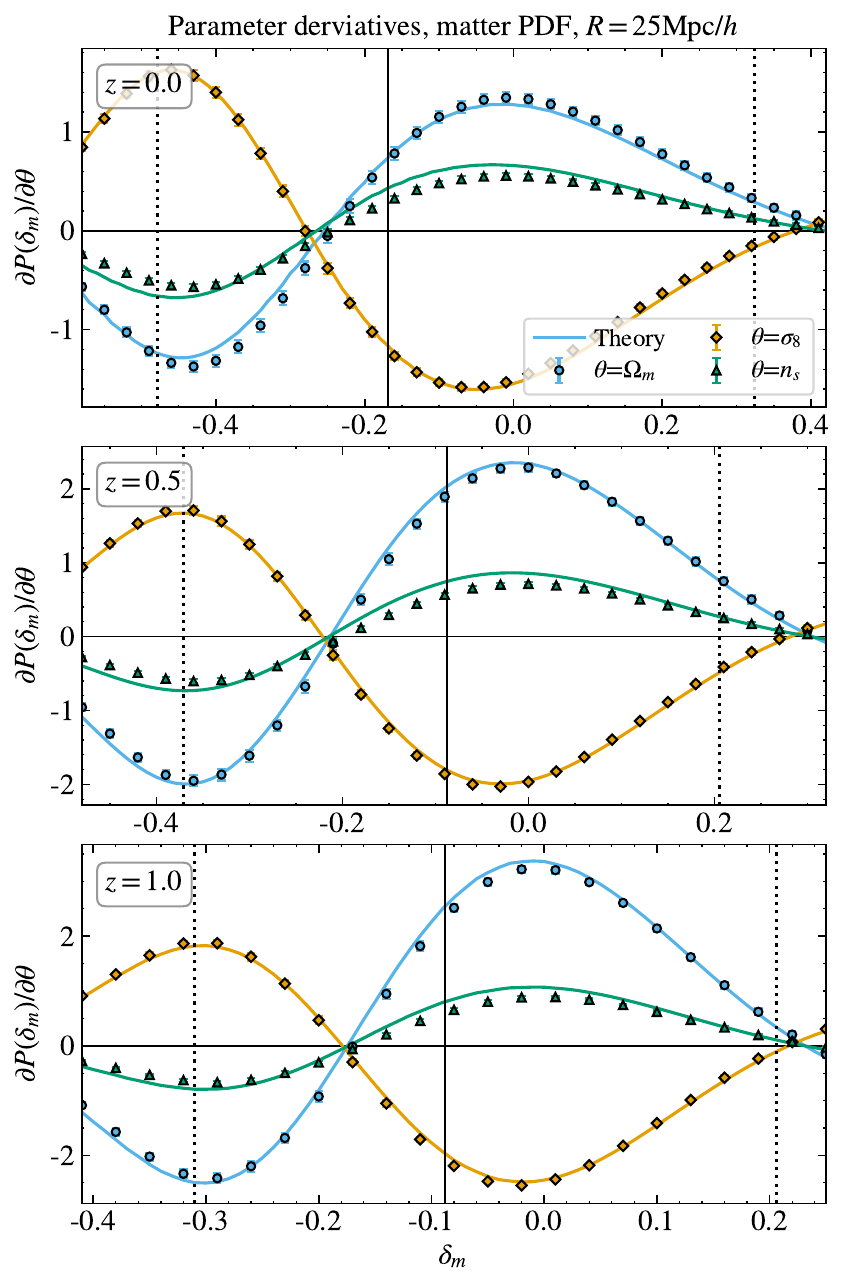}
    \caption{Derivatives of the matter PDF with respect to cosmological parameters. Theory using re-scaled non-linear matter variance for each cosmology.}
    \label{fig:matterderivs}
\end{figure}
Figure~\ref{fig:matterderivs} shows a comparison of the simulated and predicted derivatives of the matter PDFs when changing cosmological parameters. The matter PDFs for all cosmologies are predicted using a non-linear matter variance $\sigma_\mathrm{NL}^2$ rescaled via the ratio between that for the fiducial cosmology predicted in CosMomentum and measured from \Quijote. We see a good agreement, in line with previous results using a saddle-point approximation for the log-density \citep{uhlemann_fisher_2020}.

\subsection{Tracer PDF residuals}

\begin{figure}
    \centering
\includegraphics[width=\columnwidth]{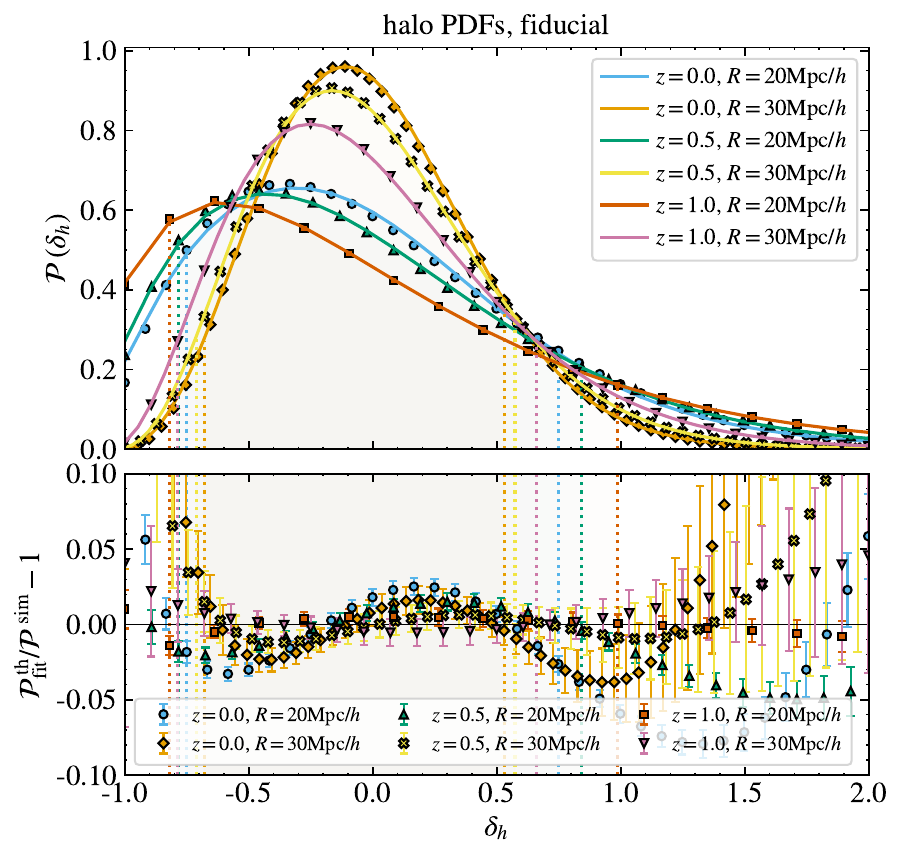}
\caption{Halo PDF as measured from Quijote (data points) and predicted from CosMomentum with renormalised Gaussian Lagrangian bias and quadratic shot noise fits.}
    \label{fig:haloPDF_fid}
\end{figure}

Figure~\ref{fig:haloPDF_fid} shows a model validation for the fiducial halo PDFs across different scales and redshifts. The theory PDFs are predicted using the fitted renormalised Gaussian Lagrangian parameters and quadratic shot noise parameters, both fitted over smoothing scales $R=20,25,30$Mpc/$h$ simultaneously.

\subsection{Tracer PDF derivatives}

\begin{figure}
    \centering
    \includegraphics[width=0.95\columnwidth]{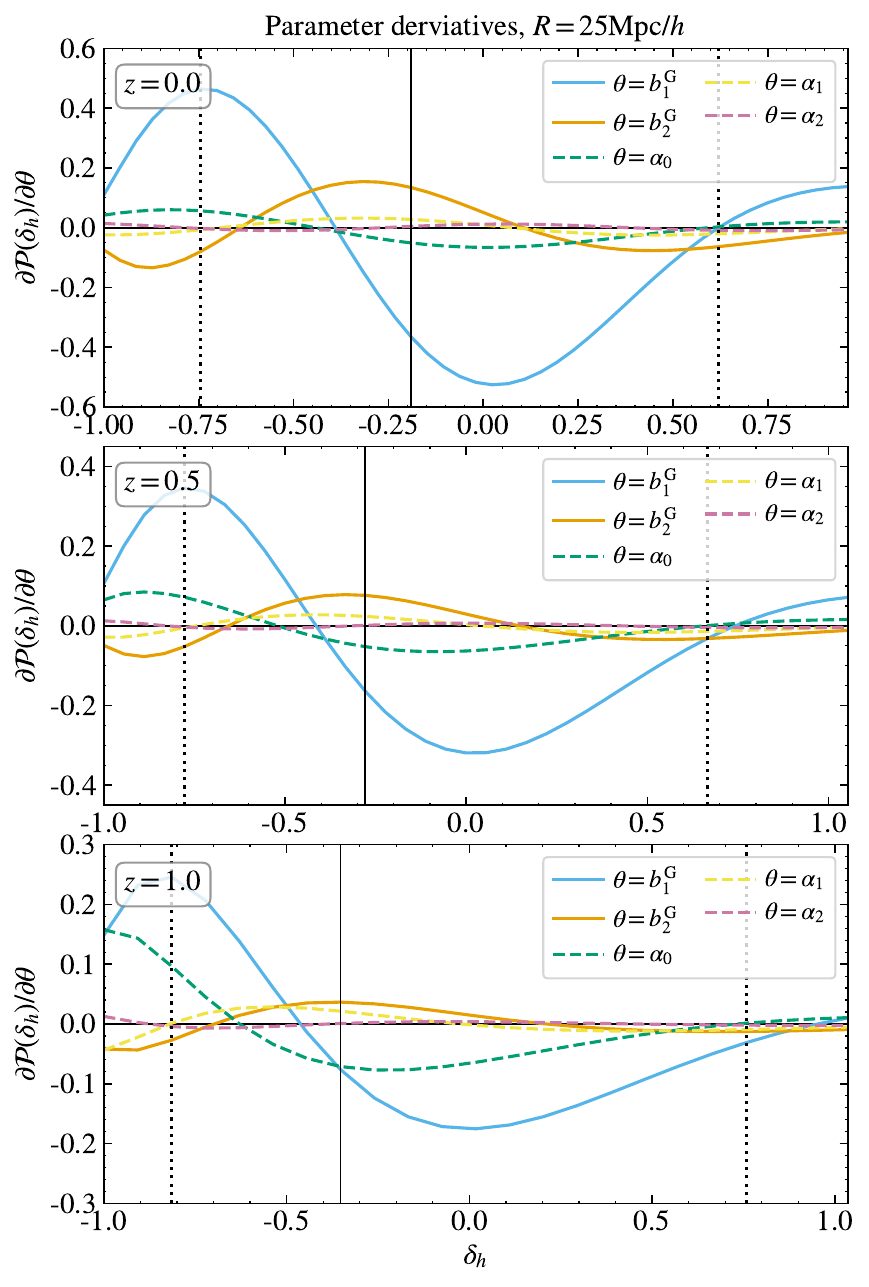}
    \caption{Halo PDF derivatives w.r.t. the renormalised Lagrangian Gaussian bias and quadratic shot noise parameters.}
    \label{fig:haloPDF_derivatives_bias}
\end{figure}

In Figure~\ref{fig:haloPDF_derivatives_bias} we show the derivatives of the tracer PDF with respect to the different PDF bias and stochasticity parameters, evaluated around the values found for the fiducial halo PDFs. The derivative with respect to the leading-order Gaussian bias $b_1^{\rm G}$ is the largest. We observe similar signatures for $\alpha_0$ and $b_1^{\rm G}$, as well as $\alpha_1$ and $b_2^{\rm G}$, which is expected from their coupling through the conditional variance $\langle\delta_h^2|\delta_m\rangle_c=\alpha(\delta_m)(1+b(\delta_m))$, as discussed for the leading order moments in a simpler Eulerian bias model in Appendix~\ref{app:crosscov}.

\subsection{Tracer power spectra}
\label{app:Pk_validation}

\begin{figure}
    \centering
\includegraphics[width=\columnwidth]{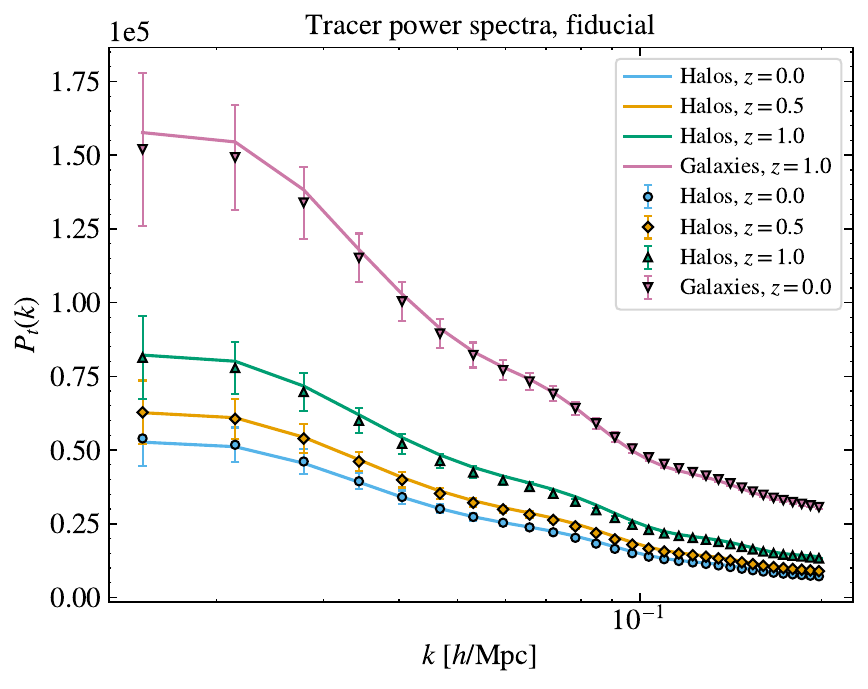}
    \caption{Halo and galaxy power spectra as measured from the fiducial \Quijote simulations or \Molino catalogues (data points) and predicted from the model~\eqref{eq:haloPk_model} (solid lines).}
    \label{fig:haloPk_model}
\end{figure}

\begin{figure}
    \centering
    \includegraphics[width=1\columnwidth]{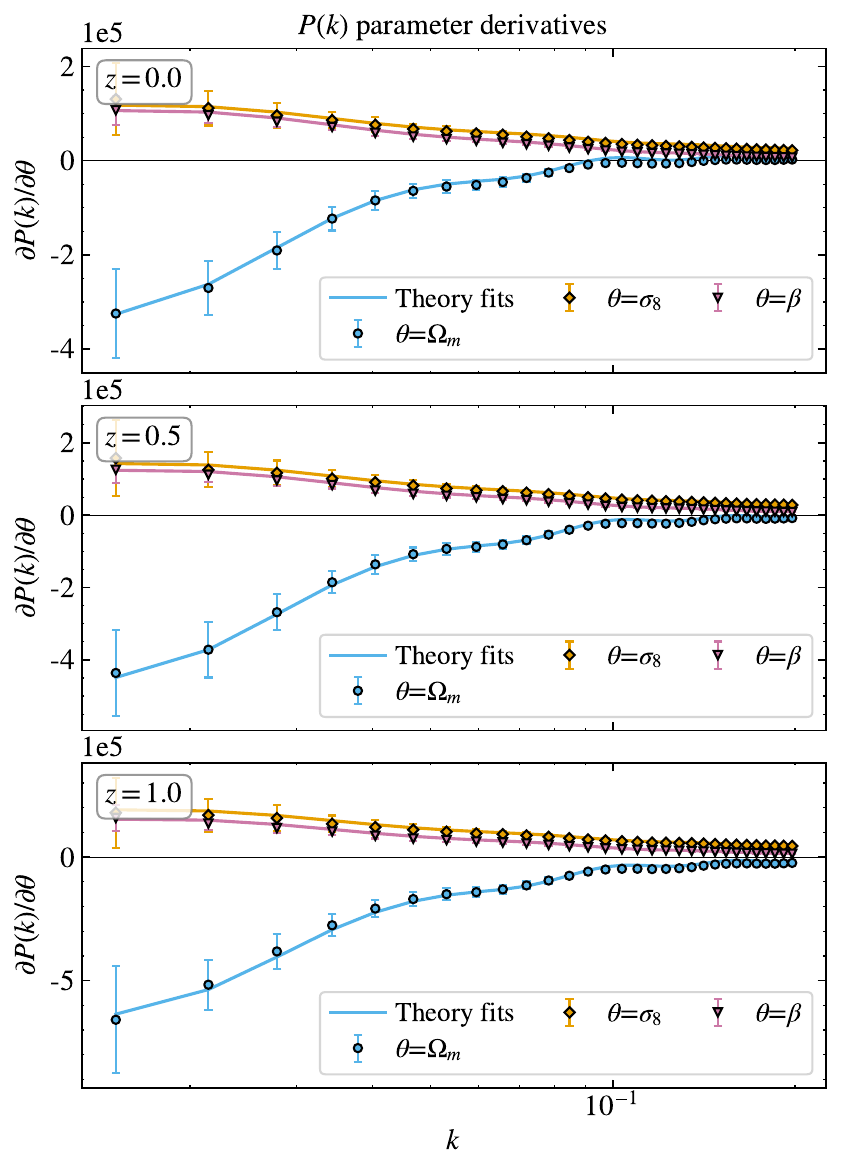}
        \caption{Halo power spectrum derivatives w.r.t. cosmological and combined bias and shot noise parameters (data points) and prediction from the simplistic two-parameter model~\eqref{eq:haloPk_model} with fixed parameters across cosmologies (solid lines).}
    \label{fig:halo_Pkderivs}
        \centering
    \includegraphics[width=1\columnwidth]{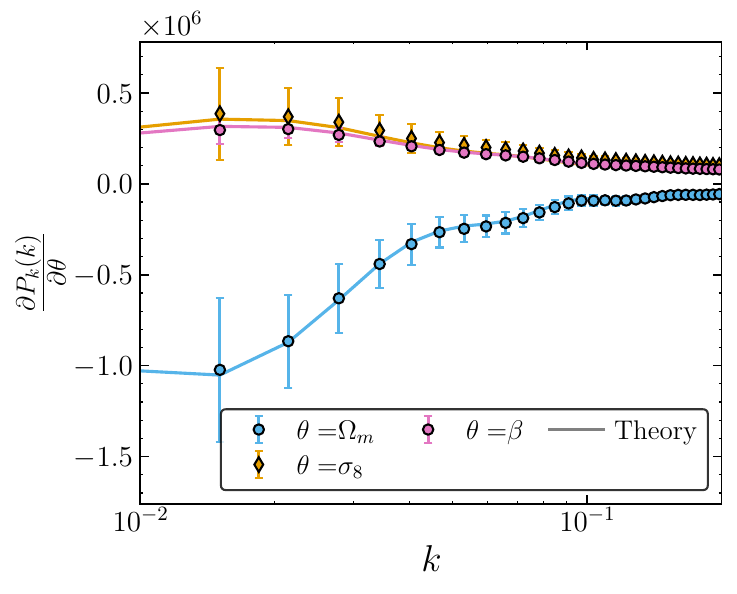}
        \caption{Galaxy power spectrum derivatives w.r.t. cosmological and combined bias and shot noise parameters. The data points represent the derivatives computed from \Molino measurements of the galaxy power spectrum. The solid lines represent the predictions from the theoretical model presented in equation \eqref{eq:haloPk_model}.}
    \label{fig:galaxy_Pkderivs}
\end{figure}

We show the performance of our simplistic halo power spectrum modelling for the fiducial cosmology in Figure~\ref{fig:haloPk_model}. The derivatives of the halo and mock galaxy power spectra with respect to cosmological and bias parameters are shown in Figures~\ref{fig:halo_Pkderivs}~and~\ref{fig:galaxy_Pkderivs}, respectively. As expected, the parameters $\sigma_8$ and $\beta$ control the amplitude of the power spectrum such that their shape closely reflects the shape of the power spectrum. The spectral index $n_s$ additionally leads to a change in the scale dependence. The matter density $\Omega_m$ modulates the large-scale power through the growth factor leading to larger differences at larger redshifts. Additionally, it modifies the Baryon Acoustic Oscillation wiggles in the matter power spectrum, which get washed out by the scale-dependent bias and shot noise. We checked that despite its simplicity, our tracer power spectrum model~\eqref{eq:haloPk_model} reproduces the cosmology and bias constraints obtained from the simulated power spectra. For the case of constraints at a single redshift, the constraints on the amplitude parameters $\sigma_8$ and $\beta$ are very weak, while the theory provides 
marginally stronger results. This becomes insignificant when adding several redshifts, as the shared cosmological parameters become more tightly constrained. This cross-validation demonstrates the convergence of the simulated power spectrum derivatives and the sufficiency of the simplistic modelling given the parameters of interest.

\newpage

\subsection{Fisher forecast: Gaussianity of data vector}
\begin{figure}
    \centering
\includegraphics[width=\columnwidth]{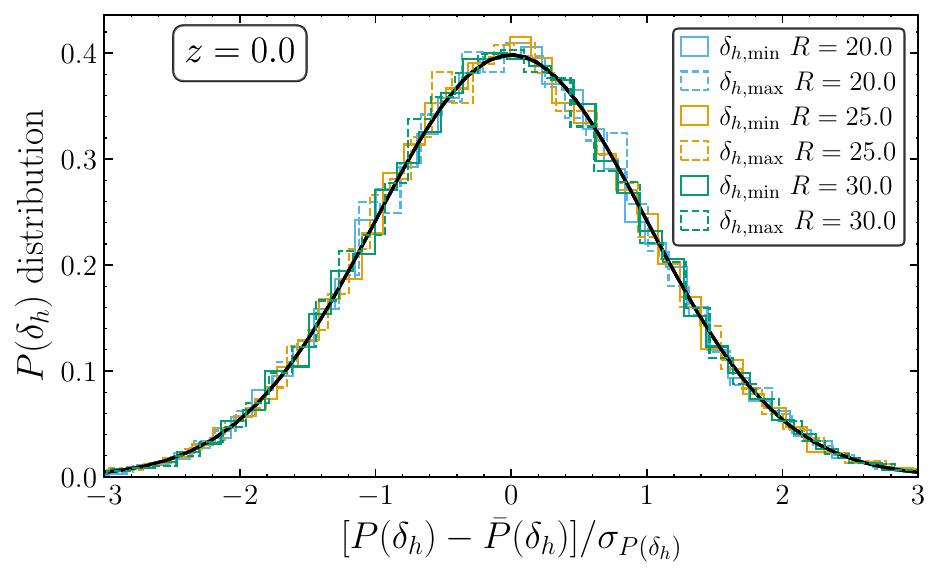}
    \caption{Gaussianity test of the halo PDF data bins. Histogram of the fluctuations of the bin values relative to the mean, divided by the standard deviation across $15,000$ realisations of the halo \Quijote simulations at redshift $z=0$. Histogram of the distribution of halo PDF values for the minimum (solid lines) and maximum (dashed lines) $\delta_h$ for three smoothing scales$R=20.0,25.0,30.0$ Mpc$/h$ (blue, orange and green lines respectively). The solid black line represents a Gaussian distribution with mean zero and unit variance. }
    \label{fig:Gaussianity_Quijote}
\end{figure}

The validity of the Fisher analysis relies on the assumption of Gaussian likelihood. Here, we verify the validity of our Fisher analysis by testing the Gaussian distribution of the bins in the halo \Quijote PDF across different realisations. We achieve this by computing histograms of the fluctuation of the bin values relative to the mean, divided by the standard deviation. Since bins located at the low and high density regions can potentially deviate from Gaussianity, Figure \ref{fig:Gaussianity_Quijote} illustrates how the halo PDF  bins for the minimum and maximum $\delta_h$ values we consider at redshift $z=0.0$ and for the three smoothing scales $R=20.0,25.0,30.0$Mpc$/h$ follow a Gaussian distribution with a mean given by the average over realisations and a variance determined by the diagonal of the covariance matrix. We also verify that other bins follow a Gaussian distribution, obtaining similar results for the other redshifts. 
To validate the full data vector and the inverse covariance used in the Fisher forecast we perform a $\chi^2-$test. We employ the following expression
\begin{equation}
\label{eq:chi2}
    \chi^2 = \frac{1}{2} \left(S_\alpha-\langle S_{\alpha} \rangle\right) C_{\alpha\beta}^{-1} \left(S_{\beta}-\langle S_{\beta} \rangle\right)\,,
\end{equation} 
and plot histograms of the $\chi^2$ values. 
\begin{figure}
    \centering
\includegraphics[width=\columnwidth]{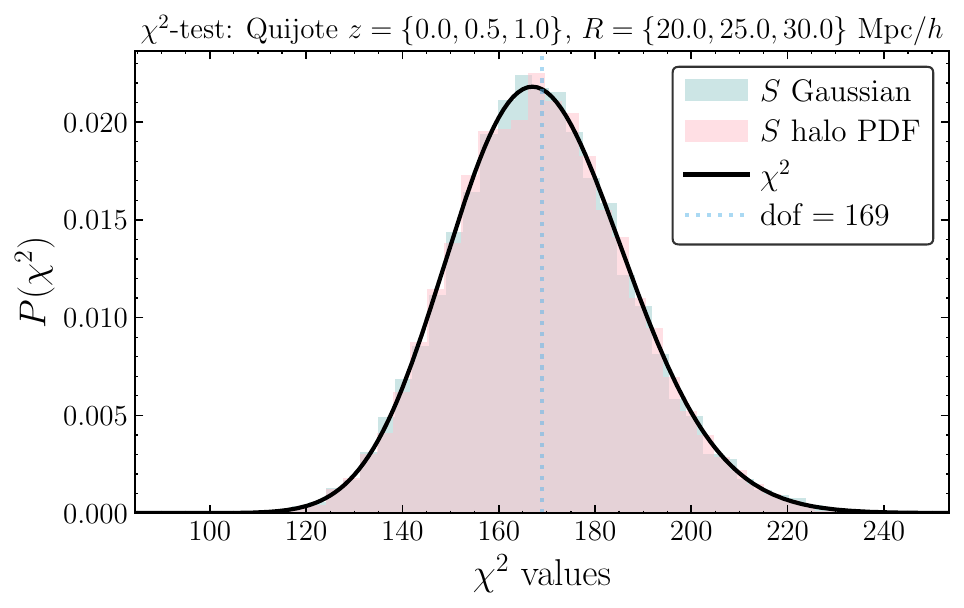}
    \caption{Distribution of $\chi^2-$values combining three redshifts $z=0.0,0.5,1.0$ and three smoothing scales $R=20.0,25.0,30.0$Mpc$/h$ for each redshift (pink histogram). The distribution follows a $\chi^2-$distribution with the same number of degrees of freedom (dof) (black line). $\chi^2-$values distribution for a data vector, build from a Gaussian distribution with same mean and covariance as data vector for the halo PDF (blue histograms).}
    \label{fig:chi2_Quijote}
\end{figure}
The data vector $\vec{S}$ combines the \Quijote halo one-point PDF at three redshifts $z=0.0,0.5,1.0$ and three smoothing scales $R=20.0,25.0,30.0$Mpc$/h$ for each redshift. The covariance is estimated using $15,000$ realisations of the fiducial \Quijote halo PDF. Figure \ref{fig:chi2_Quijote} confirms that the $\chi^2$-values follow a $\chi^2$  distribution, validating the accuracy of the inverse covariance.  Additionally, we show the $\chi^2$-values distribution considering a data vector build from a Gaussian distribution with the same mean and covariance as the \Quijote halo PDF data vector which also follows the $\chi^2$ distribution supporting the applicability of the Fisher forecast in section \ref{sec:Fisher_forecast}.

\subsection{Convergence of Fisher forecasts}
\begin{figure}
    \centering
\includegraphics[width=\columnwidth]{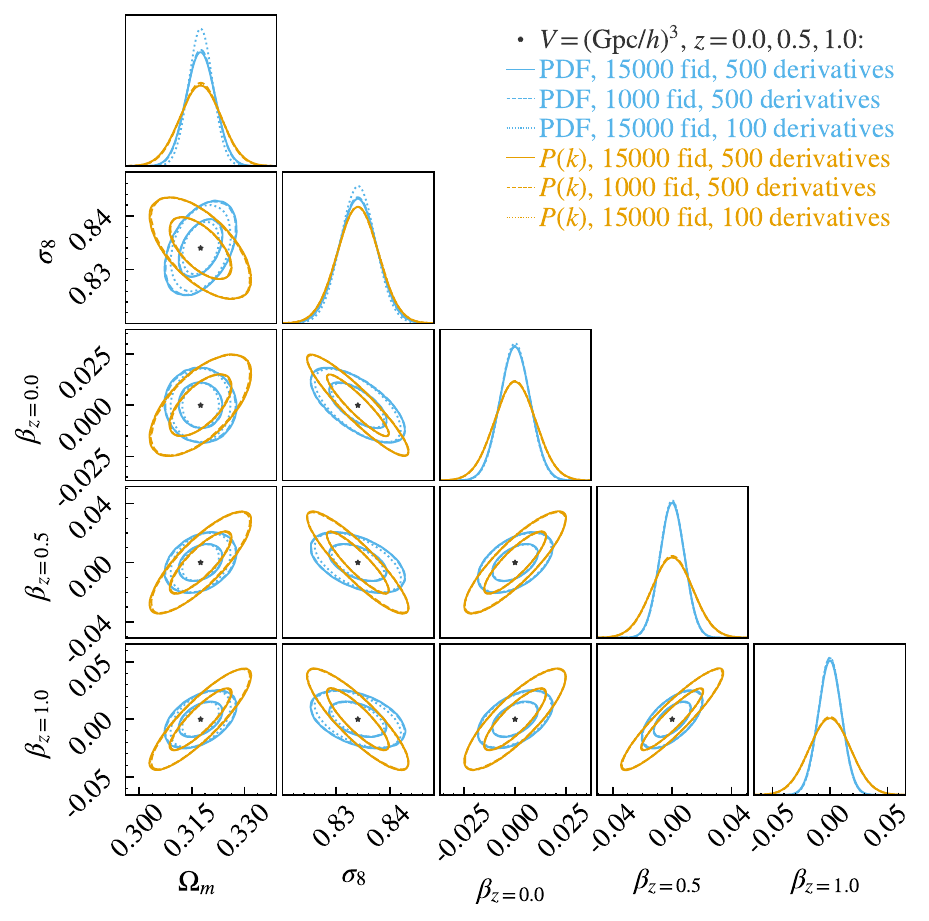}
    \caption{Fisher forecast for halo PDF (blue) and power spectrum (red) as measured from Quijote using 15000 fiducial realisations and 500 derivative realisations (solid), 1000 fiducial realisations and 500 derivative realisations (dashed) and 5000 fiducial realisations and 100 derivative realisations (dotted).}
    \label{fig:Fisher_haloPDF_Pk_convergence}
\end{figure}

In Figure~\ref{fig:Fisher_haloPDF_Pk_convergence} we check the convergence of the Fisher forecasts for the parameters $\{\Omega_m,\sigma_8,\beta_z\}$. We vary both the number of realisations used for computing the data covariance and the number of realisations used for computing the data derivatives. The effect of super-sample covariance is not included here, which accounts for the slight difference to Figure~\ref{fig:fisher_PDF_Pk}.
We find robustness with respect to reducing the realisations used for estimating the covariance. In contrast, we find a significant dependence of the obtained constraints on the number of realisations used to estimate the derivatives indicating artificial degeneracy breaking due to numerical noise. This effect would be further exacerbated when marginalising over the bias parameters, which is why we decide to reduce the number of cosmological parameters we vary. An alternative avenue would be to use the approach of \cite{CoultonWandelt2023} to obtain conservative estimates and use compression to limit the impact of numerical noise.

\label{lastpage}
\end{document}